\documentclass[reprint,twocolumn,aps,prl,amsmath,amssymb,floatfix,superscriptaddress,longbibliography]{revtex4-2}
\usepackage{graphicx}
\usepackage{epsfig}
\usepackage{bm}
\usepackage{dsfont}
\usepackage{dcolumn}
\usepackage{color}
\usepackage{physics}
\usepackage{float}
\usepackage[colorlinks,urlcolor=cyan,citecolor=blue,linkcolor=magenta]{hyperref}
\makeatletter
\newcommand*{\rom}[1]{\expandafter\@slowromancap\romannumeral #1@}
\makeatother
\usepackage{tikz}
\usepackage{subfigure}

\newcommand{\jyd}{\textcolor[rgb]{0,0.0,1}}

\begin{document}

\title{Enhanced many-body quantum scars from the non-Hermitian Fock skin effect}

\author{Ruizhe Shen}
\email{ruizhe20@u.nus.edu}
\affiliation{Department of Physics, National University of Singapore, Singapore 117542}

\author{Fang Qin}
\affiliation{Department of Physics, National University of Singapore, Singapore 117542}

\author{Jean-Yves Desaules}
\affiliation{School of Physics and Astronomy, University of Leeds, Leeds LS2 9JT, United Kingdom}
\affiliation{Institute of Science and Technology Austria (ISTA), Am Campus 1, 3400 Klosterneuburg, Austria}

\author{Zlatko Papi\'c}
\affiliation{School of Physics and Astronomy, University of Leeds, Leeds LS2 9JT, United Kingdom}

\author{Ching Hua Lee}
\email{phylch@nus.edu.sg}
\affiliation{Department of Physics, National University of Singapore, Singapore 117542}
\date{\today}

\begin{abstract}
In contrast with extended Bloch waves, a single particle can become spatially localized due to the so-called skin effect originating from non-Hermitian pumping. Here we show that in kinetically-constrained many-body systems, the skin effect can instead manifest as dynamical amplification within the Fock space, beyond the intuitively expected and previously studied particle localization and clustering. We exemplify this non-Hermitian Fock skin effect in an asymmetric version of the PXP model and show that it gives rise to ergodicity-breaking eigenstates – the non-Hermitian analogs of quantum many-body scars. A distinguishing feature of these non-Hermitian scars is their enhanced robustness against external disorders. We propose an experimental realization of the non-Hermitian scar enhancement in a tilted Bose-Hubbard optical lattice with laser-induced loss. Additionally, we implement digital simulations of such scar enhancement on the IBM quantum processor. Our results show that the Fock skin effect provides a powerful tool for creating robust non-ergodic states in generic open quantum systems.
\end{abstract}
\maketitle


{\bf \em Introduction.---}The physics of quantum thermalization, which describes how a system reaches an equilibrium state, has garnered substantial attention in recent years~\cite{srednicki1994chaos,rigol2008thermalization,rigol2009breakdown,banuls2011strong,d2016quantum,mori2018thermalization,deutsch2018eigenstate,mallayya2019prethermalization}. A key concept -- the eigenstate thermalization hypothesis~\cite{deutsch1991quantum,srednicki1994chaos,dymarsky2018subsystem,deutsch2018eigenstate} -- indicates that closed systems with unbroken ergodicity invariably reach an equilibrium state at late times. However, not all systems behave this way, a notable exception being the observation of persistent oscillations in a 51 Rydberg atom quantum simulator~\cite{bernien2017probing}. This form of ergodicity breaking has been attributed to non-thermalizing eigenstates known as quantum many-body scars (QMBSs)~\cite{turner2018weak,ho2019periodic} (see also the reviews~\cite{serbyn2021quantum,MoudgalyaReview,ChandranReview}).
In contrast to integrable models~\cite{sutherland2004beautiful} or strongly disordered systems displaying many-body localization~\cite{Nandkishore_review,Abanin_review}, typical QMBS systems weakly break ergodicity due to the non-thermal eigenstates forming a small-dimensional representation of a ``restricted spectrum generating algebra''~\cite{moudgalya2018entanglement,Iadecola2019_2,choi2019emergent,MarkLinMotrunich,Omiya2022}, although more general mechanisms continue to be explored~\cite{Dea2020, Pakrouski2020, MoudgalyaCommutant, Buca2023}.

The experimental observations of QMBSs in synthetic platforms ranging from ultracold atoms to superconducting circuits~\cite{bernien2017probing, bluvstein2021controlling, jepsen2020spin, scherg2021observing, su2022observation, zhang2022many,kohlert2021experimental}
have revealed the intrinsic fragility of QMBS dynamical revivals.  In some cases, this has been mitigated by periodic driving, which can extend the lifetime of QMBS oscillations~\cite{bluvstein2021controlling, su2022observation, maskara2021discrete, hudomal2022driving,Halimeh2023robustquantummany}. Nevertheless, thermalization or decay to a steady state, due to inherent decoherence from atom loss or thermal fluctuations, ultimately sets in. Thus, achieving stable QMBSs in real-world settings, with unavoidable non-unitarity, remains a challenge~\cite{shishkov2018zeroth,reichental2018thermalization,shirai2020thermalization,chen2022non,suthar2022non}.

In this paper, we show that non-Hermiticity presents a new route for enhancing ergodicity breaking and many-body scarring. Borrowing inspiration from non-Hermitian pumping, which breaks the conventional bulk-edge correspondence~\cite{longhi2019probing,song2019non,lee2020unraveling,li2020topological,okuma2020topological,zhang2021observation,zhang2021acoustic,lee2021many,li2021impurity,li2022non,jiang2022dimensional,tai2022zoology,zhang2022review,lei2023pt}, we demonstrate that a class of many-body systems with kinetic constraints admit a form of ``Fock skin effect'' that amplifies the revivals of special initial states and makes them resilient against spatial disorder. Unlike the conventional non-Hermitian skin effect (NHSE), which can induce nontrivial edge localization in real space, the Fock skin effect is characterized 
by persistent dynamical recurrences in the Fock space of a many-body system. We demonstrate this mechanism of QMBS enhancement in a non-Hermitian variant of the PXP model, which can be realized in a Bose–Hubbard optical lattice with laser-induced loss~\cite{lapp2019engineering,ren2022chiral,liang2022observation,qin2022non,shen2023proposal}. 
Connection with conventional non-Hermitian pumping is made through the ``forward scattering approximation''~\cite{turner2018weak, turner2018quantum, choi2019emergent}, which effectively maps the many-body system to a spatially-inhomogeneous tight-binding chain with non-reciprocal hopping amplitudes. Such QMBS enhancement is further demonstrated in the digital simulation conducted on the IBM Q device without deliberate error mitigation. Our results offer a new perspective on obtaining stable many-body revivals through dissipative perturbations, extending beyond recent approaches based on decoherence-free subspaces~\cite{Buca2019,wang2023embedding}.

\begin{figure}[tb]
	\centering
	\includegraphics[width=0.99\linewidth]{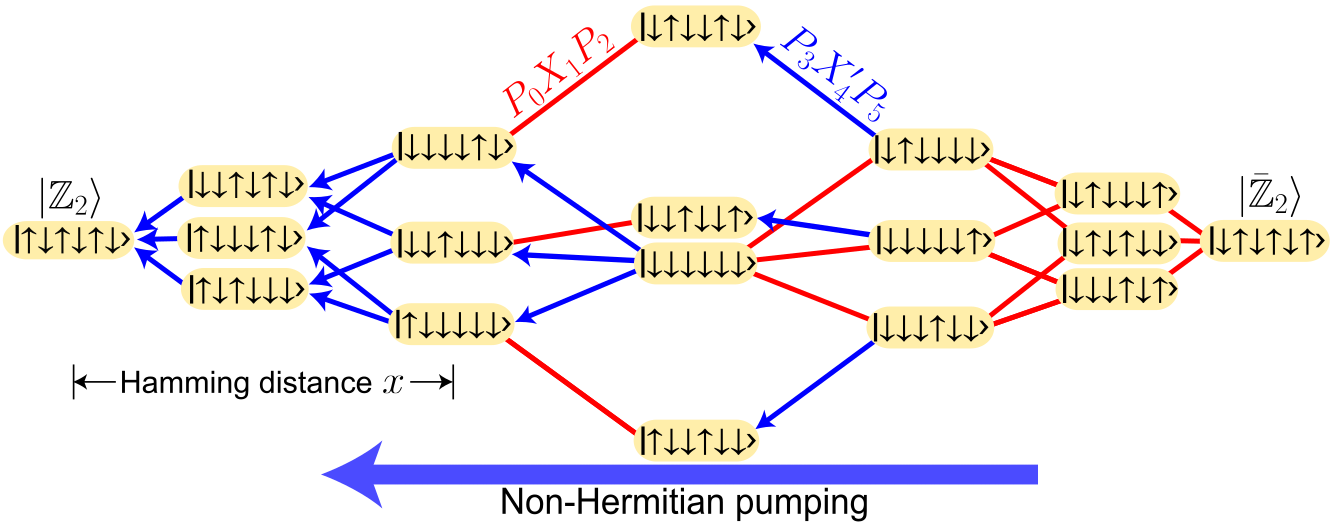}
	\caption{
     The directed Fock-space graph of the non-Hermitian PXP model in Eq.~\eqref{nhpxp} with $L{=}6$ sites under PBCs. Each vertex represents a Fock state, organized into layers by the Hamming distance $x$ from the state $\ket{\mathbb{Z}_2}$ on the left. The translated $\ket{\bar{\mathbb{Z}}_2}$ state is on the opposite end. Edges represent the transitions allowed by the Hamiltonian in Eq.~\eqref{nhpxp}. 
     Red links represent the symmetric coupling $\hat{P}_{j-1} \hat{X}_{j} \hat{P}_{j+1}$ (example of $P_0X_1P_2$ shown). 
     Blue directed arrows label the asymmetric couplings $\hat{P}_{j-1} \hat{X}^{\prime}_{j} \hat{P}_{j+1}$ (example of $P_3 X_4^\prime P_5$ shown), with the arrow denoting enhanced spin flipping $(1{+}u)\hat{X}^{+}_{j}$. 
     All directed edges point towards $x{=}0$, denoting asymmetric pumping from the Fock skin effect. 
	}
	\label{fig:figure1}
\end{figure}

{\bf \em Fock skin effect.---}We consider the following non-Hermitian generalization of the one-dimensional (1D) PXP model~\cite{FendleySachdev,Lesanovsky2012}, a paradigmatic model of QMBSs~\cite{turner2018weak,ho2019periodic}:
\begin{equation}\label{nhpxp}
	\hat{H}{=}\hspace{-0.25cm}\sum_{j\in {\rm even}}\hspace{-0.18cm}\hat{P}_{j-1} \hat{X}^{\prime}_{j} \hat{P}_{j+1}{+}\hspace{-0.2cm}\sum_{j\in {\rm odd}}\hspace{-0.15cm}\hat{P}_{j-1} \hat{X}_{j} \hat{P}_{j+1},
\end{equation}
where $\hat{X}$, $\hat{Y}$, $\hat{Z}$ denote the standard Pauli matrices, with  $\hat{P}{=}(\mathds{1}{-}\hat{Z})/2$ projecting onto the spin-$\downarrow$ state, $j=0,1,\ldots,L-1$ and $L $ is the number of spins. Non-Hermiticity enters through the asymmetric spin flip operator, $\hat{X}^{\prime}{=}\hat{X}{+}iu\hat{Y}$, acting on the even sites, FIG.~\ref{fig:figure1}, where $u$ controls the strength of non-Hermiticity. Each $\hat X$ or $\hat X'$ spin flip acts only when its neighboring spins are both spin-$\downarrow$.
To distinguish our Fock skin effect from conventional skin effects with spatial boundary accumulation, we use periodic boundary conditions (PBCs). We also restrict $0{<}u{<}1$ to keep the energy spectrum real and avoid wavefunction decay. More results on  the irrelevance of physical boundaries and level spacing distribution are given in Supplementary Material (SM)~\cite{SuppMat}.

To provide insight behind the non-Hermitian scar enhancement, we analyze the wavefunction dynamics within the Fock space. The latter can be visualized as a quasi-1D network structure by grouping Fock states according to their Hamming distance, which measures the length of the shortest path from any Fock state and the N\'eel state, $\ket{\mathbb{Z}_2} = \ket{\uparrow\downarrow\uparrow\downarrow\cdots}$, see an example in FIG.~\ref{fig:figure1}. Note that the $\ket{\mathbb{Z}_2}$ state is the one that exhibits QMBS revivals in the Hermitian PXP model~\cite{turner2018weak}. The Hamming distance indicates how many applications of the Hamiltonian are needed to reach the states in a given layer starting from the $\ket{\mathbb{Z}_2}$ state.  This is because the Hamiltonian only connects layer $x$ with $x\pm 1$, as shown by edges in FIG.~\ref{fig:figure1},
making apparent the connection with a tight-binding chain. 
Crucially, this picture reveals a consistent direction in the asymmetry of the transitions between Fock states. As illustrated, the blue directional transitions are \emph{all} directed towards the $\ket{\mathbb{Z}_2}$ state at $x{=}0$, reminiscent of a conventional NHSE chain but with Fock states instead of physical sites. However, with multiple Fock states in each layer instead of a single site, the ``boundary'' skin localization would be weak~\cite{SuppMat}, and we instead expect more pronounced QMBS revivals, rather than the directional bulk current in conventional NHSE systems~\cite{song2019non,okuma2020topological,qin2023universal,qin2023kinked}.

The stratification of the Fock space into Hamming layers can be formally expressed by generalizing the forward scattering approximation (FSA)~\cite{turner2018weak,turner2018quantum}, which allowed to understand the origin of QMBSs in the Hermitian case~\cite{choi2019emergent}. Here, we demonstrate that this method can approximate the behavior of scar states in our non-Hermitian setup.
We first decompose our non-Hermitian PXP Hamiltonian in Eq.~\eqref{nhpxp} as $\hat H {=} \hat H^+ {+} \hat H^-$, where
\begin{equation}\label{lad}
	\begin{aligned}
		\hat{H}^{+}=&\sum_{j\in {\rm even}}\hat{P}_{j-1} \hat{X}^{\prime-}_{j} \hat{P}_{j+1}\!+\!\sum_{j\in {\rm odd}}\hat{P}_{j-1} \hat{X}^{+}_{j} \hat{P}_{j+1},\\
		\hat{H}^{-}=&\sum_{j\in {\rm even}}\hat{P}_{j-1} \hat{X}^{\prime+}_{j} \hat{P}_{j+1}\!+\!\sum_{j\in {\rm odd}}\hat{P}_{j-1} \hat{X}^{-}_{j} \hat{P}_{j+1},
	\end{aligned}
\end{equation}
with $\hat{X}^{\pm}{\equiv}(\hat{X}\pm i\hat{Y})/2$ and $\hat{X}^{\prime\pm}{\equiv}(1\pm u)(\hat{X}\pm i\hat{Y})/2$. By successively acting with $\hat H^+$ on the $\ket{\mathbb{Z}_{2}}$ state, which we label as $\ket{0}_{\rm FSA}$, we can generate a tower of $L+1$ FSA basis states $\ket{n}_\mathrm{FSA} {\propto} (H^{+})^n \ket{0}_\mathrm{FSA}$, with $n=0,1,2,\ldots, L$~\cite{turner2018quantum}. These states span the Fock-space layers in FIG.~\ref{fig:figure1}. In the FSA basis, the projected Hamiltonian~\eqref{nhpxp} is given by
\begin{equation}\label{fsah}
	\hat{H}_{\rm F S A}\!\!=\!\!\sum_{x=0}^{L}t^{-}_{x}|x\rangle_{\rm FSA}\langle x+1|_{\rm FSA}+t^{+}_{x}|x+1\rangle_{\rm FSA}\langle x|_{\rm FSA},
\end{equation}
with the non-uniform and asymmetric hopping amplitudes
$t^{+}_{x} = \bra{x\!+\!1}_{\rm FSA}\!\!\hat{H}^{+}\!\!\ket{x}_{\rm FSA}$ and $t^{-}_{x} = \bra{x}_{\rm FSA}\!\!\hat{H}^{-}\!\!\ket{x\!+\!1}_{\rm FSA}$.
Despite a formal analogy between the model~\eqref{fsah} and a tight-binding chain, each FSA basis state $\ket{x}_\mathrm{FSA}$ represents a superposition of all Fock states in the Hamming layer $x$. Moreover, both the basis states and the effective hoppings $t^\pm$ intricately depend on the non-hermiticity $u$ of the Hamiltonian, as we show below. 

\begin{figure*}
	\centering
	\includegraphics[width=1\linewidth]{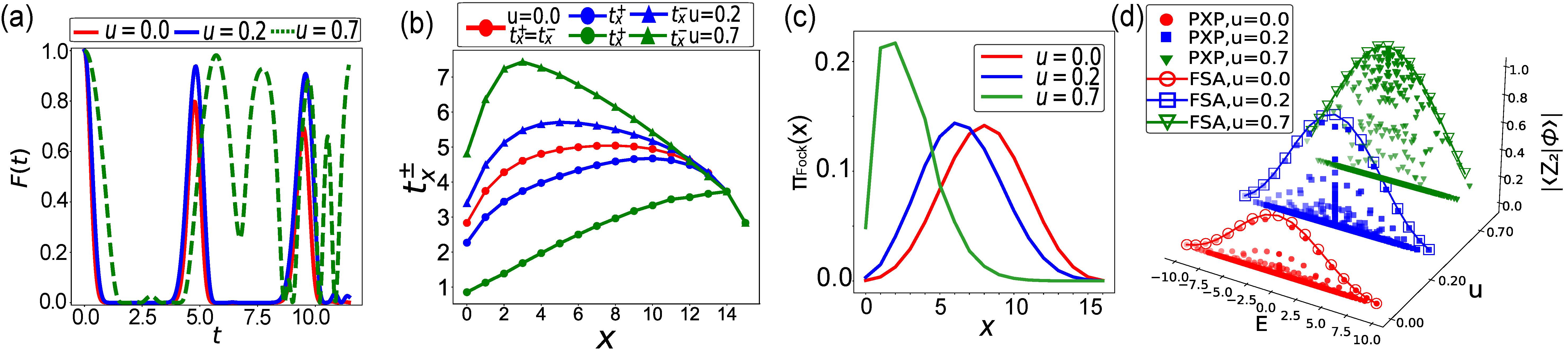}
	\caption{ (a) Normalized fidelity $F(t)$ for the initial state $\ket{\mathbb{Z}_{2}}$ in the non-Hermitian PXP model \eqref{nhpxp}. 
Moderate non-Hermiticity $u\neq 0$ significantly enhances the return probability to the initial state. For large $u$ (dashed green curve), robust but irregular revivals occur due to multiple frequencies in the dynamics.
		(b) Amplitudes of effective hoppings $t_x^\pm$. The asymmetry between $t^{\pm}_{x}$ for $u{>}0$ (circles vs. triangles) leads to the non-Hermitian Fock skin effect in our model. 
		(c) The concentration of eigenstates in the Hamming layer $x$, Eq.~\eqref{ipr}. A sharp Fock skin mode emerges under strong non-Hermiticity (green curve), corresponding to the persistent revivals in (a). (d) Comparison between the FSA left-boundary localization, $|\bra{\mathbb{Z}_{2}}\ket{\phi_{j}}_{\rm FSA}|$, and overlaps of $\ket{\mathbb{Z}_{2}}$ with all eigenstates of the Hamiltonian in Eq.\eqref{nhpxp}. 
		From $u{=}0.0$ to $u{=}0.2$, the FSA boundary localization closely matches the overlap of $\mathbb{Z}_{2}$ state with the equally-spaced QMBS states. All data is for a chain of size $L{=}16$ with PBCs.
	}
	\label{fig:fig3}
\end{figure*}

{\bf \em QMBS enhancement due to Fock skin effect.---}QMBSs manifest as pronounced revivals in the fidelity, $F(t)\equiv |{\bra{\psi(0)}\ket{\psi(t)}}|^{2}/{\bra{\psi(t)}\ket{\psi(t)}}$ with $\ket{\psi(t)}=\exp(-i\hat H t)\ket{\psi(0)}$, from the $\ket{\psi(0)}=\ket{\mathbb{Z}_{2}}$ state for PXP-type models~\cite{bernien2017probing,turner2018weak}. 
Strong revivals in $F(t)$ indicate broken ergodicity since the fidelity should decay rapidly under thermalizing dynamics~\cite{turner2018weak,ho2019periodic}. In FIG.~\ref{fig:fig3}(a), the fidelity $F(t)$ is plotted for the PXP model \eqref{nhpxp} with various degrees of non-Hermiticity $u$. The decay of peaks in $F(t)$ is noticeably slower for $u{=}0.2$ (blue) compared to the Hermitian $u{=}0$ case (red). With much larger non-Hermiticity $u{=}0.7$ (green dashed), both frequency and magnitude of the revivals are significantly enhanced, showing no significant decay even after $t{=}10$.

To explicitly attribute this non-Hermitian scar enhancement to the Fock skin effect, we plot the hopping amplitudes $t_x^\pm$ in FIG.~\ref{fig:fig3}(b). Saliently, the FSA hoppings are asymmetric, with the disparity between $t_x^+$ (round markers) and $t_x^-$ (triangle markers) becoming more pronounced with greater non-Hermiticity $u$. This asymmetry is markedly different from that in the physical Hamiltonian~\eqref{nhpxp}, since $x$ is the Hamming distance from the $\ket{\mathbb{Z}_2}$ state in Fock space and not a spatial coordinate. Importantly, these hopping amplitudes are non-trivially dependent on $x$ but consistently asymmetric ($t_x^+ \geq t_x^-$) along the FSA chain. This property is not generic, see SM~\cite{SuppMat} for the discussion of a related model where this is not the case, with crucial implications on the existence of Fock skin accumulation.

The FSA non-uniformity and the large dimensionality of each layer $x$ lead to a unique Fock skin effect, distinct from typical skin localization. This is elucidated through the Hamming distance-resolved Fock space density $\Pi_{\rm Fock}(x)$ which quantifies the total weight of all eigenstates at a Hamming distance $x$ from the $\ket{\mathbb{Z}_2}$ state:
\begin{equation}\label{ipr}
	\begin{aligned}
		\Pi_{\rm Fock}(x)=\frac{1}{\mathcal{D}}\sum_{\ket{\varphi} \in \mathcal{L}(x)}\sum_{i}|\bra{\varphi}\ket{\phi_{i}}|^{2},
	\end{aligned}
\end{equation}
with $\mathcal{L}(x)$ denoting the set of all Fock states in the Hamming layer $x$~\cite{yao2023observation}, $\ket{\phi_{i}}$ is the $i$-th eigenstate, and $\mathcal{D}$ is the dimension of the Hilbert space.

As seen in FIG.~\ref{fig:fig3}(c), the Fock skin pumping of $u{=}0.2$ shifts the symmetric $\Pi_{\rm Fock}(x)$ profile of $u{=}0$ slightly leftward, yet without pronounced ``boundary'' localization. Nevertheless, this increased proximity to the $\ket{\mathbb{Z}_2}$ state is sufficient to noticeably enhance the scarred revivals [FIG.~\ref{fig:fig3}(a), blue]. Stronger non-Hermiticity  $u{=}0.7$  results in a more skewed  $\Pi_{\rm Fock}(x)$, but less localized than conventional skin states with $t_x^+/t_x^-{\approx} 3$ [FIG.~\ref{fig:fig3} (b), green], with localization lengths around  $[\log 3]^{-1}{\approx} 1$ site. Crucially, this lack of conventional exponential boundary skin localization allows for significant state diffusion in the Hilbert space, but with enhanced return probability to the $\ket{\mathbb{Z}_2}$ state.

We next explore how shifted Fock skin profiles amplify QMBS revivals. Within the FSA framework, FIG.~\ref{fig:figure1}, scarred revivals oscillate from the left FSA boundary $\ket{0}_{\rm FSA}\equiv \ket{\mathbb{Z}_2}$ to the opposite end, $\ket{L}_{\rm FSA}\equiv \ket{\bar{\mathbb{Z}}_2}$. The revivals are driven by a branch of eigenstates $\ket{\phi}$ with approximately equally spaced eigenenergies $E$ and high overlaps $|\bra{\mathbb{Z}_{2}}\ket{\phi}_{\rm FSA}|$, as depicted in FIG.~\ref{fig:fig3}(d). These FSA overlaps (hollow markers) closely align with the overlaps of $\ket{\mathbb{Z}_2}$ with the eigenstates of our original PXP Hamiltonian \eqref{nhpxp} (solid markers), demonstrating the accuracy of the FSA description. In particular, the Fock skin effect boosts the $\ket{\mathbb{Z}_2}$ overlap of the FSA branch at $u{=}0.2$, consequently enhancing revivals over the Hermitian case at $u{=}0.0$. Greater non-Hermiticity, e.g., $u{=}0.7$, enhances the $\ket{\mathbb{Z}_{2}}$ overlap even more but also introduces other non-thermal states with high overlaps, such that additional frequencies appear in the revival dynamics -- see the green curve in FIG.~\ref{fig:fig3}(a).

\begin{figure}[b]
	\centering
	\includegraphics[width=1\linewidth]{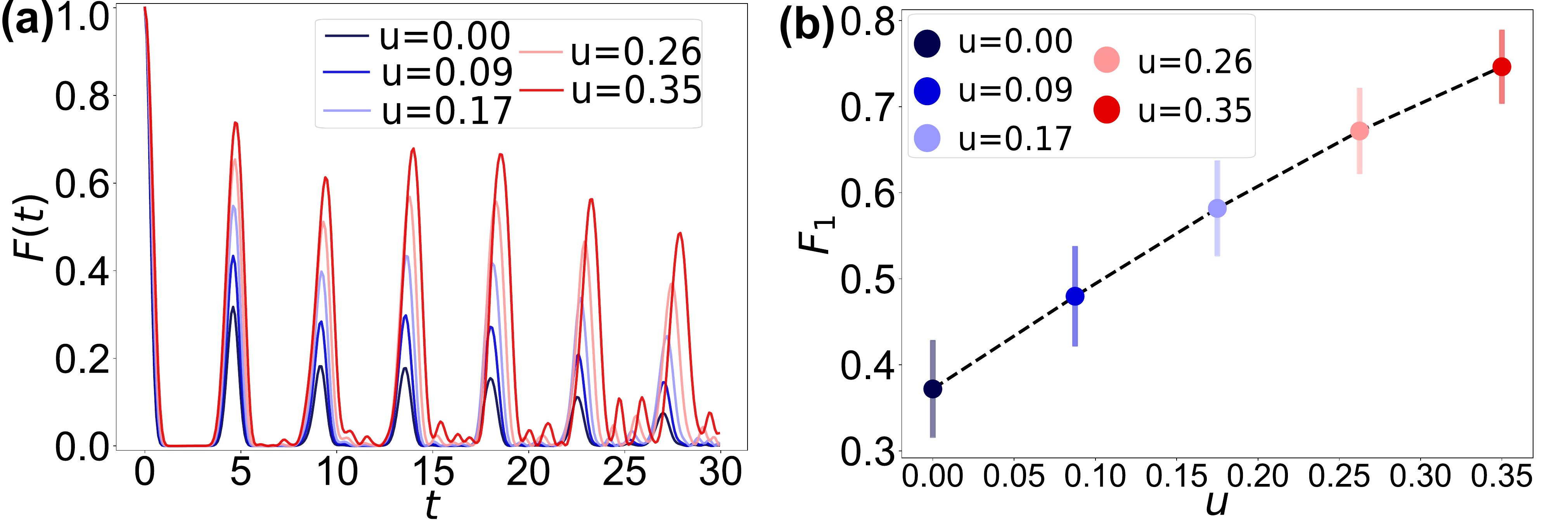}
	\caption{(a) Fidelity dynamics for the $\ket{\mathbb{Z}_{2}}$ initial state evolving under the disordered Hamiltonian $\hat{H}_{\rm dis}$ at several values of $u$. At $u{=}0$, scarring signatures swiftly collapse at moderate disorders, while the Fock skin effect leads to prolonged and robust scarring recurrences up to $u{=}0.35$. (b) The first fidelity peak $F_{1}$ within $t\in[2.5,7.5]$, which is significantly shifted by non-Hermiticity $u$. Error bars represent the standard deviation over $200$ realizations. All data is for system size $L{=}14$ and uniform disorder strength of $W=0.8$.
	}
	\label{fig:fig4}
\end{figure}
{\bf \em Robustness to disorder.---} We have established that the Fock skin effect can significantly enhance QMBSs. Given that the real-space skin effect can result in robust, directed dynamical currents, we posit that the non-Hermiticity in our system can not only amplify the quantum scar but also protect it against external (Hermitian) perturbations. 
To check this, we investigate the robustness of quantum scar under  spatial disorder, 
$\hat{H}_{\rm dis}=\hat{H}+\sum_{j}w_{j}\hat{Z}_{j}$,
where $w_{j}$ denotes the strength of on-site random potential drawn from a uniform distribution within the range $[-W/2,W/2]$. It was previously shown that Hermitian QMBS revivals decay with increasing disorder strength~\cite{mondragon2021fate}. For our model with the Fock skin effect at up to $u{=}0.35$, this resilience under moderate disorders is remarkably highlighted by the fidelity shown in FIG.~\ref{fig:fig4}(a), where robust scarring signatures are still present, in contrast to the rapid collapse at Hermitian $u{=}0$ without the Fock skin effect. The enhanced stability is demonstrated through the first fidelity peak, $F_{1}={\rm Max}[F(t)]$ with $t\in[2.5,7.5]$, as shown in FIG.~\ref{fig:fig4} (b). $F_1$ as well as subsequent peaks are markedly enhanced with increasing $u$.

\begin{figure}[tb]
	\centering
	\includegraphics[width=1\linewidth]{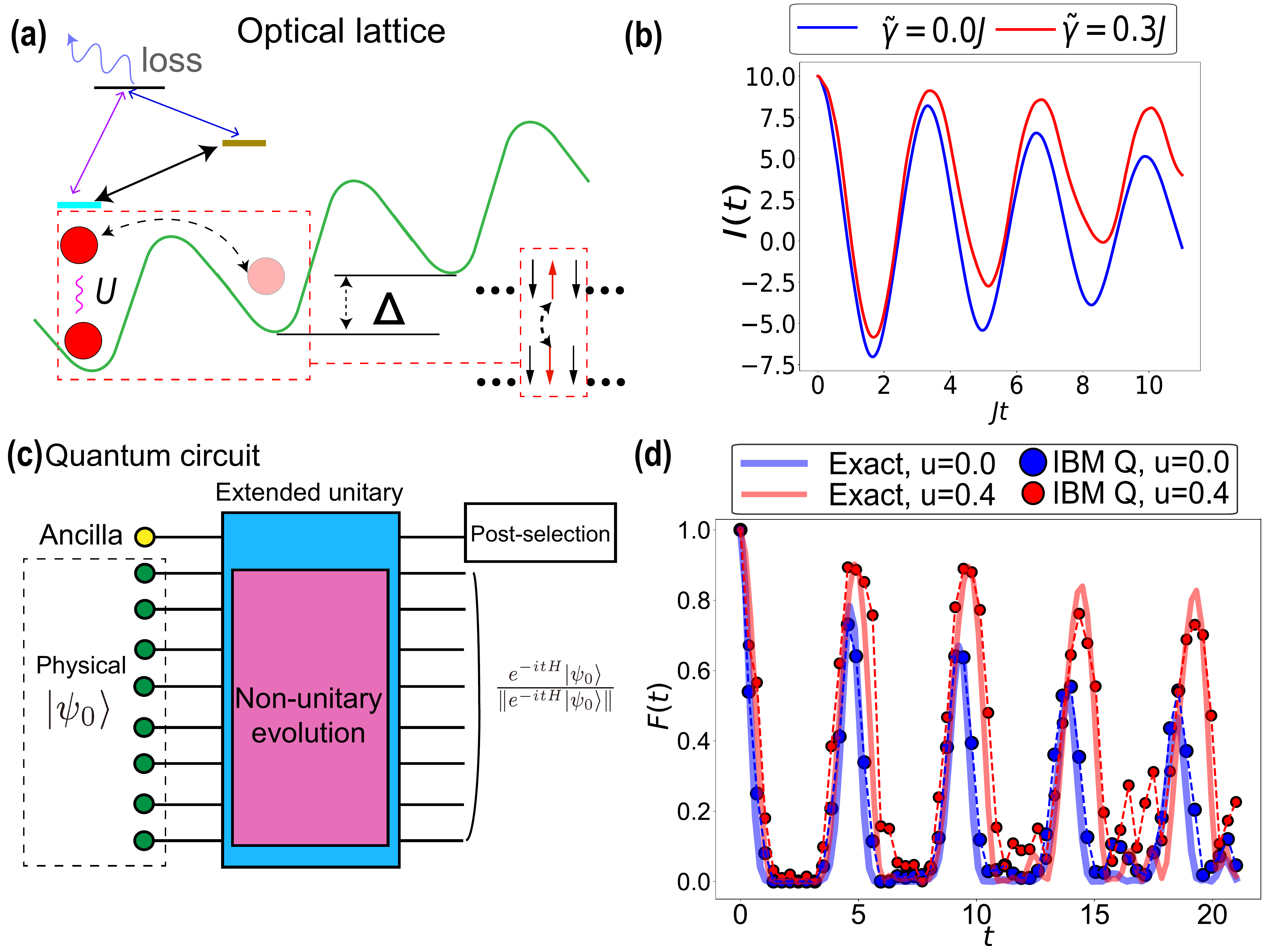}
	\caption{
(a) Schematics of the non-Hermitian Bose-Hubbard model~\eqref{BH} in a tilted optical lattice.
		The non-reciprocal hopping $J\pm \tilde{\gamma}$ (black arrow) is realized by coupling (blue and violet double arrows) two nearest ground states with an excited state with the laser-induced loss (wavy arrow) on the excited state. The on-site interaction $U$ and tilt potential $\Delta$ lead to kinetic constraints under $U = \Delta \gg J,\tilde{\gamma}$~\cite{su2022observation}. This results in the effective non-Hermitian PXP model in Eq.\eqref{nhpxp}, and boson hopping is mapped to spin flipping (red block) \cite{SuppMat}.
		(b) Classical simulations of population imbalance $I(t)$ for the non-Hermitian Bose-Hubbard model~\eqref{BH}. The initial state is $\ket{20202020201}$ with $N=L=11$. 
		Other parameters are $V=\Delta=10J$, and $\tilde{\gamma}/J=0.0$ (blue) and $0.3$ (red). The red curve exhibits pronounced revivals due to enhanced quantum scars. (c) Quantum circuit for simulating the model Eq.~\eqref{ibmqmodelmain}. The non-unitary operation $e^{-itH}$ (pink block) is embedded in an extended unitary (blue block), with physical qubits (green) coupled to the yellow ancilla qubit. The post-selection of $\ket{\uparrow}$ on this ancilla qubit gives the final normalization.  (d) Digital simulations of normalized fidelity $F(t)$ using the circuit in (c). We use the IBM Q Brisbane device, with noise conditions shown in SM~\cite{SuppMat}.  The initial state is $\ket{\uparrow\downarrow\uparrow\downarrow\uparrow\downarrow\uparrow\downarrow}$.  Both numerical results (solid curves) and simulation data (circles) show excellent agreement. Notably, the blue squares ($u=0.4$) demonstrate the robustness of enhanced scars under device noise.
	} 
	\label{fig:fig1}
\end{figure}

{\bf \em Optical lattice proposal.---}The Fock skin effect and non-Hermitian QMBSs can be observed in a Bose-Hubbard (BH) quantum simulator with additional laser-induced loss~\cite{baier2016extended,tomita2017observation,yang2020,su2022observation}. The setup is illustrated in FIG.~\ref{fig:fig1} (a) and it corresponds to the Hamiltonian
\begin{equation}\label{BH}
	\begin{aligned}
		\hat{H}_{\mathrm{BH}}\!\!&=\!\!-\!\!\!\!\!\sum_{j=0}^{L/2-1}\!\![\!\left.\left(J\!\!+\!\tilde{\gamma}\right) \hat{b}_{2 j}^{\dagger} \hat{b}_{2 j\!+\!1}\!+\!\!\right. 
		\left(J\!\!-\!\tilde{\gamma}\right) \hat{b}_{2 j\!+\!1}^{\dagger} \hat{b}_{2 j}
		\!\!+\!\!J\hat{b}_{2 j\!+\!2}^{\dagger} \hat{b}_{2 j\!+\!1}\!\!\\&+J\hat{b}_{2 j+1}^{\dagger} \hat{b}_{2 j+2}]+\!\!\!\sum_{j=0}^{L-1} [\Delta j \hat{n}_{j}\!\!+\!\!\frac{U}{2}\hat{n}_{j}\left(\hat{n}_{j}\!-\!1\right)],
	\end{aligned}
\end{equation}
where $\hat{b}_{j}^{\dagger}$ ($\hat{b}_{j}$) denotes the boson creation (annihilation) operator on
site $j$, the occupation number $\hat{n}_{j} = \hat{b}^{\dagger}_{j}\hat{b}_{j}$, and we use open boundary conditions (OBCs), see SM~\cite{SuppMat} for a derivation. The nonreciprocity $J \pm \tilde\gamma$ results from coupling a dissipative excited state (wavy arrow in FIG.~\ref{fig:fig1} for laser-induced atom loss) to its nearest ground states~\cite{qin2022non}.

The tilt $\Delta$ and on-site repulsion $U$ induce kinetic constraints under the condition $U = \Delta \gg J,\tilde \gamma$ and with the boson filling factor set to $\nu = N/L = 1$, where $N$ is the total number of bosons~\cite{su2022observation,SuppMat}. In this regime, the three-boson occupancy of any site is forbidden and the only allowed dynamical processes consist of local hops $\cdots 20 \cdots \leftrightarrow \cdots 11 \cdots$. The latter can be mapped to the Pauli $\hat X$ operator~\cite{sachdev2002mott,sengupta2021phases}, as shown by the red box in Fig.~\ref{fig:fig1} (a). In the resulting effective non-Hermitian PXP model~\eqref{nhpxp}, the $\ket{\mathbb{Z}_2}$ state simply maps to a product state of doublons separated by empty sites, $\ket{2020\ldots}$.

In FIG.~\ref{fig:fig1}(b) we numerically compute the dynamics of boson imbalance, $I(t)=(1/N){\sum_{i} \left(n_{2i}(t){-}n_{2i+1}(t)\right)}$, when the system is prepared in the $\ket{2020\ldots}$ state, with the normalized occupations 
$n_{i}(t)=\bra{\psi(t)}\hat{n}_i\ket{\psi(t)}/\bra{\psi(t)}\ket{\psi(t)}$.
$I(t)$ is proportional to the 
$\ket{\mathbb{Z}_2}$ echo of the effective spins. The results in FIG.~\ref{fig:fig1}(b) with $U = \Delta = 10J$ show that the non-Hermitian case with $\tilde{\gamma} = 0.3 J$ (red) exhibits stronger revivals of $I(t)$ than the Hermitian case with vanishing $\tilde{\gamma}$ (blue).

{\bf \em Digital simulation.---}Quantum computers emerge as a promising platform due to their great programmability  ~\cite{preskill2018quantum,preskill2023quantum}, allowing for the design of sophisticated models.
As such, utilizing the IBM quantum processor \cite{santos2016ibm}, we implement the digital simulation of the dynamics of the following Ising model under:
\begin{equation}\label{ibmqmodelmain}
	\hat{H}_{V}= V \sum^{L-2}_{j=0} (\mathds{1}-\hat P_j) (\mathds{1}-\hat P_{j+1}) + \sum_{j\in\mathrm{even}}\hat{X}^{\prime}_{j} +  \sum_{j\in\mathrm{odd}} \hat{X}_{j},
\end{equation}
where a strong interaction $V=10$ is chosen to suppress neighboring $\uparrow$-spins, allowing us to recover the non-Hermitian PXP model. 
We use non-unitary postselection to embed non-unitary evolution $e^{_it \hat H_V}$ in an extended unitary with an additional ancilla qubit \cite{lin2021real,chen2022high,shen2023observation}, see FIG.~\ref{fig:fig1} (c) and SM~\cite{SuppMat} for further details.
The circuit comprises $8$ physical qubits prepared in the $\ket{\mathbb{Z}_2}$ state and one ancilla qubit, decomposed through variational optimization~\cite{chen2022high,koh2022stabilizing,shen2023observation,koh2023observation,chen2023robust}
. The final post-selection on the ancilla $\ket{\uparrow}$ lead to the normalization $\ket{\psi(t)}=e^{-itH}\ket{\psi_{0}}/\left\| e^{-itH}\ket{\psi_{0}} \right\| $. The signature of scars can be measured through the overlap between $\ket{\psi(t)}$ and $\ket{\mathbb{Z}_2}$. The simulation results, shown in FIG.~\ref{fig:fig1} (d), demonstrate excellent agreement between exact, noiseless results (solid curves) and noisy simulation data (circles). Importantly, the results in blue ($u=0.4$) highlight the robustness of enhanced scars, showing pronounced revivals throughout this period despite the inherent noise in quantum hardware. Such simulations, without any deliberate effort in error mitigation, demonstrate the capability of the Fock skin effect in improving robustness.

{\bf \em Discussion.---}We have shown that asymmetric Fock-state transitions can significantly enhance quantum many-body scarring. Unlike the conventional NHSE, which leads to robust localizations at physical boundaries, our Fock skin effect leads to many-body state revivals with periodic pumping through delocalized Fock skin accumulation. This is enabled by asymmetric transitions across Hamming layers from a specific initial state. Compared to the periodic driving method, the Fock skin effect enables tunable enhancement. We have argued that the non-Hermitian scar enhancement can be realized in existing Bose-Hubbard quantum simulators~\cite{yang2020,su2022observation}, and further confirmed its robustness with digital simulations on IBMQ hardware.

The mechanisms for stabilizing scar states in disordered systems have recently attracted much attention~\cite{OnsagerScars,MondragonShem2020,Huang2021, Voorden21, Zhang2022Clusters, Srivatsa2022,Chen23,Iversen23}. 
The robustness of QMBSs in our model is protected by the Fock skin effect, absent in other non-Hermitian deformations of the PXP model~\cite{chen2023weak}, see SM~\cite{SuppMat}. Moreover, compared to previous works on disordered scarred systems~\cite{MondragonShem2020}, the non-Hermiticity in our case can ``shield'' all characteristic QMBS signatures, including wavefunction revivals, from disorder. 
Finally, exploring the impact of Fock skin effect on entanglement phase transitions~\cite{kawabata2023entanglement,gliozzi2024many} in various non-Hermitian many-body systems~\cite{hamazaki2019non,mu2020emergent,kawabata2022many,alsallom2022fate,shen2022non,shen2023observation,de2023stable} would be also intriguing.

\begin{acknowledgements}
	{\bf \em Acknowledgements.---}Some of the exact diagonalization simulations have been performed using QuSpin~\cite{weinberg2017quspin,weinberg2019quspin}. F.Q. and C.H.L acknowledges support from the QEP2.0 Grant from the Singapore National Research Foundation (Grant No. NRF2021-QEP2-02-P09) and the Singapore MOE Tier-II Grant (Award number: MOE-T2EP50222-0003). J.-Y.D.~and Z.P.~acknowledge support by the Leverhulme Trust Research Leadership Award RL-2019-015. This project has received funding from the European Union’s Horizon 2020 research and innovation programme under the Marie Skłodowska-Curie Grant Agreement No.101034413. This research was supported in part by grant NSF PHY-2309135 to the Kavli Institute for Theoretical Physics (KITP).
    We acknowledge the use of IBM Quantum services for this work. The views expressed are those of the authors and do not reflect the official policy or position of IBM or the IBM Quantum team.
\end{acknowledgements}

\bibliography{references}

\begin{thebibliography}{124}%
\makeatletter
\providecommand \@ifxundefined [1]{%
 \@ifx{#1\undefined}
}%
\providecommand \@ifnum [1]{%
 \ifnum #1\expandafter \@firstoftwo
 \else \expandafter \@secondoftwo
 \fi
}%
\providecommand \@ifx [1]{%
 \ifx #1\expandafter \@firstoftwo
 \else \expandafter \@secondoftwo
 \fi
}%
\providecommand \natexlab [1]{#1}%
\providecommand \enquote  [1]{``#1''}%
\providecommand \bibnamefont  [1]{#1}%
\providecommand \bibfnamefont [1]{#1}%
\providecommand \citenamefont [1]{#1}%
\providecommand \href@noop [0]{\@secondoftwo}%
\providecommand \href [0]{\begingroup \@sanitize@url \@href}%
\providecommand \@href[1]{\@@startlink{#1}\@@href}%
\providecommand \@@href[1]{\endgroup#1\@@endlink}%
\providecommand \@sanitize@url [0]{\catcode `\\12\catcode `\$12\catcode
  `\&12\catcode `\#12\catcode `\^12\catcode `\_12\catcode `\%12\relax}%
\providecommand \@@startlink[1]{}%
\providecommand \@@endlink[0]{}%
\providecommand \url  [0]{\begingroup\@sanitize@url \@url }%
\providecommand \@url [1]{\endgroup\@href {#1}{\urlprefix }}%
\providecommand \urlprefix  [0]{URL }%
\providecommand \Eprint [0]{\href }%
\providecommand \doibase [0]{https://doi.org/}%
\providecommand \selectlanguage [0]{\@gobble}%
\providecommand \bibinfo  [0]{\@secondoftwo}%
\providecommand \bibfield  [0]{\@secondoftwo}%
\providecommand \translation [1]{[#1]}%
\providecommand \BibitemOpen [0]{}%
\providecommand \bibitemStop [0]{}%
\providecommand \bibitemNoStop [0]{.\EOS\space}%
\providecommand \EOS [0]{\spacefactor3000\relax}%
\providecommand \BibitemShut  [1]{\csname bibitem#1\endcsname}%
\let\auto@bib@innerbib\@empty
\bibitem [{\citenamefont {Srednicki}(1994)}]{srednicki1994chaos}%
  \BibitemOpen
  \bibfield  {author} {\bibinfo {author} {\bibfnamefont {M.}~\bibnamefont
  {Srednicki}},\ }\bibfield  {title} {\bibinfo {title} {Chaos and quantum
  thermalization},\ }\href {https://doi.org/10.1103/PhysRevE.50.888} {\bibfield
   {journal} {\bibinfo  {journal} {Phys. Rev. E}\ }\textbf {\bibinfo {volume}
  {50}},\ \bibinfo {pages} {888} (\bibinfo {year} {1994})}\BibitemShut
  {NoStop}%
\bibitem [{\citenamefont {Rigol}\ \emph {et~al.}(2008)\citenamefont {Rigol},
  \citenamefont {Dunjko},\ and\ \citenamefont
  {Olshanii}}]{rigol2008thermalization}%
  \BibitemOpen
  \bibfield  {author} {\bibinfo {author} {\bibfnamefont {M.}~\bibnamefont
  {Rigol}}, \bibinfo {author} {\bibfnamefont {V.}~\bibnamefont {Dunjko}},\ and\
  \bibinfo {author} {\bibfnamefont {M.}~\bibnamefont {Olshanii}},\ }\bibfield
  {title} {\bibinfo {title} {Thermalization and its mechanism for generic
  isolated quantum systems},\ }\href {https://doi.org/10.1038/nature06838}
  {\bibfield  {journal} {\bibinfo  {journal} {Nature}\ }\textbf {\bibinfo
  {volume} {452}},\ \bibinfo {pages} {854} (\bibinfo {year}
  {2008})}\BibitemShut {NoStop}%
\bibitem [{\citenamefont {Rigol}(2009)}]{rigol2009breakdown}%
  \BibitemOpen
  \bibfield  {author} {\bibinfo {author} {\bibfnamefont {M.}~\bibnamefont
  {Rigol}},\ }\bibfield  {title} {\bibinfo {title} {Breakdown of thermalization
  in finite one-dimensional systems},\ }\href
  {https://doi.org/10.1103/PhysRevLett.103.100403} {\bibfield  {journal}
  {\bibinfo  {journal} {Phys. Rev. Lett.}\ }\textbf {\bibinfo {volume} {103}},\
  \bibinfo {pages} {100403} (\bibinfo {year} {2009})}\BibitemShut {NoStop}%
\bibitem [{\citenamefont {Ba\~nuls}\ \emph {et~al.}(2011)\citenamefont
  {Ba\~nuls}, \citenamefont {Cirac},\ and\ \citenamefont
  {Hastings}}]{banuls2011strong}%
  \BibitemOpen
  \bibfield  {author} {\bibinfo {author} {\bibfnamefont {M.~C.}\ \bibnamefont
  {Ba\~nuls}}, \bibinfo {author} {\bibfnamefont {J.~I.}\ \bibnamefont
  {Cirac}},\ and\ \bibinfo {author} {\bibfnamefont {M.~B.}\ \bibnamefont
  {Hastings}},\ }\bibfield  {title} {\bibinfo {title} {Strong and weak
  thermalization of infinite nonintegrable quantum systems},\ }\href
  {https://doi.org/10.1103/PhysRevLett.106.050405} {\bibfield  {journal}
  {\bibinfo  {journal} {Phys. Rev. Lett.}\ }\textbf {\bibinfo {volume} {106}},\
  \bibinfo {pages} {050405} (\bibinfo {year} {2011})}\BibitemShut {NoStop}%
\bibitem [{\citenamefont {D'Alessio}\ \emph {et~al.}(2016)\citenamefont
  {D'Alessio}, \citenamefont {Kafri}, \citenamefont {Polkovnikov},\ and\
  \citenamefont {Rigol}}]{d2016quantum}%
  \BibitemOpen
  \bibfield  {author} {\bibinfo {author} {\bibfnamefont {L.}~\bibnamefont
  {D'Alessio}}, \bibinfo {author} {\bibfnamefont {Y.}~\bibnamefont {Kafri}},
  \bibinfo {author} {\bibfnamefont {A.}~\bibnamefont {Polkovnikov}},\ and\
  \bibinfo {author} {\bibfnamefont {M.}~\bibnamefont {Rigol}},\ }\bibfield
  {title} {\bibinfo {title} {From quantum chaos and eigenstate thermalization
  to statistical mechanics and thermodynamics},\ }\href
  {https://doi.org/10.1080/00018732.2016.1198134} {\bibfield  {journal}
  {\bibinfo  {journal} {Advances in Physics}\ }\textbf {\bibinfo {volume}
  {65}},\ \bibinfo {pages} {239} (\bibinfo {year} {2016})}\BibitemShut
  {NoStop}%
\bibitem [{\citenamefont {Mori}\ \emph {et~al.}(2018)\citenamefont {Mori},
  \citenamefont {Ikeda}, \citenamefont {Kaminishi},\ and\ \citenamefont
  {Ueda}}]{mori2018thermalization}%
  \BibitemOpen
  \bibfield  {author} {\bibinfo {author} {\bibfnamefont {T.}~\bibnamefont
  {Mori}}, \bibinfo {author} {\bibfnamefont {T.~N.}\ \bibnamefont {Ikeda}},
  \bibinfo {author} {\bibfnamefont {E.}~\bibnamefont {Kaminishi}},\ and\
  \bibinfo {author} {\bibfnamefont {M.}~\bibnamefont {Ueda}},\ }\bibfield
  {title} {\bibinfo {title} {Thermalization and prethermalization in isolated
  quantum systems: a theoretical overview},\ }\href
  {https://doi.org/10.1088/1361-6455/aabcdf} {\bibfield  {journal} {\bibinfo
  {journal} {Journal of Physics B: Atomic, Molecular and Optical Physics}\
  }\textbf {\bibinfo {volume} {51}},\ \bibinfo {pages} {112001} (\bibinfo
  {year} {2018})}\BibitemShut {NoStop}%
\bibitem [{\citenamefont {Deutsch}(2018)}]{deutsch2018eigenstate}%
  \BibitemOpen
  \bibfield  {author} {\bibinfo {author} {\bibfnamefont {J.~M.}\ \bibnamefont
  {Deutsch}},\ }\bibfield  {title} {\bibinfo {title} {Eigenstate thermalization
  hypothesis},\ }\href {https://doi.org/10.1088/1361-6633/aac9f1} {\bibfield
  {journal} {\bibinfo  {journal} {Reports on Progress in Physics}\ }\textbf
  {\bibinfo {volume} {81}},\ \bibinfo {pages} {082001} (\bibinfo {year}
  {2018})}\BibitemShut {NoStop}%
\bibitem [{\citenamefont {Mallayya}\ \emph {et~al.}(2019)\citenamefont
  {Mallayya}, \citenamefont {Rigol},\ and\ \citenamefont
  {De~Roeck}}]{mallayya2019prethermalization}%
  \BibitemOpen
  \bibfield  {author} {\bibinfo {author} {\bibfnamefont {K.}~\bibnamefont
  {Mallayya}}, \bibinfo {author} {\bibfnamefont {M.}~\bibnamefont {Rigol}},\
  and\ \bibinfo {author} {\bibfnamefont {W.}~\bibnamefont {De~Roeck}},\
  }\bibfield  {title} {\bibinfo {title} {Prethermalization and thermalization
  in isolated quantum systems},\ }\href
  {https://doi.org/10.1103/PhysRevX.9.021027} {\bibfield  {journal} {\bibinfo
  {journal} {Phys. Rev. X}\ }\textbf {\bibinfo {volume} {9}},\ \bibinfo {pages}
  {021027} (\bibinfo {year} {2019})}\BibitemShut {NoStop}%
\bibitem [{\citenamefont {Deutsch}(1991)}]{deutsch1991quantum}%
  \BibitemOpen
  \bibfield  {author} {\bibinfo {author} {\bibfnamefont {J.~M.}\ \bibnamefont
  {Deutsch}},\ }\bibfield  {title} {\bibinfo {title} {Quantum statistical
  mechanics in a closed system},\ }\href
  {https://doi.org/10.1103/PhysRevA.43.2046} {\bibfield  {journal} {\bibinfo
  {journal} {Phys. Rev. A}\ }\textbf {\bibinfo {volume} {43}},\ \bibinfo
  {pages} {2046} (\bibinfo {year} {1991})}\BibitemShut {NoStop}%
\bibitem [{\citenamefont {Dymarsky}\ \emph {et~al.}(2018)\citenamefont
  {Dymarsky}, \citenamefont {Lashkari},\ and\ \citenamefont
  {Liu}}]{dymarsky2018subsystem}%
  \BibitemOpen
  \bibfield  {author} {\bibinfo {author} {\bibfnamefont {A.}~\bibnamefont
  {Dymarsky}}, \bibinfo {author} {\bibfnamefont {N.}~\bibnamefont {Lashkari}},\
  and\ \bibinfo {author} {\bibfnamefont {H.}~\bibnamefont {Liu}},\ }\bibfield
  {title} {\bibinfo {title} {Subsystem eigenstate thermalization hypothesis},\
  }\href {https://doi.org/10.1103/PhysRevE.97.012140} {\bibfield  {journal}
  {\bibinfo  {journal} {Phys. Rev. E}\ }\textbf {\bibinfo {volume} {97}},\
  \bibinfo {pages} {012140} (\bibinfo {year} {2018})}\BibitemShut {NoStop}%
\bibitem [{\citenamefont {Bernien}\ \emph {et~al.}(2017)\citenamefont
  {Bernien}, \citenamefont {Schwartz}, \citenamefont {Keesling}, \citenamefont
  {Levine}, \citenamefont {Omran}, \citenamefont {Pichler}, \citenamefont
  {Choi}, \citenamefont {Zibrov}, \citenamefont {Endres}, \citenamefont
  {Greiner}, \citenamefont {Vuleti{\'{c}}},\ and\ \citenamefont
  {Lukin}}]{bernien2017probing}%
  \BibitemOpen
  \bibfield  {author} {\bibinfo {author} {\bibfnamefont {H.}~\bibnamefont
  {Bernien}}, \bibinfo {author} {\bibfnamefont {S.}~\bibnamefont {Schwartz}},
  \bibinfo {author} {\bibfnamefont {A.}~\bibnamefont {Keesling}}, \bibinfo
  {author} {\bibfnamefont {H.}~\bibnamefont {Levine}}, \bibinfo {author}
  {\bibfnamefont {A.}~\bibnamefont {Omran}}, \bibinfo {author} {\bibfnamefont
  {H.}~\bibnamefont {Pichler}}, \bibinfo {author} {\bibfnamefont
  {S.}~\bibnamefont {Choi}}, \bibinfo {author} {\bibfnamefont {A.~S.}\
  \bibnamefont {Zibrov}}, \bibinfo {author} {\bibfnamefont {M.}~\bibnamefont
  {Endres}}, \bibinfo {author} {\bibfnamefont {M.}~\bibnamefont {Greiner}},
  \bibinfo {author} {\bibfnamefont {V.}~\bibnamefont {Vuleti{\'{c}}}},\ and\
  \bibinfo {author} {\bibfnamefont {M.~D.}\ \bibnamefont {Lukin}},\ }\bibfield
  {title} {\bibinfo {title} {Probing many-body dynamics on a 51-atom quantum
  simulator},\ }\href {https://doi.org/10.1038/nature24622} {\bibfield
  {journal} {\bibinfo  {journal} {Nature}\ }\textbf {\bibinfo {volume} {551}},\
  \bibinfo {pages} {579} (\bibinfo {year} {2017})}\BibitemShut {NoStop}%
\bibitem [{\citenamefont {Turner}\ \emph
  {et~al.}(2018{\natexlab{a}})\citenamefont {Turner}, \citenamefont
  {Michailidis}, \citenamefont {Abanin}, \citenamefont {Serbyn},\ and\
  \citenamefont {Papi{\'c}}}]{turner2018weak}%
  \BibitemOpen
  \bibfield  {author} {\bibinfo {author} {\bibfnamefont {C.~J.}\ \bibnamefont
  {Turner}}, \bibinfo {author} {\bibfnamefont {A.~A.}\ \bibnamefont
  {Michailidis}}, \bibinfo {author} {\bibfnamefont {D.~A.}\ \bibnamefont
  {Abanin}}, \bibinfo {author} {\bibfnamefont {M.}~\bibnamefont {Serbyn}},\
  and\ \bibinfo {author} {\bibfnamefont {Z.}~\bibnamefont {Papi{\'c}}},\
  }\bibfield  {title} {\bibinfo {title} {Weak ergodicity breaking from quantum
  many-body scars},\ }\href
  {https://doi.org/https://doi.org/10.1038/s41567-018-0137-5} {\bibfield
  {journal} {\bibinfo  {journal} {Nature Physics}\ }\textbf {\bibinfo {volume}
  {14}},\ \bibinfo {pages} {745} (\bibinfo {year}
  {2018}{\natexlab{a}})}\BibitemShut {NoStop}%
\bibitem [{\citenamefont {Ho}\ \emph {et~al.}(2019)\citenamefont {Ho},
  \citenamefont {Choi}, \citenamefont {Pichler},\ and\ \citenamefont
  {Lukin}}]{ho2019periodic}%
  \BibitemOpen
  \bibfield  {author} {\bibinfo {author} {\bibfnamefont {W.~W.}\ \bibnamefont
  {Ho}}, \bibinfo {author} {\bibfnamefont {S.}~\bibnamefont {Choi}}, \bibinfo
  {author} {\bibfnamefont {H.}~\bibnamefont {Pichler}},\ and\ \bibinfo {author}
  {\bibfnamefont {M.~D.}\ \bibnamefont {Lukin}},\ }\bibfield  {title} {\bibinfo
  {title} {Periodic orbits, entanglement, and quantum many-body scars in
  constrained models: Matrix product state approach},\ }\href
  {https://doi.org/10.1103/PhysRevLett.122.040603} {\bibfield  {journal}
  {\bibinfo  {journal} {Phys. Rev. Lett.}\ }\textbf {\bibinfo {volume} {122}},\
  \bibinfo {pages} {040603} (\bibinfo {year} {2019})}\BibitemShut {NoStop}%
\bibitem [{\citenamefont {Serbyn}\ \emph {et~al.}(2021)\citenamefont {Serbyn},
  \citenamefont {Abanin},\ and\ \citenamefont {Papi{\'c}}}]{serbyn2021quantum}%
  \BibitemOpen
  \bibfield  {author} {\bibinfo {author} {\bibfnamefont {M.}~\bibnamefont
  {Serbyn}}, \bibinfo {author} {\bibfnamefont {D.~A.}\ \bibnamefont {Abanin}},\
  and\ \bibinfo {author} {\bibfnamefont {Z.}~\bibnamefont {Papi{\'c}}},\
  }\bibfield  {title} {\bibinfo {title} {Quantum many-body scars and weak
  breaking of ergodicity},\ }\href
  {https://doi.org/https://doi.org/10.1038/s41567-021-01230-2} {\bibfield
  {journal} {\bibinfo  {journal} {Nature Physics}\ }\textbf {\bibinfo {volume}
  {17}},\ \bibinfo {pages} {675} (\bibinfo {year} {2021})}\BibitemShut
  {NoStop}%
\bibitem [{\citenamefont {Moudgalya}\ \emph {et~al.}(2022)\citenamefont
  {Moudgalya}, \citenamefont {Bernevig},\ and\ \citenamefont
  {Regnault}}]{MoudgalyaReview}%
  \BibitemOpen
  \bibfield  {author} {\bibinfo {author} {\bibfnamefont {S.}~\bibnamefont
  {Moudgalya}}, \bibinfo {author} {\bibfnamefont {B.~A.}\ \bibnamefont
  {Bernevig}},\ and\ \bibinfo {author} {\bibfnamefont {N.}~\bibnamefont
  {Regnault}},\ }\bibfield  {title} {\bibinfo {title} {Quantum many-body scars
  and {H}ilbert space fragmentation: a review of exact results},\ }\href
  {https://doi.org/10.1088/1361-6633/ac73a0} {\bibfield  {journal} {\bibinfo
  {journal} {Reports on Progress in Physics}\ }\textbf {\bibinfo {volume}
  {85}},\ \bibinfo {pages} {086501} (\bibinfo {year} {2022})}\BibitemShut
  {NoStop}%
\bibitem [{\citenamefont {Chandran}\ \emph {et~al.}(2023)\citenamefont
  {Chandran}, \citenamefont {Iadecola}, \citenamefont {Khemani},\ and\
  \citenamefont {Moessner}}]{ChandranReview}%
  \BibitemOpen
  \bibfield  {author} {\bibinfo {author} {\bibfnamefont {A.}~\bibnamefont
  {Chandran}}, \bibinfo {author} {\bibfnamefont {T.}~\bibnamefont {Iadecola}},
  \bibinfo {author} {\bibfnamefont {V.}~\bibnamefont {Khemani}},\ and\ \bibinfo
  {author} {\bibfnamefont {R.}~\bibnamefont {Moessner}},\ }\bibfield  {title}
  {\bibinfo {title} {Quantum many-body scars: A quasiparticle perspective},\
  }\href {https://doi.org/10.1146/annurev-conmatphys-031620-101617} {\bibfield
  {journal} {\bibinfo  {journal} {Annual Review of Condensed Matter Physics}\
  }\textbf {\bibinfo {volume} {14}},\ \bibinfo {pages} {443} (\bibinfo {year}
  {2023})}\BibitemShut {NoStop}%
\bibitem [{\citenamefont {Sutherland}(2004)}]{sutherland2004beautiful}%
  \BibitemOpen
  \bibfield  {author} {\bibinfo {author} {\bibfnamefont {B.}~\bibnamefont
  {Sutherland}},\ }\href@noop {} {\emph {\bibinfo {title} {Beautiful models: 70
  years of exactly solved quantum many-body problems}}}\ (\bibinfo  {publisher}
  {World Scientific Publishing Company},\ \bibinfo {year} {2004})\BibitemShut
  {NoStop}%
\bibitem [{\citenamefont {Nandkishore}\ and\ \citenamefont
  {Huse}(2015)}]{Nandkishore_review}%
  \BibitemOpen
  \bibfield  {author} {\bibinfo {author} {\bibfnamefont {R.}~\bibnamefont
  {Nandkishore}}\ and\ \bibinfo {author} {\bibfnamefont {D.~A.}\ \bibnamefont
  {Huse}},\ }\bibfield  {title} {\bibinfo {title} {Many-body localization and
  thermalization in quantum statistical mechanics},\ }\href
  {https://doi.org/10.1146/annurev-conmatphys-031214-014726} {\bibfield
  {journal} {\bibinfo  {journal} {Annual Review of Condensed Matter Physics}\
  }\textbf {\bibinfo {volume} {6}},\ \bibinfo {pages} {15} (\bibinfo {year}
  {2015})}\BibitemShut {NoStop}%
\bibitem [{\citenamefont {Abanin}\ \emph {et~al.}(2019)\citenamefont {Abanin},
  \citenamefont {Altman}, \citenamefont {Bloch},\ and\ \citenamefont
  {Serbyn}}]{Abanin_review}%
  \BibitemOpen
  \bibfield  {author} {\bibinfo {author} {\bibfnamefont {D.~A.}\ \bibnamefont
  {Abanin}}, \bibinfo {author} {\bibfnamefont {E.}~\bibnamefont {Altman}},
  \bibinfo {author} {\bibfnamefont {I.}~\bibnamefont {Bloch}},\ and\ \bibinfo
  {author} {\bibfnamefont {M.}~\bibnamefont {Serbyn}},\ }\bibfield  {title}
  {\bibinfo {title} {Colloquium: Many-body localization, thermalization, and
  entanglement},\ }\href {https://doi.org/10.1103/RevModPhys.91.021001}
  {\bibfield  {journal} {\bibinfo  {journal} {Rev. Mod. Phys.}\ }\textbf
  {\bibinfo {volume} {91}},\ \bibinfo {pages} {021001} (\bibinfo {year}
  {2019})}\BibitemShut {NoStop}%
\bibitem [{\citenamefont {Moudgalya}\ \emph {et~al.}(2018)\citenamefont
  {Moudgalya}, \citenamefont {Regnault},\ and\ \citenamefont
  {Bernevig}}]{moudgalya2018entanglement}%
  \BibitemOpen
  \bibfield  {author} {\bibinfo {author} {\bibfnamefont {S.}~\bibnamefont
  {Moudgalya}}, \bibinfo {author} {\bibfnamefont {N.}~\bibnamefont
  {Regnault}},\ and\ \bibinfo {author} {\bibfnamefont {B.~A.}\ \bibnamefont
  {Bernevig}},\ }\bibfield  {title} {\bibinfo {title} {Entanglement of exact
  excited states of {Affleck-Kennedy-Lieb-Tasaki} models: Exact results,
  many-body scars, and violation of the strong eigenstate thermalization
  hypothesis},\ }\href {https://doi.org/10.1103/PhysRevB.98.235156} {\bibfield
  {journal} {\bibinfo  {journal} {Phys. Rev. B}\ }\textbf {\bibinfo {volume}
  {98}},\ \bibinfo {pages} {235156} (\bibinfo {year} {2018})}\BibitemShut
  {NoStop}%
\bibitem [{\citenamefont {Schecter}\ and\ \citenamefont
  {Iadecola}(2019)}]{Iadecola2019_2}%
  \BibitemOpen
  \bibfield  {author} {\bibinfo {author} {\bibfnamefont {M.}~\bibnamefont
  {Schecter}}\ and\ \bibinfo {author} {\bibfnamefont {T.}~\bibnamefont
  {Iadecola}},\ }\bibfield  {title} {\bibinfo {title} {Weak ergodicity breaking
  and quantum many-body scars in spin-1 {XY} magnets},\ }\href
  {https://doi.org/10.1103/PhysRevLett.123.147201} {\bibfield  {journal}
  {\bibinfo  {journal} {Phys. Rev. Lett.}\ }\textbf {\bibinfo {volume} {123}},\
  \bibinfo {pages} {147201} (\bibinfo {year} {2019})}\BibitemShut {NoStop}%
\bibitem [{\citenamefont {Choi}\ \emph {et~al.}(2019)\citenamefont {Choi},
  \citenamefont {Turner}, \citenamefont {Pichler}, \citenamefont {Ho},
  \citenamefont {Michailidis}, \citenamefont {Papi\ifmmode~\acute{c}\else
  \'{c}\fi{}}, \citenamefont {Serbyn}, \citenamefont {Lukin},\ and\
  \citenamefont {Abanin}}]{choi2019emergent}%
  \BibitemOpen
  \bibfield  {author} {\bibinfo {author} {\bibfnamefont {S.}~\bibnamefont
  {Choi}}, \bibinfo {author} {\bibfnamefont {C.~J.}\ \bibnamefont {Turner}},
  \bibinfo {author} {\bibfnamefont {H.}~\bibnamefont {Pichler}}, \bibinfo
  {author} {\bibfnamefont {W.~W.}\ \bibnamefont {Ho}}, \bibinfo {author}
  {\bibfnamefont {A.~A.}\ \bibnamefont {Michailidis}}, \bibinfo {author}
  {\bibfnamefont {Z.}~\bibnamefont {Papi\ifmmode~\acute{c}\else \'{c}\fi{}}},
  \bibinfo {author} {\bibfnamefont {M.}~\bibnamefont {Serbyn}}, \bibinfo
  {author} {\bibfnamefont {M.~D.}\ \bibnamefont {Lukin}},\ and\ \bibinfo
  {author} {\bibfnamefont {D.~A.}\ \bibnamefont {Abanin}},\ }\bibfield  {title}
  {\bibinfo {title} {Emergent {SU(2)} dynamics and perfect quantum many-body
  scars},\ }\href {https://doi.org/10.1103/PhysRevLett.122.220603} {\bibfield
  {journal} {\bibinfo  {journal} {Phys. Rev. Lett.}\ }\textbf {\bibinfo
  {volume} {122}},\ \bibinfo {pages} {220603} (\bibinfo {year}
  {2019})}\BibitemShut {NoStop}%
\bibitem [{\citenamefont {Mark}\ \emph {et~al.}(2020)\citenamefont {Mark},
  \citenamefont {Lin},\ and\ \citenamefont {Motrunich}}]{MarkLinMotrunich}%
  \BibitemOpen
  \bibfield  {author} {\bibinfo {author} {\bibfnamefont {D.~K.}\ \bibnamefont
  {Mark}}, \bibinfo {author} {\bibfnamefont {C.-J.}\ \bibnamefont {Lin}},\ and\
  \bibinfo {author} {\bibfnamefont {O.~I.}\ \bibnamefont {Motrunich}},\
  }\bibfield  {title} {\bibinfo {title} {Unified structure for exact towers of
  scar states in the {Affleck-Kennedy-Lieb-Tasaki} and other models},\ }\href
  {https://doi.org/10.1103/PhysRevB.101.195131} {\bibfield  {journal} {\bibinfo
   {journal} {Phys. Rev. B}\ }\textbf {\bibinfo {volume} {101}},\ \bibinfo
  {pages} {195131} (\bibinfo {year} {2020})}\BibitemShut {NoStop}%
\bibitem [{\citenamefont {Omiya}\ and\ \citenamefont
  {M\"uller}(2023)}]{Omiya2022}%
  \BibitemOpen
  \bibfield  {author} {\bibinfo {author} {\bibfnamefont {K.}~\bibnamefont
  {Omiya}}\ and\ \bibinfo {author} {\bibfnamefont {M.}~\bibnamefont
  {M\"uller}},\ }\bibfield  {title} {\bibinfo {title} {Quantum many-body scars
  in bipartite {Rydberg} arrays originating from hidden projector embedding},\
  }\href {https://doi.org/10.1103/PhysRevA.107.023318} {\bibfield  {journal}
  {\bibinfo  {journal} {Phys. Rev. A}\ }\textbf {\bibinfo {volume} {107}},\
  \bibinfo {pages} {023318} (\bibinfo {year} {2023})}\BibitemShut {NoStop}%
\bibitem [{\citenamefont {O'Dea}\ \emph {et~al.}(2020)\citenamefont {O'Dea},
  \citenamefont {Burnell}, \citenamefont {Chandran},\ and\ \citenamefont
  {Khemani}}]{Dea2020}%
  \BibitemOpen
  \bibfield  {author} {\bibinfo {author} {\bibfnamefont {N.}~\bibnamefont
  {O'Dea}}, \bibinfo {author} {\bibfnamefont {F.}~\bibnamefont {Burnell}},
  \bibinfo {author} {\bibfnamefont {A.}~\bibnamefont {Chandran}},\ and\
  \bibinfo {author} {\bibfnamefont {V.}~\bibnamefont {Khemani}},\ }\bibfield
  {title} {\bibinfo {title} {From tunnels to towers: Quantum scars from {Lie}
  algebras and $q$-deformed {Lie} algebras},\ }\href
  {https://doi.org/10.1103/PhysRevResearch.2.043305} {\bibfield  {journal}
  {\bibinfo  {journal} {Phys. Rev. Research}\ }\textbf {\bibinfo {volume}
  {2}},\ \bibinfo {pages} {043305} (\bibinfo {year} {2020})}\BibitemShut
  {NoStop}%
\bibitem [{\citenamefont {Pakrouski}\ \emph {et~al.}(2020)\citenamefont
  {Pakrouski}, \citenamefont {Pallegar}, \citenamefont {Popov},\ and\
  \citenamefont {Klebanov}}]{Pakrouski2020}%
  \BibitemOpen
  \bibfield  {author} {\bibinfo {author} {\bibfnamefont {K.}~\bibnamefont
  {Pakrouski}}, \bibinfo {author} {\bibfnamefont {P.~N.}\ \bibnamefont
  {Pallegar}}, \bibinfo {author} {\bibfnamefont {F.~K.}\ \bibnamefont
  {Popov}},\ and\ \bibinfo {author} {\bibfnamefont {I.~R.}\ \bibnamefont
  {Klebanov}},\ }\bibfield  {title} {\bibinfo {title} {Many-body scars as a
  group invariant sector of {Hilbert} space},\ }\href
  {https://doi.org/10.1103/PhysRevLett.125.230602} {\bibfield  {journal}
  {\bibinfo  {journal} {Phys. Rev. Lett.}\ }\textbf {\bibinfo {volume} {125}},\
  \bibinfo {pages} {230602} (\bibinfo {year} {2020})}\BibitemShut {NoStop}%
\bibitem [{\citenamefont {Moudgalya}\ and\ \citenamefont
  {Motrunich}(2022)}]{MoudgalyaCommutant}%
  \BibitemOpen
  \bibfield  {author} {\bibinfo {author} {\bibfnamefont {S.}~\bibnamefont
  {Moudgalya}}\ and\ \bibinfo {author} {\bibfnamefont {O.~I.}\ \bibnamefont
  {Motrunich}},\ }\bibfield  {title} {\bibinfo {title} {Hilbert space
  fragmentation and commutant algebras},\ }\href
  {https://doi.org/10.1103/PhysRevX.12.011050} {\bibfield  {journal} {\bibinfo
  {journal} {Phys. Rev. X}\ }\textbf {\bibinfo {volume} {12}},\ \bibinfo
  {pages} {011050} (\bibinfo {year} {2022})}\BibitemShut {NoStop}%
\bibitem [{\citenamefont {Bu\ifmmode~\check{c}\else
  \v{c}\fi{}a}(2023)}]{Buca2023}%
  \BibitemOpen
  \bibfield  {author} {\bibinfo {author} {\bibfnamefont {B.}~\bibnamefont
  {Bu\ifmmode~\check{c}\else \v{c}\fi{}a}},\ }\bibfield  {title} {\bibinfo
  {title} {Unified theory of local quantum many-body dynamics: Eigenoperator
  thermalization theorems},\ }\href
  {https://doi.org/10.1103/PhysRevX.13.031013} {\bibfield  {journal} {\bibinfo
  {journal} {Phys. Rev. X}\ }\textbf {\bibinfo {volume} {13}},\ \bibinfo
  {pages} {031013} (\bibinfo {year} {2023})}\BibitemShut {NoStop}%
\bibitem [{\citenamefont {Bluvstein}\ \emph {et~al.}(2021)\citenamefont
  {Bluvstein}, \citenamefont {Omran}, \citenamefont {Levine}, \citenamefont
  {Keesling}, \citenamefont {Semeghini}, \citenamefont {Ebadi}, \citenamefont
  {Wang}, \citenamefont {Michailidis}, \citenamefont {Maskara}, \citenamefont
  {Ho}, \citenamefont {Choi}, \citenamefont {Serbyn}, \citenamefont {Greiner},
  \citenamefont {Vuletić},\ and\ \citenamefont
  {Lukin}}]{bluvstein2021controlling}%
  \BibitemOpen
  \bibfield  {author} {\bibinfo {author} {\bibfnamefont {D.}~\bibnamefont
  {Bluvstein}}, \bibinfo {author} {\bibfnamefont {A.}~\bibnamefont {Omran}},
  \bibinfo {author} {\bibfnamefont {H.}~\bibnamefont {Levine}}, \bibinfo
  {author} {\bibfnamefont {A.}~\bibnamefont {Keesling}}, \bibinfo {author}
  {\bibfnamefont {G.}~\bibnamefont {Semeghini}}, \bibinfo {author}
  {\bibfnamefont {S.}~\bibnamefont {Ebadi}}, \bibinfo {author} {\bibfnamefont
  {T.~T.}\ \bibnamefont {Wang}}, \bibinfo {author} {\bibfnamefont {A.~A.}\
  \bibnamefont {Michailidis}}, \bibinfo {author} {\bibfnamefont
  {N.}~\bibnamefont {Maskara}}, \bibinfo {author} {\bibfnamefont {W.~W.}\
  \bibnamefont {Ho}}, \bibinfo {author} {\bibfnamefont {S.}~\bibnamefont
  {Choi}}, \bibinfo {author} {\bibfnamefont {M.}~\bibnamefont {Serbyn}},
  \bibinfo {author} {\bibfnamefont {M.}~\bibnamefont {Greiner}}, \bibinfo
  {author} {\bibfnamefont {V.}~\bibnamefont {Vuletić}},\ and\ \bibinfo
  {author} {\bibfnamefont {M.~D.}\ \bibnamefont {Lukin}},\ }\bibfield  {title}
  {\bibinfo {title} {Controlling quantum many-body dynamics in driven {Rydberg}
  atom arrays},\ }\href {https://doi.org/10.1126/science.abg2530} {\bibfield
  {journal} {\bibinfo  {journal} {Science}\ }\textbf {\bibinfo {volume}
  {371}},\ \bibinfo {pages} {1355} (\bibinfo {year} {2021})}\BibitemShut
  {NoStop}%
\bibitem [{\citenamefont {Jepsen}\ \emph {et~al.}(2020)\citenamefont {Jepsen},
  \citenamefont {Amato-Grill}, \citenamefont {Dimitrova}, \citenamefont {Ho},
  \citenamefont {Demler},\ and\ \citenamefont {Ketterle}}]{jepsen2020spin}%
  \BibitemOpen
  \bibfield  {author} {\bibinfo {author} {\bibfnamefont {P.~N.}\ \bibnamefont
  {Jepsen}}, \bibinfo {author} {\bibfnamefont {J.}~\bibnamefont {Amato-Grill}},
  \bibinfo {author} {\bibfnamefont {I.}~\bibnamefont {Dimitrova}}, \bibinfo
  {author} {\bibfnamefont {W.~W.}\ \bibnamefont {Ho}}, \bibinfo {author}
  {\bibfnamefont {E.}~\bibnamefont {Demler}},\ and\ \bibinfo {author}
  {\bibfnamefont {W.}~\bibnamefont {Ketterle}},\ }\bibfield  {title} {\bibinfo
  {title} {Spin transport in a tunable {Heisenberg} model realized with
  ultracold atoms},\ }\href {https://doi.org/10.1038/s41586-020-3033-y}
  {\bibfield  {journal} {\bibinfo  {journal} {Nature}\ }\textbf {\bibinfo
  {volume} {588}},\ \bibinfo {pages} {403} (\bibinfo {year}
  {2020})}\BibitemShut {NoStop}%
\bibitem [{\citenamefont {Scherg}\ \emph {et~al.}(2021)\citenamefont {Scherg},
  \citenamefont {Kohlert}, \citenamefont {Sala}, \citenamefont {Pollmann},
  \citenamefont {Hebbe~Madhusudhana}, \citenamefont {Bloch},\ and\
  \citenamefont {Aidelsburger}}]{scherg2021observing}%
  \BibitemOpen
  \bibfield  {author} {\bibinfo {author} {\bibfnamefont {S.}~\bibnamefont
  {Scherg}}, \bibinfo {author} {\bibfnamefont {T.}~\bibnamefont {Kohlert}},
  \bibinfo {author} {\bibfnamefont {P.}~\bibnamefont {Sala}}, \bibinfo {author}
  {\bibfnamefont {F.}~\bibnamefont {Pollmann}}, \bibinfo {author}
  {\bibfnamefont {B.}~\bibnamefont {Hebbe~Madhusudhana}}, \bibinfo {author}
  {\bibfnamefont {I.}~\bibnamefont {Bloch}},\ and\ \bibinfo {author}
  {\bibfnamefont {M.}~\bibnamefont {Aidelsburger}},\ }\bibfield  {title}
  {\bibinfo {title} {Observing non-ergodicity due to kinetic constraints in
  tilted {Fermi-Hubbard} chains},\ }\href
  {https://doi.org/https://doi.org/10.1038/s41467-021-24726-0} {\bibfield
  {journal} {\bibinfo  {journal} {Nature Communications}\ }\textbf {\bibinfo
  {volume} {12}},\ \bibinfo {pages} {4490} (\bibinfo {year}
  {2021})}\BibitemShut {NoStop}%
\bibitem [{\citenamefont {Su}\ \emph {et~al.}(2023)\citenamefont {Su},
  \citenamefont {Sun}, \citenamefont {Hudomal}, \citenamefont {Desaules},
  \citenamefont {Zhou}, \citenamefont {Yang}, \citenamefont {Halimeh},
  \citenamefont {Yuan}, \citenamefont {Papi\ifmmode~\acute{c}\else
  \'{c}\fi{}},\ and\ \citenamefont {Pan}}]{su2022observation}%
  \BibitemOpen
  \bibfield  {author} {\bibinfo {author} {\bibfnamefont {G.-X.}\ \bibnamefont
  {Su}}, \bibinfo {author} {\bibfnamefont {H.}~\bibnamefont {Sun}}, \bibinfo
  {author} {\bibfnamefont {A.}~\bibnamefont {Hudomal}}, \bibinfo {author}
  {\bibfnamefont {J.-Y.}\ \bibnamefont {Desaules}}, \bibinfo {author}
  {\bibfnamefont {Z.-Y.}\ \bibnamefont {Zhou}}, \bibinfo {author}
  {\bibfnamefont {B.}~\bibnamefont {Yang}}, \bibinfo {author} {\bibfnamefont
  {J.~C.}\ \bibnamefont {Halimeh}}, \bibinfo {author} {\bibfnamefont {Z.-S.}\
  \bibnamefont {Yuan}}, \bibinfo {author} {\bibfnamefont {Z.}~\bibnamefont
  {Papi\ifmmode~\acute{c}\else \'{c}\fi{}}},\ and\ \bibinfo {author}
  {\bibfnamefont {J.-W.}\ \bibnamefont {Pan}},\ }\bibfield  {title} {\bibinfo
  {title} {Observation of many-body scarring in a {Bose-Hubbard} quantum
  simulator},\ }\href {https://doi.org/10.1103/PhysRevResearch.5.023010}
  {\bibfield  {journal} {\bibinfo  {journal} {Phys. Rev. Res.}\ }\textbf
  {\bibinfo {volume} {5}},\ \bibinfo {pages} {023010} (\bibinfo {year}
  {2023})}\BibitemShut {NoStop}%
\bibitem [{\citenamefont {Zhang}\ \emph {et~al.}(2023)\citenamefont {Zhang},
  \citenamefont {Dong}, \citenamefont {Gao}, \citenamefont {Zhao},
  \citenamefont {Hao}, \citenamefont {Desaules}, \citenamefont {Guo},
  \citenamefont {Chen}, \citenamefont {Deng}, \citenamefont {Liu},
  \citenamefont {Ren}, \citenamefont {Yao}, \citenamefont {Zhang},
  \citenamefont {Xu}, \citenamefont {Wang}, \citenamefont {Jin}, \citenamefont
  {Zhu}, \citenamefont {Zhang}, \citenamefont {Li}, \citenamefont {Song},
  \citenamefont {Wang}, \citenamefont {Liu}, \citenamefont {Papi{\'{c}}},
  \citenamefont {Ying}, \citenamefont {Wang},\ and\ \citenamefont
  {Lai}}]{zhang2022many}%
  \BibitemOpen
  \bibfield  {author} {\bibinfo {author} {\bibfnamefont {P.}~\bibnamefont
  {Zhang}}, \bibinfo {author} {\bibfnamefont {H.}~\bibnamefont {Dong}},
  \bibinfo {author} {\bibfnamefont {Y.}~\bibnamefont {Gao}}, \bibinfo {author}
  {\bibfnamefont {L.}~\bibnamefont {Zhao}}, \bibinfo {author} {\bibfnamefont
  {J.}~\bibnamefont {Hao}}, \bibinfo {author} {\bibfnamefont {J.-Y.}\
  \bibnamefont {Desaules}}, \bibinfo {author} {\bibfnamefont {Q.}~\bibnamefont
  {Guo}}, \bibinfo {author} {\bibfnamefont {J.}~\bibnamefont {Chen}}, \bibinfo
  {author} {\bibfnamefont {J.}~\bibnamefont {Deng}}, \bibinfo {author}
  {\bibfnamefont {B.}~\bibnamefont {Liu}}, \bibinfo {author} {\bibfnamefont
  {W.}~\bibnamefont {Ren}}, \bibinfo {author} {\bibfnamefont {Y.}~\bibnamefont
  {Yao}}, \bibinfo {author} {\bibfnamefont {X.}~\bibnamefont {Zhang}}, \bibinfo
  {author} {\bibfnamefont {S.}~\bibnamefont {Xu}}, \bibinfo {author}
  {\bibfnamefont {K.}~\bibnamefont {Wang}}, \bibinfo {author} {\bibfnamefont
  {F.}~\bibnamefont {Jin}}, \bibinfo {author} {\bibfnamefont {X.}~\bibnamefont
  {Zhu}}, \bibinfo {author} {\bibfnamefont {B.}~\bibnamefont {Zhang}}, \bibinfo
  {author} {\bibfnamefont {H.}~\bibnamefont {Li}}, \bibinfo {author}
  {\bibfnamefont {C.}~\bibnamefont {Song}}, \bibinfo {author} {\bibfnamefont
  {Z.}~\bibnamefont {Wang}}, \bibinfo {author} {\bibfnamefont {F.}~\bibnamefont
  {Liu}}, \bibinfo {author} {\bibfnamefont {Z.}~\bibnamefont {Papi{\'{c}}}},
  \bibinfo {author} {\bibfnamefont {L.}~\bibnamefont {Ying}}, \bibinfo {author}
  {\bibfnamefont {H.}~\bibnamefont {Wang}},\ and\ \bibinfo {author}
  {\bibfnamefont {Y.-C.}\ \bibnamefont {Lai}},\ }\bibfield  {title} {\bibinfo
  {title} {Many-body {Hilbert} space scarring on a superconducting processor},\
  }\href {https://doi.org/10.1038/s41567-022-01784-9} {\bibfield  {journal}
  {\bibinfo  {journal} {Nature Physics}\ }\textbf {\bibinfo {volume} {19}},\
  \bibinfo {pages} {120} (\bibinfo {year} {2023})}\BibitemShut {NoStop}%
\bibitem [{\citenamefont {Kohlert}\ \emph {et~al.}(2023)\citenamefont
  {Kohlert}, \citenamefont {Scherg}, \citenamefont {Sala}, \citenamefont
  {Pollmann}, \citenamefont {Hebbe~Madhusudhana}, \citenamefont {Bloch},\ and\
  \citenamefont {Aidelsburger}}]{kohlert2021experimental}%
  \BibitemOpen
  \bibfield  {author} {\bibinfo {author} {\bibfnamefont {T.}~\bibnamefont
  {Kohlert}}, \bibinfo {author} {\bibfnamefont {S.}~\bibnamefont {Scherg}},
  \bibinfo {author} {\bibfnamefont {P.}~\bibnamefont {Sala}}, \bibinfo {author}
  {\bibfnamefont {F.}~\bibnamefont {Pollmann}}, \bibinfo {author}
  {\bibfnamefont {B.}~\bibnamefont {Hebbe~Madhusudhana}}, \bibinfo {author}
  {\bibfnamefont {I.}~\bibnamefont {Bloch}},\ and\ \bibinfo {author}
  {\bibfnamefont {M.}~\bibnamefont {Aidelsburger}},\ }\bibfield  {title}
  {\bibinfo {title} {Exploring the regime of fragmentation in strongly tilted
  {Fermi-Hubbard Chains}},\ }\href
  {https://doi.org/10.1103/PhysRevLett.130.010201} {\bibfield  {journal}
  {\bibinfo  {journal} {Phys. Rev. Lett.}\ }\textbf {\bibinfo {volume} {130}},\
  \bibinfo {pages} {010201} (\bibinfo {year} {2023})}\BibitemShut {NoStop}%
\bibitem [{\citenamefont {Maskara}\ \emph {et~al.}(2021)\citenamefont
  {Maskara}, \citenamefont {Michailidis}, \citenamefont {Ho}, \citenamefont
  {Bluvstein}, \citenamefont {Choi}, \citenamefont {Lukin},\ and\ \citenamefont
  {Serbyn}}]{maskara2021discrete}%
  \BibitemOpen
  \bibfield  {author} {\bibinfo {author} {\bibfnamefont {N.}~\bibnamefont
  {Maskara}}, \bibinfo {author} {\bibfnamefont {A.~A.}\ \bibnamefont
  {Michailidis}}, \bibinfo {author} {\bibfnamefont {W.~W.}\ \bibnamefont {Ho}},
  \bibinfo {author} {\bibfnamefont {D.}~\bibnamefont {Bluvstein}}, \bibinfo
  {author} {\bibfnamefont {S.}~\bibnamefont {Choi}}, \bibinfo {author}
  {\bibfnamefont {M.~D.}\ \bibnamefont {Lukin}},\ and\ \bibinfo {author}
  {\bibfnamefont {M.}~\bibnamefont {Serbyn}},\ }\bibfield  {title} {\bibinfo
  {title} {Discrete time-crystalline order enabled by quantum many-body scars:
  Entanglement steering via periodic driving},\ }\href
  {https://doi.org/10.1103/PhysRevLett.127.090602} {\bibfield  {journal}
  {\bibinfo  {journal} {Phys. Rev. Lett.}\ }\textbf {\bibinfo {volume} {127}},\
  \bibinfo {pages} {090602} (\bibinfo {year} {2021})}\BibitemShut {NoStop}%
\bibitem [{\citenamefont {Hudomal}\ \emph {et~al.}(2022)\citenamefont
  {Hudomal}, \citenamefont {Desaules}, \citenamefont {Mukherjee}, \citenamefont
  {Su}, \citenamefont {Halimeh},\ and\ \citenamefont
  {Papi\ifmmode~\acute{c}\else \'{c}\fi{}}}]{hudomal2022driving}%
  \BibitemOpen
  \bibfield  {author} {\bibinfo {author} {\bibfnamefont {A.}~\bibnamefont
  {Hudomal}}, \bibinfo {author} {\bibfnamefont {J.-Y.}\ \bibnamefont
  {Desaules}}, \bibinfo {author} {\bibfnamefont {B.}~\bibnamefont {Mukherjee}},
  \bibinfo {author} {\bibfnamefont {G.-X.}\ \bibnamefont {Su}}, \bibinfo
  {author} {\bibfnamefont {J.~C.}\ \bibnamefont {Halimeh}},\ and\ \bibinfo
  {author} {\bibfnamefont {Z.}~\bibnamefont {Papi\ifmmode~\acute{c}\else
  \'{c}\fi{}}},\ }\bibfield  {title} {\bibinfo {title} {Driving quantum
  many-body scars in the {PXP} model},\ }\href
  {https://doi.org/10.1103/PhysRevB.106.104302} {\bibfield  {journal} {\bibinfo
   {journal} {Phys. Rev. B}\ }\textbf {\bibinfo {volume} {106}},\ \bibinfo
  {pages} {104302} (\bibinfo {year} {2022})}\BibitemShut {NoStop}%
\bibitem [{\citenamefont {Halimeh}\ \emph {et~al.}(2023)\citenamefont
  {Halimeh}, \citenamefont {Barbiero}, \citenamefont {Hauke}, \citenamefont
  {Grusdt},\ and\ \citenamefont {Bohrdt}}]{Halimeh2023robustquantummany}%
  \BibitemOpen
  \bibfield  {author} {\bibinfo {author} {\bibfnamefont {J.~C.}\ \bibnamefont
  {Halimeh}}, \bibinfo {author} {\bibfnamefont {L.}~\bibnamefont {Barbiero}},
  \bibinfo {author} {\bibfnamefont {P.}~\bibnamefont {Hauke}}, \bibinfo
  {author} {\bibfnamefont {F.}~\bibnamefont {Grusdt}},\ and\ \bibinfo {author}
  {\bibfnamefont {A.}~\bibnamefont {Bohrdt}},\ }\bibfield  {title} {\bibinfo
  {title} {Robust quantum many-body scars in lattice gauge theories},\ }\href
  {https://doi.org/10.22331/q-2023-05-15-1004} {\bibfield  {journal} {\bibinfo
  {journal} {{Quantum}}\ }\textbf {\bibinfo {volume} {7}},\ \bibinfo {pages}
  {1004} (\bibinfo {year} {2023})}\BibitemShut {NoStop}%
\bibitem [{\citenamefont {Shishkov}\ \emph {et~al.}(2018)\citenamefont
  {Shishkov}, \citenamefont {Andrianov}, \citenamefont {Pukhov}, \citenamefont
  {Vinogradov},\ and\ \citenamefont {Lisyansky}}]{shishkov2018zeroth}%
  \BibitemOpen
  \bibfield  {author} {\bibinfo {author} {\bibfnamefont {V.~Y.}\ \bibnamefont
  {Shishkov}}, \bibinfo {author} {\bibfnamefont {E.~S.}\ \bibnamefont
  {Andrianov}}, \bibinfo {author} {\bibfnamefont {A.~A.}\ \bibnamefont
  {Pukhov}}, \bibinfo {author} {\bibfnamefont {A.~P.}\ \bibnamefont
  {Vinogradov}},\ and\ \bibinfo {author} {\bibfnamefont {A.~A.}\ \bibnamefont
  {Lisyansky}},\ }\bibfield  {title} {\bibinfo {title} {Zeroth law of
  thermodynamics for thermalized open quantum systems having constants of
  motion},\ }\href {https://doi.org/10.1103/PhysRevE.98.022132} {\bibfield
  {journal} {\bibinfo  {journal} {Phys. Rev. E}\ }\textbf {\bibinfo {volume}
  {98}},\ \bibinfo {pages} {022132} (\bibinfo {year} {2018})}\BibitemShut
  {NoStop}%
\bibitem [{\citenamefont {Reichental}\ \emph {et~al.}(2018)\citenamefont
  {Reichental}, \citenamefont {Klempner}, \citenamefont {Kafri},\ and\
  \citenamefont {Podolsky}}]{reichental2018thermalization}%
  \BibitemOpen
  \bibfield  {author} {\bibinfo {author} {\bibfnamefont {I.}~\bibnamefont
  {Reichental}}, \bibinfo {author} {\bibfnamefont {A.}~\bibnamefont
  {Klempner}}, \bibinfo {author} {\bibfnamefont {Y.}~\bibnamefont {Kafri}},\
  and\ \bibinfo {author} {\bibfnamefont {D.}~\bibnamefont {Podolsky}},\
  }\bibfield  {title} {\bibinfo {title} {Thermalization in open quantum
  systems},\ }\href {https://doi.org/10.1103/PhysRevB.97.134301} {\bibfield
  {journal} {\bibinfo  {journal} {Phys. Rev. B}\ }\textbf {\bibinfo {volume}
  {97}},\ \bibinfo {pages} {134301} (\bibinfo {year} {2018})}\BibitemShut
  {NoStop}%
\bibitem [{\citenamefont {Shirai}\ and\ \citenamefont
  {Mori}(2020)}]{shirai2020thermalization}%
  \BibitemOpen
  \bibfield  {author} {\bibinfo {author} {\bibfnamefont {T.}~\bibnamefont
  {Shirai}}\ and\ \bibinfo {author} {\bibfnamefont {T.}~\bibnamefont {Mori}},\
  }\bibfield  {title} {\bibinfo {title} {Thermalization in open many-body
  systems based on eigenstate thermalization hypothesis},\ }\href
  {https://doi.org/10.1103/PhysRevE.101.042116} {\bibfield  {journal} {\bibinfo
   {journal} {Phys. Rev. E}\ }\textbf {\bibinfo {volume} {101}},\ \bibinfo
  {pages} {042116} (\bibinfo {year} {2020})}\BibitemShut {NoStop}%
\bibitem [{\citenamefont {Chen}\ \emph
  {et~al.}(2023{\natexlab{a}})\citenamefont {Chen}, \citenamefont {Chen},\ and\
  \citenamefont {Zhu}}]{chen2022non}%
  \BibitemOpen
  \bibfield  {author} {\bibinfo {author} {\bibfnamefont {Q.}~\bibnamefont
  {Chen}}, \bibinfo {author} {\bibfnamefont {S.~A.}\ \bibnamefont {Chen}},\
  and\ \bibinfo {author} {\bibfnamefont {Z.}~\bibnamefont {Zhu}},\ }\bibfield
  {title} {\bibinfo {title} {Weak ergodicity breaking in non-{Hermitian}
  many-body systems},\ }\href {https://doi.org/10.21468/SciPostPhys.15.2.052}
  {\bibfield  {journal} {\bibinfo  {journal} {SciPost Phys.}\ }\textbf
  {\bibinfo {volume} {15}},\ \bibinfo {pages} {052} (\bibinfo {year}
  {2023}{\natexlab{a}})}\BibitemShut {NoStop}%
\bibitem [{\citenamefont {Suthar}\ \emph {et~al.}(2022)\citenamefont {Suthar},
  \citenamefont {Wang}, \citenamefont {Huang}, \citenamefont {Jen},\ and\
  \citenamefont {You}}]{suthar2022non}%
  \BibitemOpen
  \bibfield  {author} {\bibinfo {author} {\bibfnamefont {K.}~\bibnamefont
  {Suthar}}, \bibinfo {author} {\bibfnamefont {Y.-C.}\ \bibnamefont {Wang}},
  \bibinfo {author} {\bibfnamefont {Y.-P.}\ \bibnamefont {Huang}}, \bibinfo
  {author} {\bibfnamefont {H.~H.}\ \bibnamefont {Jen}},\ and\ \bibinfo {author}
  {\bibfnamefont {J.-S.}\ \bibnamefont {You}},\ }\bibfield  {title} {\bibinfo
  {title} {Non-{Hermitian} many-body localization with open boundaries},\
  }\href {https://doi.org/10.1103/PhysRevB.106.064208} {\bibfield  {journal}
  {\bibinfo  {journal} {Phys. Rev. B}\ }\textbf {\bibinfo {volume} {106}},\
  \bibinfo {pages} {064208} (\bibinfo {year} {2022})}\BibitemShut {NoStop}%
\bibitem [{\citenamefont {Longhi}(2019{\natexlab{a}})}]{longhi2019probing}%
  \BibitemOpen
  \bibfield  {author} {\bibinfo {author} {\bibfnamefont {S.}~\bibnamefont
  {Longhi}},\ }\bibfield  {title} {\bibinfo {title} {Probing non-{Hermitian}
  skin effect and non-{Bloch} phase transitions},\ }\href
  {https://doi.org/10.1103/PhysRevResearch.1.023013} {\bibfield  {journal}
  {\bibinfo  {journal} {Phys. Rev. Res.}\ }\textbf {\bibinfo {volume} {1}},\
  \bibinfo {pages} {023013} (\bibinfo {year} {2019}{\natexlab{a}})}\BibitemShut
  {NoStop}%
\bibitem [{\citenamefont {Song}\ \emph {et~al.}(2019)\citenamefont {Song},
  \citenamefont {Yao},\ and\ \citenamefont {Wang}}]{song2019non}%
  \BibitemOpen
  \bibfield  {author} {\bibinfo {author} {\bibfnamefont {F.}~\bibnamefont
  {Song}}, \bibinfo {author} {\bibfnamefont {S.}~\bibnamefont {Yao}},\ and\
  \bibinfo {author} {\bibfnamefont {Z.}~\bibnamefont {Wang}},\ }\bibfield
  {title} {\bibinfo {title} {Non-{Hermitian} skin effect and chiral damping in
  open quantum systems},\ }\href
  {https://doi.org/10.1103/PhysRevLett.123.170401} {\bibfield  {journal}
  {\bibinfo  {journal} {Phys. Rev. Lett.}\ }\textbf {\bibinfo {volume} {123}},\
  \bibinfo {pages} {170401} (\bibinfo {year} {2019})}\BibitemShut {NoStop}%
\bibitem [{\citenamefont {Lee}\ \emph {et~al.}(2020)\citenamefont {Lee},
  \citenamefont {Li}, \citenamefont {Thomale},\ and\ \citenamefont
  {Gong}}]{lee2020unraveling}%
  \BibitemOpen
  \bibfield  {author} {\bibinfo {author} {\bibfnamefont {C.~H.}\ \bibnamefont
  {Lee}}, \bibinfo {author} {\bibfnamefont {L.}~\bibnamefont {Li}}, \bibinfo
  {author} {\bibfnamefont {R.}~\bibnamefont {Thomale}},\ and\ \bibinfo {author}
  {\bibfnamefont {J.}~\bibnamefont {Gong}},\ }\bibfield  {title} {\bibinfo
  {title} {Unraveling non-{Hermitian} pumping: Emergent spectral singularities
  and anomalous responses},\ }\href
  {https://doi.org/10.1103/PhysRevB.102.085151} {\bibfield  {journal} {\bibinfo
   {journal} {Phys. Rev. B}\ }\textbf {\bibinfo {volume} {102}},\ \bibinfo
  {pages} {085151} (\bibinfo {year} {2020})}\BibitemShut {NoStop}%
\bibitem [{\citenamefont {Li}\ \emph {et~al.}(2020)\citenamefont {Li},
  \citenamefont {Lee},\ and\ \citenamefont {Gong}}]{li2020topological}%
  \BibitemOpen
  \bibfield  {author} {\bibinfo {author} {\bibfnamefont {L.}~\bibnamefont
  {Li}}, \bibinfo {author} {\bibfnamefont {C.~H.}\ \bibnamefont {Lee}},\ and\
  \bibinfo {author} {\bibfnamefont {J.}~\bibnamefont {Gong}},\ }\bibfield
  {title} {\bibinfo {title} {Topological switch for non-{Hermitian} skin effect
  in cold-atom systems with loss},\ }\href
  {https://doi.org/10.1103/PhysRevLett.124.250402} {\bibfield  {journal}
  {\bibinfo  {journal} {Phys. Rev. Lett.}\ }\textbf {\bibinfo {volume} {124}},\
  \bibinfo {pages} {250402} (\bibinfo {year} {2020})}\BibitemShut {NoStop}%
\bibitem [{\citenamefont {Okuma}\ \emph {et~al.}(2020)\citenamefont {Okuma},
  \citenamefont {Kawabata}, \citenamefont {Shiozaki},\ and\ \citenamefont
  {Sato}}]{okuma2020topological}%
  \BibitemOpen
  \bibfield  {author} {\bibinfo {author} {\bibfnamefont {N.}~\bibnamefont
  {Okuma}}, \bibinfo {author} {\bibfnamefont {K.}~\bibnamefont {Kawabata}},
  \bibinfo {author} {\bibfnamefont {K.}~\bibnamefont {Shiozaki}},\ and\
  \bibinfo {author} {\bibfnamefont {M.}~\bibnamefont {Sato}},\ }\bibfield
  {title} {\bibinfo {title} {Topological origin of non-{Hermitian} skin
  effects},\ }\href {https://doi.org/10.1103/PhysRevLett.124.086801} {\bibfield
   {journal} {\bibinfo  {journal} {Phys. Rev. Lett.}\ }\textbf {\bibinfo
  {volume} {124}},\ \bibinfo {pages} {086801} (\bibinfo {year}
  {2020})}\BibitemShut {NoStop}%
\bibitem [{\citenamefont {Zhang}\ \emph
  {et~al.}(2021{\natexlab{a}})\citenamefont {Zhang}, \citenamefont {Tian},
  \citenamefont {Jiang}, \citenamefont {Lu},\ and\ \citenamefont
  {Chen}}]{zhang2021observation}%
  \BibitemOpen
  \bibfield  {author} {\bibinfo {author} {\bibfnamefont {X.}~\bibnamefont
  {Zhang}}, \bibinfo {author} {\bibfnamefont {Y.}~\bibnamefont {Tian}},
  \bibinfo {author} {\bibfnamefont {J.-H.}\ \bibnamefont {Jiang}}, \bibinfo
  {author} {\bibfnamefont {M.-H.}\ \bibnamefont {Lu}},\ and\ \bibinfo {author}
  {\bibfnamefont {Y.-F.}\ \bibnamefont {Chen}},\ }\bibfield  {title} {\bibinfo
  {title} {Observation of higher-order non-{Hermitian} skin effect},\ }\href
  {https://doi.org/10.1038/s41467-021-25716-y} {\bibfield  {journal} {\bibinfo
  {journal} {Nature Communications}\ }\textbf {\bibinfo {volume} {12}},\
  \bibinfo {pages} {5377} (\bibinfo {year} {2021}{\natexlab{a}})}\BibitemShut
  {NoStop}%
\bibitem [{\citenamefont {Zhang}\ \emph
  {et~al.}(2021{\natexlab{b}})\citenamefont {Zhang}, \citenamefont {Yang},
  \citenamefont {Ge}, \citenamefont {Guan}, \citenamefont {Chen}, \citenamefont
  {Yan}, \citenamefont {Chen}, \citenamefont {Xi}, \citenamefont {Li},
  \citenamefont {Jia}, \citenamefont {Yuan}, \citenamefont {Sun}, \citenamefont
  {Chen},\ and\ \citenamefont {Zhang}}]{zhang2021acoustic}%
  \BibitemOpen
  \bibfield  {author} {\bibinfo {author} {\bibfnamefont {L.}~\bibnamefont
  {Zhang}}, \bibinfo {author} {\bibfnamefont {Y.}~\bibnamefont {Yang}},
  \bibinfo {author} {\bibfnamefont {Y.}~\bibnamefont {Ge}}, \bibinfo {author}
  {\bibfnamefont {Y.-J.}\ \bibnamefont {Guan}}, \bibinfo {author}
  {\bibfnamefont {Q.}~\bibnamefont {Chen}}, \bibinfo {author} {\bibfnamefont
  {Q.}~\bibnamefont {Yan}}, \bibinfo {author} {\bibfnamefont {F.}~\bibnamefont
  {Chen}}, \bibinfo {author} {\bibfnamefont {R.}~\bibnamefont {Xi}}, \bibinfo
  {author} {\bibfnamefont {Y.}~\bibnamefont {Li}}, \bibinfo {author}
  {\bibfnamefont {D.}~\bibnamefont {Jia}}, \bibinfo {author} {\bibfnamefont
  {S.-Q.}\ \bibnamefont {Yuan}}, \bibinfo {author} {\bibfnamefont {H.-X.}\
  \bibnamefont {Sun}}, \bibinfo {author} {\bibfnamefont {H.}~\bibnamefont
  {Chen}},\ and\ \bibinfo {author} {\bibfnamefont {B.}~\bibnamefont {Zhang}},\
  }\bibfield  {title} {\bibinfo {title} {Acoustic non-{Hermitian} skin effect
  from twisted winding topology},\ }\href
  {https://doi.org/10.1038/s41467-021-26619-8} {\bibfield  {journal} {\bibinfo
  {journal} {Nature Communications}\ }\textbf {\bibinfo {volume} {12}},\
  \bibinfo {pages} {6297} (\bibinfo {year} {2021}{\natexlab{b}})}\BibitemShut
  {NoStop}%
\bibitem [{\citenamefont {Lee}(2021)}]{lee2021many}%
  \BibitemOpen
  \bibfield  {author} {\bibinfo {author} {\bibfnamefont {C.~H.}\ \bibnamefont
  {Lee}},\ }\bibfield  {title} {\bibinfo {title} {Many-body topological and
  skin states without open boundaries},\ }\href
  {https://doi.org/10.1103/PhysRevB.104.195102} {\bibfield  {journal} {\bibinfo
   {journal} {Phys. Rev. B}\ }\textbf {\bibinfo {volume} {104}},\ \bibinfo
  {pages} {195102} (\bibinfo {year} {2021})}\BibitemShut {NoStop}%
\bibitem [{\citenamefont {Li}\ \emph {et~al.}(2021)\citenamefont {Li},
  \citenamefont {Lee},\ and\ \citenamefont {Gong}}]{li2021impurity}%
  \BibitemOpen
  \bibfield  {author} {\bibinfo {author} {\bibfnamefont {L.}~\bibnamefont
  {Li}}, \bibinfo {author} {\bibfnamefont {C.~H.}\ \bibnamefont {Lee}},\ and\
  \bibinfo {author} {\bibfnamefont {J.}~\bibnamefont {Gong}},\ }\bibfield
  {title} {\bibinfo {title} {Impurity induced scale-free localization},\ }\href
  {https://doi.org/10.1038/s42005-021-00547-x} {\bibfield  {journal} {\bibinfo
  {journal} {Communications Physics}\ }\textbf {\bibinfo {volume} {4}},\
  \bibinfo {pages} {1} (\bibinfo {year} {2021})}\BibitemShut {NoStop}%
\bibitem [{\citenamefont {Li}\ and\ \citenamefont {Lee}(2022)}]{li2022non}%
  \BibitemOpen
  \bibfield  {author} {\bibinfo {author} {\bibfnamefont {L.}~\bibnamefont
  {Li}}\ and\ \bibinfo {author} {\bibfnamefont {C.~H.}\ \bibnamefont {Lee}},\
  }\bibfield  {title} {\bibinfo {title} {Non-{Hermitian} pseudo-gaps},\ }\href
  {https://doi.org/https://doi.org/10.1016/j.scib.2022.01.017} {\bibfield
  {journal} {\bibinfo  {journal} {Science Bulletin}\ }\textbf {\bibinfo
  {volume} {67}},\ \bibinfo {pages} {685} (\bibinfo {year} {2022})}\BibitemShut
  {NoStop}%
\bibitem [{\citenamefont {Jiang}\ and\ \citenamefont
  {Lee}(2023)}]{jiang2022dimensional}%
  \BibitemOpen
  \bibfield  {author} {\bibinfo {author} {\bibfnamefont {H.}~\bibnamefont
  {Jiang}}\ and\ \bibinfo {author} {\bibfnamefont {C.~H.}\ \bibnamefont
  {Lee}},\ }\bibfield  {title} {\bibinfo {title} {Dimensional transmutation
  from non-hermiticity},\ }\href
  {https://doi.org/10.1103/PhysRevLett.131.076401} {\bibfield  {journal}
  {\bibinfo  {journal} {Phys. Rev. Lett.}\ }\textbf {\bibinfo {volume} {131}},\
  \bibinfo {pages} {076401} (\bibinfo {year} {2023})}\BibitemShut {NoStop}%
\bibitem [{\citenamefont {Tai}\ and\ \citenamefont
  {Lee}(2023)}]{tai2022zoology}%
  \BibitemOpen
  \bibfield  {author} {\bibinfo {author} {\bibfnamefont {T.}~\bibnamefont
  {Tai}}\ and\ \bibinfo {author} {\bibfnamefont {C.~H.}\ \bibnamefont {Lee}},\
  }\bibfield  {title} {\bibinfo {title} {Zoology of non-{Hermitian} spectra and
  their graph topology},\ }\href {https://doi.org/10.1103/PhysRevB.107.L220301}
  {\bibfield  {journal} {\bibinfo  {journal} {Phys. Rev. B}\ }\textbf {\bibinfo
  {volume} {107}},\ \bibinfo {pages} {L220301} (\bibinfo {year}
  {2023})}\BibitemShut {NoStop}%
\bibitem [{\citenamefont {Xiujuan~Zhang}\ and\ \citenamefont
  {Chen}(2022)}]{zhang2022review}%
  \BibitemOpen
  \bibfield  {author} {\bibinfo {author} {\bibfnamefont {M.-H.~L.}\
  \bibnamefont {Xiujuan~Zhang}, \bibfnamefont {Tian~Zhang}}\ and\ \bibinfo
  {author} {\bibfnamefont {Y.-F.}\ \bibnamefont {Chen}},\ }\bibfield  {title}
  {\bibinfo {title} {A review on non-{Hermitian} skin effect},\ }\href
  {https://doi.org/10.1080/23746149.2022.2109431} {\bibfield  {journal}
  {\bibinfo  {journal} {Advances in Physics: X}\ }\textbf {\bibinfo {volume}
  {7}},\ \bibinfo {pages} {2109431} (\bibinfo {year} {2022})}\BibitemShut
  {NoStop}%
\bibitem [{\citenamefont {Lei}\ \emph {et~al.}(2023)\citenamefont {Lei},
  \citenamefont {Lee},\ and\ \citenamefont {Li}}]{lei2023pt}%
  \BibitemOpen
  \bibfield  {author} {\bibinfo {author} {\bibfnamefont {Z.}~\bibnamefont
  {Lei}}, \bibinfo {author} {\bibfnamefont {C.~H.}\ \bibnamefont {Lee}},\ and\
  \bibinfo {author} {\bibfnamefont {L.}~\bibnamefont {Li}},\ }\bibfield
  {title} {\bibinfo {title} {{$\mathcal{PT}$}-activated non-{Hermitian} skin
  modes},\ }\href@noop {} {\bibfield  {journal} {\bibinfo  {journal} {arXiv
  e-prints}\ } (\bibinfo {year} {2023})},\ \Eprint
  {https://arxiv.org/abs/2304.13955} {arXiv:2304.13955 [cond-mat.mes-hall]}
  \BibitemShut {NoStop}%
\bibitem [{\citenamefont {Lapp}\ \emph {et~al.}(2019)\citenamefont {Lapp},
  \citenamefont {Ang’ong’a}, \citenamefont {An},\ and\ \citenamefont
  {Gadway}}]{lapp2019engineering}%
  \BibitemOpen
  \bibfield  {author} {\bibinfo {author} {\bibfnamefont {S.}~\bibnamefont
  {Lapp}}, \bibinfo {author} {\bibfnamefont {J.}~\bibnamefont {Ang’ong’a}},
  \bibinfo {author} {\bibfnamefont {F.~A.}\ \bibnamefont {An}},\ and\ \bibinfo
  {author} {\bibfnamefont {B.}~\bibnamefont {Gadway}},\ }\bibfield  {title}
  {\bibinfo {title} {Engineering tunable local loss in a synthetic lattice of
  momentum states},\ }\href {https://doi.org/10.1088/1367-2630/ab1147}
  {\bibfield  {journal} {\bibinfo  {journal} {New Journal of Physics}\ }\textbf
  {\bibinfo {volume} {21}},\ \bibinfo {pages} {045006} (\bibinfo {year}
  {2019})}\BibitemShut {NoStop}%
\bibitem [{\citenamefont {Ren}\ \emph {et~al.}(2022)\citenamefont {Ren},
  \citenamefont {Liu}, \citenamefont {Zhao}, \citenamefont {He}, \citenamefont
  {Pak}, \citenamefont {Li},\ and\ \citenamefont {Jo}}]{ren2022chiral}%
  \BibitemOpen
  \bibfield  {author} {\bibinfo {author} {\bibfnamefont {Z.}~\bibnamefont
  {Ren}}, \bibinfo {author} {\bibfnamefont {D.}~\bibnamefont {Liu}}, \bibinfo
  {author} {\bibfnamefont {E.}~\bibnamefont {Zhao}}, \bibinfo {author}
  {\bibfnamefont {C.}~\bibnamefont {He}}, \bibinfo {author} {\bibfnamefont
  {K.~K.}\ \bibnamefont {Pak}}, \bibinfo {author} {\bibfnamefont
  {J.}~\bibnamefont {Li}},\ and\ \bibinfo {author} {\bibfnamefont {G.-B.}\
  \bibnamefont {Jo}},\ }\bibfield  {title} {\bibinfo {title} {Chiral control of
  quantum states in non-{Hermitian} spin--orbit-coupled fermions},\ }\href
  {https://doi.org/10.1038/s41567-021-01491-x} {\bibfield  {journal} {\bibinfo
  {journal} {Nature Physics}\ }\textbf {\bibinfo {volume} {18}},\ \bibinfo
  {pages} {385} (\bibinfo {year} {2022})}\BibitemShut {NoStop}%
\bibitem [{\citenamefont {Liang}\ \emph {et~al.}(2022)\citenamefont {Liang},
  \citenamefont {Xie}, \citenamefont {Dong}, \citenamefont {Li}, \citenamefont
  {Li}, \citenamefont {Gadway}, \citenamefont {Yi},\ and\ \citenamefont
  {Yan}}]{liang2022observation}%
  \BibitemOpen
  \bibfield  {author} {\bibinfo {author} {\bibfnamefont {Q.}~\bibnamefont
  {Liang}}, \bibinfo {author} {\bibfnamefont {D.}~\bibnamefont {Xie}}, \bibinfo
  {author} {\bibfnamefont {Z.}~\bibnamefont {Dong}}, \bibinfo {author}
  {\bibfnamefont {H.}~\bibnamefont {Li}}, \bibinfo {author} {\bibfnamefont
  {H.}~\bibnamefont {Li}}, \bibinfo {author} {\bibfnamefont {B.}~\bibnamefont
  {Gadway}}, \bibinfo {author} {\bibfnamefont {W.}~\bibnamefont {Yi}},\ and\
  \bibinfo {author} {\bibfnamefont {B.}~\bibnamefont {Yan}},\ }\bibfield
  {title} {\bibinfo {title} {Dynamic signatures of non-{Hermitian} skin effect
  and topology in ultracold atoms},\ }\href
  {https://doi.org/10.1103/PhysRevLett.129.070401} {\bibfield  {journal}
  {\bibinfo  {journal} {Phys. Rev. Lett.}\ }\textbf {\bibinfo {volume} {129}},\
  \bibinfo {pages} {070401} (\bibinfo {year} {2022})}\BibitemShut {NoStop}%
\bibitem [{\citenamefont {Qin}\ \emph {et~al.}(2023{\natexlab{a}})\citenamefont
  {Qin}, \citenamefont {Shen},\ and\ \citenamefont {Lee}}]{qin2022non}%
  \BibitemOpen
  \bibfield  {author} {\bibinfo {author} {\bibfnamefont {F.}~\bibnamefont
  {Qin}}, \bibinfo {author} {\bibfnamefont {R.}~\bibnamefont {Shen}},\ and\
  \bibinfo {author} {\bibfnamefont {C.~H.}\ \bibnamefont {Lee}},\ }\bibfield
  {title} {\bibinfo {title} {Non-{Hermitian} squeezed polarons},\ }\href
  {https://doi.org/10.1103/PhysRevA.107.L010202} {\bibfield  {journal}
  {\bibinfo  {journal} {Phys. Rev. A}\ }\textbf {\bibinfo {volume} {107}},\
  \bibinfo {pages} {L010202} (\bibinfo {year}
  {2023}{\natexlab{a}})}\BibitemShut {NoStop}%
\bibitem [{\citenamefont {Shen}\ \emph
  {et~al.}(2023{\natexlab{a}})\citenamefont {Shen}, \citenamefont {Chen},
  \citenamefont {Aliyu}, \citenamefont {Qin}, \citenamefont {Zhong},
  \citenamefont {Loh},\ and\ \citenamefont {Lee}}]{shen2023proposal}%
  \BibitemOpen
  \bibfield  {author} {\bibinfo {author} {\bibfnamefont {R.}~\bibnamefont
  {Shen}}, \bibinfo {author} {\bibfnamefont {T.}~\bibnamefont {Chen}}, \bibinfo
  {author} {\bibfnamefont {M.~M.}\ \bibnamefont {Aliyu}}, \bibinfo {author}
  {\bibfnamefont {F.}~\bibnamefont {Qin}}, \bibinfo {author} {\bibfnamefont
  {Y.}~\bibnamefont {Zhong}}, \bibinfo {author} {\bibfnamefont
  {H.}~\bibnamefont {Loh}},\ and\ \bibinfo {author} {\bibfnamefont {C.~H.}\
  \bibnamefont {Lee}},\ }\bibfield  {title} {\bibinfo {title} {Proposal for
  observing {Yang-Lee} criticality in {Rydberg} atomic arrays},\ }\href
  {https://doi.org/10.1103/PhysRevLett.131.080403} {\bibfield  {journal}
  {\bibinfo  {journal} {Phys. Rev. Lett.}\ }\textbf {\bibinfo {volume} {131}},\
  \bibinfo {pages} {080403} (\bibinfo {year} {2023}{\natexlab{a}})}\BibitemShut
  {NoStop}%
\bibitem [{\citenamefont {Turner}\ \emph
  {et~al.}(2018{\natexlab{b}})\citenamefont {Turner}, \citenamefont
  {Michailidis}, \citenamefont {Abanin}, \citenamefont {Serbyn},\ and\
  \citenamefont {Papi\ifmmode~\acute{c}\else \'{c}\fi{}}}]{turner2018quantum}%
  \BibitemOpen
  \bibfield  {author} {\bibinfo {author} {\bibfnamefont {C.~J.}\ \bibnamefont
  {Turner}}, \bibinfo {author} {\bibfnamefont {A.~A.}\ \bibnamefont
  {Michailidis}}, \bibinfo {author} {\bibfnamefont {D.~A.}\ \bibnamefont
  {Abanin}}, \bibinfo {author} {\bibfnamefont {M.}~\bibnamefont {Serbyn}},\
  and\ \bibinfo {author} {\bibfnamefont {Z.}~\bibnamefont
  {Papi\ifmmode~\acute{c}\else \'{c}\fi{}}},\ }\bibfield  {title} {\bibinfo
  {title} {Quantum scarred eigenstates in a {Rydberg} atom chain: Entanglement,
  breakdown of thermalization, and stability to perturbations},\ }\href
  {https://doi.org/10.1103/PhysRevB.98.155134} {\bibfield  {journal} {\bibinfo
  {journal} {Phys. Rev. B}\ }\textbf {\bibinfo {volume} {98}},\ \bibinfo
  {pages} {155134} (\bibinfo {year} {2018}{\natexlab{b}})}\BibitemShut
  {NoStop}%
\bibitem [{\citenamefont {Buca}\ \emph {et~al.}(2019)\citenamefont {Buca},
  \citenamefont {Tindall},\ and\ \citenamefont {Jaksch}}]{Buca2019}%
  \BibitemOpen
  \bibfield  {author} {\bibinfo {author} {\bibfnamefont {B.}~\bibnamefont
  {Buca}}, \bibinfo {author} {\bibfnamefont {J.}~\bibnamefont {Tindall}},\ and\
  \bibinfo {author} {\bibfnamefont {D.}~\bibnamefont {Jaksch}},\ }\bibfield
  {title} {\bibinfo {title} {Non-stationary coherent quantum many-body dynamics
  through dissipation},\ }\href {https://doi.org/10.1038/s41467-019-09757-y}
  {\bibfield  {journal} {\bibinfo  {journal} {Nature Communications}\ }\textbf
  {\bibinfo {volume} {10}},\ \bibinfo {pages} {1730} (\bibinfo {year}
  {2019})}\BibitemShut {NoStop}%
\bibitem [{\citenamefont {Wang}\ \emph {et~al.}(2023)\citenamefont {Wang},
  \citenamefont {Yuan}, \citenamefont {Zhang}, \citenamefont {Wang},
  \citenamefont {Deng},\ and\ \citenamefont {Duan}}]{wang2023embedding}%
  \BibitemOpen
  \bibfield  {author} {\bibinfo {author} {\bibfnamefont {H.-R.}\ \bibnamefont
  {Wang}}, \bibinfo {author} {\bibfnamefont {D.}~\bibnamefont {Yuan}}, \bibinfo
  {author} {\bibfnamefont {S.-Y.}\ \bibnamefont {Zhang}}, \bibinfo {author}
  {\bibfnamefont {Z.}~\bibnamefont {Wang}}, \bibinfo {author} {\bibfnamefont
  {D.-L.}\ \bibnamefont {Deng}},\ and\ \bibinfo {author} {\bibfnamefont
  {L.~M.}\ \bibnamefont {Duan}},\ }\bibfield  {title} {\bibinfo {title}
  {Embedding quantum many-body scars into decoherence-free subspaces},\
  }\href@noop {} {\bibfield  {journal} {\bibinfo  {journal} {arXiv e-prints}\ }
  (\bibinfo {year} {2023})},\ \Eprint {https://arxiv.org/abs/2304.08515}
  {arXiv:2304.08515 [quant-ph]} \BibitemShut {NoStop}%
\bibitem [{\citenamefont {Fendley}\ \emph {et~al.}(2004)\citenamefont
  {Fendley}, \citenamefont {Sengupta},\ and\ \citenamefont
  {Sachdev}}]{FendleySachdev}%
  \BibitemOpen
  \bibfield  {author} {\bibinfo {author} {\bibfnamefont {P.}~\bibnamefont
  {Fendley}}, \bibinfo {author} {\bibfnamefont {K.}~\bibnamefont {Sengupta}},\
  and\ \bibinfo {author} {\bibfnamefont {S.}~\bibnamefont {Sachdev}},\
  }\bibfield  {title} {\bibinfo {title} {Competing density-wave orders in a
  one-dimensional hard-boson model},\ }\href
  {https://doi.org/10.1103/PhysRevB.69.075106} {\bibfield  {journal} {\bibinfo
  {journal} {Phys. Rev. B}\ }\textbf {\bibinfo {volume} {69}},\ \bibinfo
  {pages} {075106} (\bibinfo {year} {2004})}\BibitemShut {NoStop}%
\bibitem [{\citenamefont {Lesanovsky}\ and\ \citenamefont
  {Katsura}(2012)}]{Lesanovsky2012}%
  \BibitemOpen
  \bibfield  {author} {\bibinfo {author} {\bibfnamefont {I.}~\bibnamefont
  {Lesanovsky}}\ and\ \bibinfo {author} {\bibfnamefont {H.}~\bibnamefont
  {Katsura}},\ }\bibfield  {title} {\bibinfo {title} {Interacting {Fibonacci}
  anyons in a {Rydberg} gas},\ }\href
  {https://doi.org/10.1103/PhysRevA.86.041601} {\bibfield  {journal} {\bibinfo
  {journal} {Phys. Rev. A}\ }\textbf {\bibinfo {volume} {86}},\ \bibinfo
  {pages} {041601} (\bibinfo {year} {2012})}\BibitemShut {NoStop}%
\bibitem [{Sup()}]{SuppMat}%
  \BibitemOpen
  \href@noop {} {}\bibinfo {note} {See Supplementary Material for additional
  results and details. The Supplementary material contains
  Refs.~\cite{su2022observation,liang2022observation,qin2022non,zhou2021engineering,hu2017creation,preskill2018quantum,
  preskill2023quantum,santos2016ibm,bravyi2011schrieffer,chen2022high,shen2023observation,lin2021real,chen2023weak,
  yao2023observation,mondragon2021fate,heiss2012physics,klett2017relation,longhi2019topological,sakhdari2019experimental,
  miri2019exceptional,ozdemir2019parity,lee2022exceptional,ding2022non,yang2023percolation,meng2024exceptional,
  kawabata2023entanglement,orito2023entanglement,chang2020entanglement,lee2022exceptional,hsieh2023relating,zou2023experimental,
  fossati2023symmetry,rottoli2024entanglement,TurnerNature}}\BibitemShut
  {NoStop}%
\bibitem [{\citenamefont {Qin}\ \emph {et~al.}(2023{\natexlab{b}})\citenamefont
  {Qin}, \citenamefont {Ma}, \citenamefont {Shen},\ and\ \citenamefont
  {Lee}}]{qin2023universal}%
  \BibitemOpen
  \bibfield  {author} {\bibinfo {author} {\bibfnamefont {F.}~\bibnamefont
  {Qin}}, \bibinfo {author} {\bibfnamefont {Y.}~\bibnamefont {Ma}}, \bibinfo
  {author} {\bibfnamefont {R.}~\bibnamefont {Shen}},\ and\ \bibinfo {author}
  {\bibfnamefont {C.~H.}\ \bibnamefont {Lee}},\ }\bibfield  {title} {\bibinfo
  {title} {Universal competitive spectral scaling from the critical
  non-{Hermitian} skin effect},\ }\href
  {https://doi.org/10.1103/PhysRevB.107.155430} {\bibfield  {journal} {\bibinfo
   {journal} {Phys. Rev. B}\ }\textbf {\bibinfo {volume} {107}},\ \bibinfo
  {pages} {155430} (\bibinfo {year} {2023}{\natexlab{b}})}\BibitemShut
  {NoStop}%
\bibitem [{\citenamefont {Qin}\ \emph {et~al.}(2023{\natexlab{c}})\citenamefont
  {Qin}, \citenamefont {Shen}, \citenamefont {Li},\ and\ \citenamefont
  {Lee}}]{qin2023kinked}%
  \BibitemOpen
  \bibfield  {author} {\bibinfo {author} {\bibfnamefont {F.}~\bibnamefont
  {Qin}}, \bibinfo {author} {\bibfnamefont {R.}~\bibnamefont {Shen}}, \bibinfo
  {author} {\bibfnamefont {L.}~\bibnamefont {Li}},\ and\ \bibinfo {author}
  {\bibfnamefont {C.~H.}\ \bibnamefont {Lee}},\ }\bibfield  {title} {\bibinfo
  {title} {Kinked linear response from non-{Hermitian} pumping},\ }\href@noop
  {} {\bibfield  {journal} {\bibinfo  {journal} {arXiv e-prints}\ } (\bibinfo
  {year} {2023}{\natexlab{c}})},\ \Eprint {https://arxiv.org/abs/2306.13139}
  {arXiv:2306.13139 [cond-mat.quant-gas]} \BibitemShut {NoStop}%
\bibitem [{\citenamefont {Yao}\ \emph {et~al.}(2023)\citenamefont {Yao},
  \citenamefont {Xiang}, \citenamefont {Guo}, \citenamefont {Bao},
  \citenamefont {Yang}, \citenamefont {Song}, \citenamefont {Shi},
  \citenamefont {Zhu}, \citenamefont {Jin}, \citenamefont {Chen}, \citenamefont
  {Xu}, \citenamefont {Zhu}, \citenamefont {Shen}, \citenamefont {Wang},
  \citenamefont {Zhang}, \citenamefont {Wu}, \citenamefont {Zou}, \citenamefont
  {Zhang}, \citenamefont {Li}, \citenamefont {Wang}, \citenamefont {Song},
  \citenamefont {Cheng}, \citenamefont {Mondaini}, \citenamefont {Wang},
  \citenamefont {You}, \citenamefont {Zhu}, \citenamefont {Ying},\ and\
  \citenamefont {Guo}}]{yao2023observation}%
  \BibitemOpen
  \bibfield  {author} {\bibinfo {author} {\bibfnamefont {Y.}~\bibnamefont
  {Yao}}, \bibinfo {author} {\bibfnamefont {L.}~\bibnamefont {Xiang}}, \bibinfo
  {author} {\bibfnamefont {Z.}~\bibnamefont {Guo}}, \bibinfo {author}
  {\bibfnamefont {Z.}~\bibnamefont {Bao}}, \bibinfo {author} {\bibfnamefont
  {Y.-F.}\ \bibnamefont {Yang}}, \bibinfo {author} {\bibfnamefont
  {Z.}~\bibnamefont {Song}}, \bibinfo {author} {\bibfnamefont {H.}~\bibnamefont
  {Shi}}, \bibinfo {author} {\bibfnamefont {X.}~\bibnamefont {Zhu}}, \bibinfo
  {author} {\bibfnamefont {F.}~\bibnamefont {Jin}}, \bibinfo {author}
  {\bibfnamefont {J.}~\bibnamefont {Chen}}, \bibinfo {author} {\bibfnamefont
  {S.}~\bibnamefont {Xu}}, \bibinfo {author} {\bibfnamefont {Z.}~\bibnamefont
  {Zhu}}, \bibinfo {author} {\bibfnamefont {F.}~\bibnamefont {Shen}}, \bibinfo
  {author} {\bibfnamefont {N.}~\bibnamefont {Wang}}, \bibinfo {author}
  {\bibfnamefont {C.}~\bibnamefont {Zhang}}, \bibinfo {author} {\bibfnamefont
  {Y.}~\bibnamefont {Wu}}, \bibinfo {author} {\bibfnamefont {Y.}~\bibnamefont
  {Zou}}, \bibinfo {author} {\bibfnamefont {P.}~\bibnamefont {Zhang}}, \bibinfo
  {author} {\bibfnamefont {H.}~\bibnamefont {Li}}, \bibinfo {author}
  {\bibfnamefont {Z.}~\bibnamefont {Wang}}, \bibinfo {author} {\bibfnamefont
  {C.}~\bibnamefont {Song}}, \bibinfo {author} {\bibfnamefont {C.}~\bibnamefont
  {Cheng}}, \bibinfo {author} {\bibfnamefont {R.}~\bibnamefont {Mondaini}},
  \bibinfo {author} {\bibfnamefont {H.}~\bibnamefont {Wang}}, \bibinfo {author}
  {\bibfnamefont {J.~Q.}\ \bibnamefont {You}}, \bibinfo {author} {\bibfnamefont
  {S.-Y.}\ \bibnamefont {Zhu}}, \bibinfo {author} {\bibfnamefont
  {L.}~\bibnamefont {Ying}},\ and\ \bibinfo {author} {\bibfnamefont
  {Q.}~\bibnamefont {Guo}},\ }\bibfield  {title} {\bibinfo {title} {Observation
  of many-body {Fock} space dynamics in two dimensions},\ }\href
  {https://doi.org/10.1038/s41567-023-02133-0} {\bibfield  {journal} {\bibinfo
  {journal} {Nature Physics}\ }\textbf {\bibinfo {volume} {19}},\ \bibinfo
  {pages} {1459} (\bibinfo {year} {2023})}\BibitemShut {NoStop}%
\bibitem [{\citenamefont {Mondragon-Shem}\ \emph
  {et~al.}(2021{\natexlab{a}})\citenamefont {Mondragon-Shem}, \citenamefont
  {Vavilov},\ and\ \citenamefont {Martin}}]{mondragon2021fate}%
  \BibitemOpen
  \bibfield  {author} {\bibinfo {author} {\bibfnamefont {I.}~\bibnamefont
  {Mondragon-Shem}}, \bibinfo {author} {\bibfnamefont {M.~G.}\ \bibnamefont
  {Vavilov}},\ and\ \bibinfo {author} {\bibfnamefont {I.}~\bibnamefont
  {Martin}},\ }\bibfield  {title} {\bibinfo {title} {Fate of quantum many-body
  scars in the presence of disorder},\ }\href
  {https://doi.org/10.1103/PRXQuantum.2.030349} {\bibfield  {journal} {\bibinfo
   {journal} {PRX Quantum}\ }\textbf {\bibinfo {volume} {2}},\ \bibinfo {pages}
  {030349} (\bibinfo {year} {2021}{\natexlab{a}})}\BibitemShut {NoStop}%
\bibitem [{\citenamefont {Baier}\ \emph {et~al.}(2016)\citenamefont {Baier},
  \citenamefont {Mark}, \citenamefont {Petter}, \citenamefont {Aikawa},
  \citenamefont {Chomaz}, \citenamefont {Cai}, \citenamefont {Baranov},
  \citenamefont {Zoller},\ and\ \citenamefont {Ferlaino}}]{baier2016extended}%
  \BibitemOpen
  \bibfield  {author} {\bibinfo {author} {\bibfnamefont {S.}~\bibnamefont
  {Baier}}, \bibinfo {author} {\bibfnamefont {M.~J.}\ \bibnamefont {Mark}},
  \bibinfo {author} {\bibfnamefont {D.}~\bibnamefont {Petter}}, \bibinfo
  {author} {\bibfnamefont {K.}~\bibnamefont {Aikawa}}, \bibinfo {author}
  {\bibfnamefont {L.}~\bibnamefont {Chomaz}}, \bibinfo {author} {\bibfnamefont
  {Z.}~\bibnamefont {Cai}}, \bibinfo {author} {\bibfnamefont {M.}~\bibnamefont
  {Baranov}}, \bibinfo {author} {\bibfnamefont {P.}~\bibnamefont {Zoller}},\
  and\ \bibinfo {author} {\bibfnamefont {F.}~\bibnamefont {Ferlaino}},\
  }\bibfield  {title} {\bibinfo {title} {Extended bose-hubbard models with
  ultracold magnetic atoms},\ }\href {https://doi.org/10.1126/science.aac9812}
  {\bibfield  {journal} {\bibinfo  {journal} {Science}\ }\textbf {\bibinfo
  {volume} {352}},\ \bibinfo {pages} {201} (\bibinfo {year}
  {2016})}\BibitemShut {NoStop}%
\bibitem [{\citenamefont {Tomita}\ \emph {et~al.}(2017)\citenamefont {Tomita},
  \citenamefont {Nakajima}, \citenamefont {Danshita}, \citenamefont {Takasu},\
  and\ \citenamefont {Takahashi}}]{tomita2017observation}%
  \BibitemOpen
  \bibfield  {author} {\bibinfo {author} {\bibfnamefont {T.}~\bibnamefont
  {Tomita}}, \bibinfo {author} {\bibfnamefont {S.}~\bibnamefont {Nakajima}},
  \bibinfo {author} {\bibfnamefont {I.}~\bibnamefont {Danshita}}, \bibinfo
  {author} {\bibfnamefont {Y.}~\bibnamefont {Takasu}},\ and\ \bibinfo {author}
  {\bibfnamefont {Y.}~\bibnamefont {Takahashi}},\ }\bibfield  {title} {\bibinfo
  {title} {Observation of the {Mott} insulator to superfluid crossover of a
  driven-dissipative {Bose-Hubbard} system},\ }\href
  {https://doi.org/10.1126/sciadv.1701513} {\bibfield  {journal} {\bibinfo
  {journal} {Science Advances}\ }\textbf {\bibinfo {volume} {3}},\ \bibinfo
  {pages} {e1701513} (\bibinfo {year} {2017})}\BibitemShut {NoStop}%
\bibitem [{\citenamefont {Yang}\ \emph {et~al.}(2020)\citenamefont {Yang},
  \citenamefont {Sun}, \citenamefont {Ott}, \citenamefont {Wang}, \citenamefont
  {Zache}, \citenamefont {Halimeh}, \citenamefont {Yuan}, \citenamefont
  {Hauke},\ and\ \citenamefont {Pan}}]{yang2020}%
  \BibitemOpen
  \bibfield  {author} {\bibinfo {author} {\bibfnamefont {B.}~\bibnamefont
  {Yang}}, \bibinfo {author} {\bibfnamefont {H.}~\bibnamefont {Sun}}, \bibinfo
  {author} {\bibfnamefont {R.}~\bibnamefont {Ott}}, \bibinfo {author}
  {\bibfnamefont {H.-Y.}\ \bibnamefont {Wang}}, \bibinfo {author}
  {\bibfnamefont {T.~V.}\ \bibnamefont {Zache}}, \bibinfo {author}
  {\bibfnamefont {J.~C.}\ \bibnamefont {Halimeh}}, \bibinfo {author}
  {\bibfnamefont {Z.-S.}\ \bibnamefont {Yuan}}, \bibinfo {author}
  {\bibfnamefont {P.}~\bibnamefont {Hauke}},\ and\ \bibinfo {author}
  {\bibfnamefont {J.-W.}\ \bibnamefont {Pan}},\ }\bibfield  {title} {\bibinfo
  {title} {Observation of gauge invariance in a 71-site {Bose--Hubbard} quantum
  simulator},\ }\href {https://doi.org/10.1038/s41586-020-2910-8} {\bibfield
  {journal} {\bibinfo  {journal} {Nature}\ }\textbf {\bibinfo {volume} {587}},\
  \bibinfo {pages} {392} (\bibinfo {year} {2020})}\BibitemShut {NoStop}%
\bibitem [{\citenamefont {Sachdev}\ \emph {et~al.}(2002)\citenamefont
  {Sachdev}, \citenamefont {Sengupta},\ and\ \citenamefont
  {Girvin}}]{sachdev2002mott}%
  \BibitemOpen
  \bibfield  {author} {\bibinfo {author} {\bibfnamefont {S.}~\bibnamefont
  {Sachdev}}, \bibinfo {author} {\bibfnamefont {K.}~\bibnamefont {Sengupta}},\
  and\ \bibinfo {author} {\bibfnamefont {S.~M.}\ \bibnamefont {Girvin}},\
  }\bibfield  {title} {\bibinfo {title} {Mott insulators in strong electric
  fields},\ }\href {https://doi.org/10.1103/PhysRevB.66.075128} {\bibfield
  {journal} {\bibinfo  {journal} {Phys. Rev. B}\ }\textbf {\bibinfo {volume}
  {66}},\ \bibinfo {pages} {075128} (\bibinfo {year} {2002})}\BibitemShut
  {NoStop}%
\bibitem [{\citenamefont {Sengupta}(2022)}]{sengupta2021phases}%
  \BibitemOpen
  \bibfield  {author} {\bibinfo {author} {\bibfnamefont {K.}~\bibnamefont
  {Sengupta}},\ }\bibinfo {title} {Phases and dynamics of ultracold {Bosons} in
  a tilted optical lattice},\ in\ \href
  {https://doi.org/10.1007/978-3-031-03998-0_15} {\emph {\bibinfo {booktitle}
  {Entanglement in Spin Chains: From Theory to Quantum Technology
  Applications}}},\ \bibinfo {editor} {edited by\ \bibinfo {editor}
  {\bibfnamefont {A.}~\bibnamefont {Bayat}}, \bibinfo {editor} {\bibfnamefont
  {S.}~\bibnamefont {Bose}},\ and\ \bibinfo {editor} {\bibfnamefont
  {H.}~\bibnamefont {Johannesson}}}\ (\bibinfo  {publisher} {Springer
  International Publishing},\ \bibinfo {address} {Cham},\ \bibinfo {year}
  {2022})\ pp.\ \bibinfo {pages} {425--458}\BibitemShut {NoStop}%
\bibitem [{\citenamefont {Preskill}(2018)}]{preskill2018quantum}%
  \BibitemOpen
  \bibfield  {author} {\bibinfo {author} {\bibfnamefont {J.}~\bibnamefont
  {Preskill}},\ }\bibfield  {title} {\bibinfo {title} {Quantum {C}omputing in
  the {NISQ} era and beyond},\ }\href
  {https://doi.org/10.22331/q-2018-08-06-79} {\bibfield  {journal} {\bibinfo
  {journal} {{Quantum}}\ }\textbf {\bibinfo {volume} {2}},\ \bibinfo {pages}
  {79} (\bibinfo {year} {2018})}\BibitemShut {NoStop}%
\bibitem [{\citenamefont {Preskill}(2023)}]{preskill2023quantum}%
  \BibitemOpen
  \bibfield  {author} {\bibinfo {author} {\bibfnamefont {J.}~\bibnamefont
  {Preskill}},\ }\bibfield  {title} {\bibinfo {title} {Quantum computing 40
  years later},\ }in\ \href@noop {} {\emph {\bibinfo {booktitle} {Feynman
  Lectures on Computation}}}\ (\bibinfo  {publisher} {CRC Press},\ \bibinfo
  {year} {2023})\ pp.\ \bibinfo {pages} {193--244}\BibitemShut {NoStop}%
\bibitem [{\citenamefont {Santos}(2016)}]{santos2016ibm}%
  \BibitemOpen
  \bibfield  {author} {\bibinfo {author} {\bibfnamefont {A.~C.}\ \bibnamefont
  {Santos}},\ }\bibfield  {title} {\bibinfo {title} {The ibm quantum computer
  and the ibm quantum experience},\ }\href@noop {} {\bibfield  {journal}
  {\bibinfo  {journal} {arXiv preprint arXiv:1610.06980}\ } (\bibinfo {year}
  {2016})}\BibitemShut {NoStop}%
\bibitem [{\citenamefont {Lin}\ \emph {et~al.}(2021)\citenamefont {Lin},
  \citenamefont {Dilip}, \citenamefont {Green}, \citenamefont {Smith},\ and\
  \citenamefont {Pollmann}}]{lin2021real}%
  \BibitemOpen
  \bibfield  {author} {\bibinfo {author} {\bibfnamefont {S.-H.}\ \bibnamefont
  {Lin}}, \bibinfo {author} {\bibfnamefont {R.}~\bibnamefont {Dilip}}, \bibinfo
  {author} {\bibfnamefont {A.~G.}\ \bibnamefont {Green}}, \bibinfo {author}
  {\bibfnamefont {A.}~\bibnamefont {Smith}},\ and\ \bibinfo {author}
  {\bibfnamefont {F.}~\bibnamefont {Pollmann}},\ }\bibfield  {title} {\bibinfo
  {title} {Real- and imaginary-time evolution with compressed quantum
  circuits},\ }\href {https://doi.org/10.1103/PRXQuantum.2.010342} {\bibfield
  {journal} {\bibinfo  {journal} {PRX Quantum}\ }\textbf {\bibinfo {volume}
  {2}},\ \bibinfo {pages} {010342} (\bibinfo {year} {2021})}\BibitemShut
  {NoStop}%
\bibitem [{\citenamefont {Chen}\ \emph
  {et~al.}(2023{\natexlab{b}})\citenamefont {Chen}, \citenamefont {Shen},
  \citenamefont {Lee},\ and\ \citenamefont {Yang}}]{chen2022high}%
  \BibitemOpen
  \bibfield  {author} {\bibinfo {author} {\bibfnamefont {T.}~\bibnamefont
  {Chen}}, \bibinfo {author} {\bibfnamefont {R.}~\bibnamefont {Shen}}, \bibinfo
  {author} {\bibfnamefont {C.~H.}\ \bibnamefont {Lee}},\ and\ \bibinfo {author}
  {\bibfnamefont {B.}~\bibnamefont {Yang}},\ }\bibfield  {title} {\bibinfo
  {title} {{High-fidelity realization of the {AKLT} state on a {NISQ}-era
  quantum processor}},\ }\href {https://doi.org/10.21468/SciPostPhys.15.4.170}
  {\bibfield  {journal} {\bibinfo  {journal} {SciPost Phys.}\ }\textbf
  {\bibinfo {volume} {15}},\ \bibinfo {pages} {170} (\bibinfo {year}
  {2023}{\natexlab{b}})}\BibitemShut {NoStop}%
\bibitem [{\citenamefont {Shen}\ \emph
  {et~al.}(2023{\natexlab{b}})\citenamefont {Shen}, \citenamefont {Chen},
  \citenamefont {Yang},\ and\ \citenamefont {Lee}}]{shen2023observation}%
  \BibitemOpen
  \bibfield  {author} {\bibinfo {author} {\bibfnamefont {R.}~\bibnamefont
  {Shen}}, \bibinfo {author} {\bibfnamefont {T.}~\bibnamefont {Chen}}, \bibinfo
  {author} {\bibfnamefont {B.}~\bibnamefont {Yang}},\ and\ \bibinfo {author}
  {\bibfnamefont {C.~H.}\ \bibnamefont {Lee}},\ }\bibfield  {title} {\bibinfo
  {title} {Observation of the non-{Hermitian} skin effect and {Fermi} skin on a
  digital quantum computer},\ }\href@noop {} {\bibfield  {journal} {\bibinfo
  {journal} {arXiv e-prints}\ } (\bibinfo {year} {2023}{\natexlab{b}})},\
  \Eprint {https://arxiv.org/abs/2311.10143} {arXiv:2311.10143 [quant-ph]}
  \BibitemShut {NoStop}%
\bibitem [{\citenamefont {Koh}\ \emph {et~al.}(2022)\citenamefont {Koh},
  \citenamefont {Tai}, \citenamefont {Phee}, \citenamefont {Ng},\ and\
  \citenamefont {Lee}}]{koh2022stabilizing}%
  \BibitemOpen
  \bibfield  {author} {\bibinfo {author} {\bibfnamefont {J.~M.}\ \bibnamefont
  {Koh}}, \bibinfo {author} {\bibfnamefont {T.}~\bibnamefont {Tai}}, \bibinfo
  {author} {\bibfnamefont {Y.~H.}\ \bibnamefont {Phee}}, \bibinfo {author}
  {\bibfnamefont {W.~E.}\ \bibnamefont {Ng}},\ and\ \bibinfo {author}
  {\bibfnamefont {C.~H.}\ \bibnamefont {Lee}},\ }\bibfield  {title} {\bibinfo
  {title} {Stabilizing multiple topological fermions on a quantum computer},\
  }\href {https://doi.org/10.1038/s41534-022-00527-1} {\bibfield  {journal}
  {\bibinfo  {journal} {npj Quantum Information}\ }\textbf {\bibinfo {volume}
  {8}},\ \bibinfo {pages} {16} (\bibinfo {year} {2022})}\BibitemShut {NoStop}%
\bibitem [{\citenamefont {Koh}\ \emph {et~al.}(2023)\citenamefont {Koh},
  \citenamefont {Tai},\ and\ \citenamefont {Lee}}]{koh2023observation}%
  \BibitemOpen
  \bibfield  {author} {\bibinfo {author} {\bibfnamefont {J.~M.}\ \bibnamefont
  {Koh}}, \bibinfo {author} {\bibfnamefont {T.}~\bibnamefont {Tai}},\ and\
  \bibinfo {author} {\bibfnamefont {C.~H.}\ \bibnamefont {Lee}},\ }\bibfield
  {title} {\bibinfo {title} {Observation of higher-order topological states on
  a quantum computer},\ }\href@noop {} {\bibfield  {journal} {\bibinfo
  {journal} {arXiv e-prints}\ } (\bibinfo {year} {2023})},\ \Eprint
  {https://arxiv.org/abs/2303.02179} {arXiv:2303.02179 [cond-mat.str-el]}
  \BibitemShut {NoStop}%
\bibitem [{\citenamefont {Chen}\ \emph
  {et~al.}(2023{\natexlab{c}})\citenamefont {Chen}, \citenamefont {Shen},
  \citenamefont {Lee}, \citenamefont {Yang},\ and\ \citenamefont
  {Bomantara}}]{chen2023robust}%
  \BibitemOpen
  \bibfield  {author} {\bibinfo {author} {\bibfnamefont {T.}~\bibnamefont
  {Chen}}, \bibinfo {author} {\bibfnamefont {R.}~\bibnamefont {Shen}}, \bibinfo
  {author} {\bibfnamefont {C.~H.}\ \bibnamefont {Lee}}, \bibinfo {author}
  {\bibfnamefont {B.}~\bibnamefont {Yang}},\ and\ \bibinfo {author}
  {\bibfnamefont {R.~W.}\ \bibnamefont {Bomantara}},\ }\bibfield  {title}
  {\bibinfo {title} {A robust large-period discrete time crystal and its
  signature in a digital quantum computer},\ }\href@noop {} {\bibfield
  {journal} {\bibinfo  {journal} {arXiv e-prints}\ } (\bibinfo {year}
  {2023}{\natexlab{c}})},\ \Eprint {https://arxiv.org/abs/2309.11560}
  {arXiv:2309.11560 [quant-ph]} \BibitemShut {NoStop}%
\bibitem [{\citenamefont {Shibata}\ \emph {et~al.}(2020)\citenamefont
  {Shibata}, \citenamefont {Yoshioka},\ and\ \citenamefont
  {Katsura}}]{OnsagerScars}%
  \BibitemOpen
  \bibfield  {author} {\bibinfo {author} {\bibfnamefont {N.}~\bibnamefont
  {Shibata}}, \bibinfo {author} {\bibfnamefont {N.}~\bibnamefont {Yoshioka}},\
  and\ \bibinfo {author} {\bibfnamefont {H.}~\bibnamefont {Katsura}},\
  }\bibfield  {title} {\bibinfo {title} {Onsager's scars in disordered spin
  chains},\ }\href {https://doi.org/10.1103/PhysRevLett.124.180604} {\bibfield
  {journal} {\bibinfo  {journal} {Phys. Rev. Lett.}\ }\textbf {\bibinfo
  {volume} {124}},\ \bibinfo {pages} {180604} (\bibinfo {year}
  {2020})}\BibitemShut {NoStop}%
\bibitem [{\citenamefont {Mondragon-Shem}\ \emph
  {et~al.}(2021{\natexlab{b}})\citenamefont {Mondragon-Shem}, \citenamefont
  {Vavilov},\ and\ \citenamefont {Martin}}]{MondragonShem2020}%
  \BibitemOpen
  \bibfield  {author} {\bibinfo {author} {\bibfnamefont {I.}~\bibnamefont
  {Mondragon-Shem}}, \bibinfo {author} {\bibfnamefont {M.~G.}\ \bibnamefont
  {Vavilov}},\ and\ \bibinfo {author} {\bibfnamefont {I.}~\bibnamefont
  {Martin}},\ }\bibfield  {title} {\bibinfo {title} {Fate of quantum many-body
  scars in the presence of disorder},\ }\href
  {https://doi.org/10.1103/PRXQuantum.2.030349} {\bibfield  {journal} {\bibinfo
   {journal} {PRX Quantum}\ }\textbf {\bibinfo {volume} {2}},\ \bibinfo {pages}
  {030349} (\bibinfo {year} {2021}{\natexlab{b}})}\BibitemShut {NoStop}%
\bibitem [{\citenamefont {Huang}\ \emph {et~al.}(2021)\citenamefont {Huang},
  \citenamefont {Wang},\ and\ \citenamefont {Li}}]{Huang2021}%
  \BibitemOpen
  \bibfield  {author} {\bibinfo {author} {\bibfnamefont {K.}~\bibnamefont
  {Huang}}, \bibinfo {author} {\bibfnamefont {Y.}~\bibnamefont {Wang}},\ and\
  \bibinfo {author} {\bibfnamefont {X.}~\bibnamefont {Li}},\ }\bibfield
  {title} {\bibinfo {title} {Stability of scar states in the two-dimensional
  {PXP} model against random disorder},\ }\href
  {https://doi.org/10.1103/PhysRevB.104.214305} {\bibfield  {journal} {\bibinfo
   {journal} {Phys. Rev. B}\ }\textbf {\bibinfo {volume} {104}},\ \bibinfo
  {pages} {214305} (\bibinfo {year} {2021})}\BibitemShut {NoStop}%
\bibitem [{\citenamefont {van Voorden}\ \emph {et~al.}(2021)\citenamefont {van
  Voorden}, \citenamefont {Marcuzzi}, \citenamefont {Schoutens},\ and\
  \citenamefont {Min\'a\ifmmode~\check{r}\else \v{r}\fi{}}}]{Voorden21}%
  \BibitemOpen
  \bibfield  {author} {\bibinfo {author} {\bibfnamefont {B.}~\bibnamefont {van
  Voorden}}, \bibinfo {author} {\bibfnamefont {M.}~\bibnamefont {Marcuzzi}},
  \bibinfo {author} {\bibfnamefont {K.}~\bibnamefont {Schoutens}},\ and\
  \bibinfo {author} {\bibfnamefont {J.~c.~v.}\ \bibnamefont
  {Min\'a\ifmmode~\check{r}\else \v{r}\fi{}}},\ }\bibfield  {title} {\bibinfo
  {title} {Disorder enhanced quantum many-body scars in {Hilbert} hypercubes},\
  }\href {https://doi.org/10.1103/PhysRevB.103.L220301} {\bibfield  {journal}
  {\bibinfo  {journal} {Phys. Rev. B}\ }\textbf {\bibinfo {volume} {103}},\
  \bibinfo {pages} {L220301} (\bibinfo {year} {2021})}\BibitemShut {NoStop}%
\bibitem [{\citenamefont {Zhang}\ and\ \citenamefont
  {Song}(2023)}]{Zhang2022Clusters}%
  \BibitemOpen
  \bibfield  {author} {\bibinfo {author} {\bibfnamefont {G.}~\bibnamefont
  {Zhang}}\ and\ \bibinfo {author} {\bibfnamefont {Z.}~\bibnamefont {Song}},\
  }\bibfield  {title} {\bibinfo {title} {Quantum scars in spin- isotropic
  {Heisenberg} clusters},\ }\href {https://doi.org/10.1088/1367-2630/acd492}
  {\bibfield  {journal} {\bibinfo  {journal} {New Journal of Physics}\ }\textbf
  {\bibinfo {volume} {25}},\ \bibinfo {pages} {053025} (\bibinfo {year}
  {2023})}\BibitemShut {NoStop}%
\bibitem [{\citenamefont {Srivatsa}\ \emph {et~al.}(2023)\citenamefont
  {Srivatsa}, \citenamefont {Yarloo}, \citenamefont {Moessner},\ and\
  \citenamefont {Nielsen}}]{Srivatsa2022}%
  \BibitemOpen
  \bibfield  {author} {\bibinfo {author} {\bibfnamefont {N.~S.}\ \bibnamefont
  {Srivatsa}}, \bibinfo {author} {\bibfnamefont {H.}~\bibnamefont {Yarloo}},
  \bibinfo {author} {\bibfnamefont {R.}~\bibnamefont {Moessner}},\ and\
  \bibinfo {author} {\bibfnamefont {A.~E.~B.}\ \bibnamefont {Nielsen}},\
  }\bibfield  {title} {\bibinfo {title} {Mobility edges through inverted
  quantum many-body scarring},\ }\href
  {https://doi.org/10.1103/PhysRevB.108.L100202} {\bibfield  {journal}
  {\bibinfo  {journal} {Phys. Rev. B}\ }\textbf {\bibinfo {volume} {108}},\
  \bibinfo {pages} {L100202} (\bibinfo {year} {2023})}\BibitemShut {NoStop}%
\bibitem [{\citenamefont {Chen}\ and\ \citenamefont {Zhu}(2023)}]{Chen23}%
  \BibitemOpen
  \bibfield  {author} {\bibinfo {author} {\bibfnamefont {Q.}~\bibnamefont
  {Chen}}\ and\ \bibinfo {author} {\bibfnamefont {Z.}~\bibnamefont {Zhu}},\
  }\bibfield  {title} {\bibinfo {title} {Inverting multiple quantum many-body
  scars via disorder},\ }\href@noop {} {\bibfield  {journal} {\bibinfo
  {journal} {arXiv e-prints}\ } (\bibinfo {year} {2023})},\ \Eprint
  {https://arxiv.org/abs/2301.03405} {arXiv:2301.03405 [cond-mat.dis-nn]}
  \BibitemShut {NoStop}%
\bibitem [{\citenamefont {Iversen}\ and\ \citenamefont
  {Nielsen}(2023)}]{Iversen23}%
  \BibitemOpen
  \bibfield  {author} {\bibinfo {author} {\bibfnamefont {M.}~\bibnamefont
  {Iversen}}\ and\ \bibinfo {author} {\bibfnamefont {A.~E.~B.}\ \bibnamefont
  {Nielsen}},\ }\bibfield  {title} {\bibinfo {title} {Tower of quantum scars in
  a partially many-body localized system},\ }\href
  {https://doi.org/10.1103/PhysRevB.107.205140} {\bibfield  {journal} {\bibinfo
   {journal} {Phys. Rev. B}\ }\textbf {\bibinfo {volume} {107}},\ \bibinfo
  {pages} {205140} (\bibinfo {year} {2023})}\BibitemShut {NoStop}%
\bibitem [{\citenamefont {Chen}\ \emph
  {et~al.}(2023{\natexlab{d}})\citenamefont {Chen}, \citenamefont {Chen},\ and\
  \citenamefont {Zhu}}]{chen2023weak}%
  \BibitemOpen
  \bibfield  {author} {\bibinfo {author} {\bibfnamefont {Q.}~\bibnamefont
  {Chen}}, \bibinfo {author} {\bibfnamefont {S.~A.}\ \bibnamefont {Chen}},\
  and\ \bibinfo {author} {\bibfnamefont {Z.}~\bibnamefont {Zhu}},\ }\bibfield
  {title} {\bibinfo {title} {Weak ergodicity breaking in non-{Hermitian}
  many-body systems},\ }\href {https://doi.org/10.21468/SciPostPhys.15.2.052}
  {\bibfield  {journal} {\bibinfo  {journal} {SciPost Phys.}\ }\textbf
  {\bibinfo {volume} {15}},\ \bibinfo {pages} {052} (\bibinfo {year}
  {2023}{\natexlab{d}})}\BibitemShut {NoStop}%
\bibitem [{\citenamefont {Kawabata}\ \emph {et~al.}(2023)\citenamefont
  {Kawabata}, \citenamefont {Numasawa},\ and\ \citenamefont
  {Ryu}}]{kawabata2023entanglement}%
  \BibitemOpen
  \bibfield  {author} {\bibinfo {author} {\bibfnamefont {K.}~\bibnamefont
  {Kawabata}}, \bibinfo {author} {\bibfnamefont {T.}~\bibnamefont {Numasawa}},\
  and\ \bibinfo {author} {\bibfnamefont {S.}~\bibnamefont {Ryu}},\ }\bibfield
  {title} {\bibinfo {title} {Entanglement phase transition induced by the
  non-{Hermitian} skin effect},\ }\href
  {https://doi.org/10.1103/PhysRevX.13.021007} {\bibfield  {journal} {\bibinfo
  {journal} {Phys. Rev. X}\ }\textbf {\bibinfo {volume} {13}},\ \bibinfo
  {pages} {021007} (\bibinfo {year} {2023})}\BibitemShut {NoStop}%
\bibitem [{\citenamefont {Gliozzi}\ \emph {et~al.}(2024)\citenamefont
  {Gliozzi}, \citenamefont {Tomasi},\ and\ \citenamefont
  {Hughes}}]{gliozzi2024many}%
  \BibitemOpen
  \bibfield  {author} {\bibinfo {author} {\bibfnamefont {J.}~\bibnamefont
  {Gliozzi}}, \bibinfo {author} {\bibfnamefont {G.~D.}\ \bibnamefont
  {Tomasi}},\ and\ \bibinfo {author} {\bibfnamefont {T.~L.}\ \bibnamefont
  {Hughes}},\ }\bibfield  {title} {\bibinfo {title} {Many-body non-{Hermitian}
  skin effect for multipoles},\ }\href@noop {} {\bibfield  {journal} {\bibinfo
  {journal} {arXiv e-prints}\ } (\bibinfo {year} {2024})},\ \Eprint
  {https://arxiv.org/abs/2401.04162} {arXiv:2401.04162 [cond-mat.str-el]}
  \BibitemShut {NoStop}%
\bibitem [{\citenamefont {Hamazaki}\ \emph {et~al.}(2019)\citenamefont
  {Hamazaki}, \citenamefont {Kawabata},\ and\ \citenamefont
  {Ueda}}]{hamazaki2019non}%
  \BibitemOpen
  \bibfield  {author} {\bibinfo {author} {\bibfnamefont {R.}~\bibnamefont
  {Hamazaki}}, \bibinfo {author} {\bibfnamefont {K.}~\bibnamefont {Kawabata}},\
  and\ \bibinfo {author} {\bibfnamefont {M.}~\bibnamefont {Ueda}},\ }\bibfield
  {title} {\bibinfo {title} {Non-{Hermitian} many-body localization},\ }\href
  {https://doi.org/10.1103/PhysRevLett.123.090603} {\bibfield  {journal}
  {\bibinfo  {journal} {Phys. Rev. Lett.}\ }\textbf {\bibinfo {volume} {123}},\
  \bibinfo {pages} {090603} (\bibinfo {year} {2019})}\BibitemShut {NoStop}%
\bibitem [{\citenamefont {Mu}\ \emph {et~al.}(2020)\citenamefont {Mu},
  \citenamefont {Lee}, \citenamefont {Li},\ and\ \citenamefont
  {Gong}}]{mu2020emergent}%
  \BibitemOpen
  \bibfield  {author} {\bibinfo {author} {\bibfnamefont {S.}~\bibnamefont
  {Mu}}, \bibinfo {author} {\bibfnamefont {C.~H.}\ \bibnamefont {Lee}},
  \bibinfo {author} {\bibfnamefont {L.}~\bibnamefont {Li}},\ and\ \bibinfo
  {author} {\bibfnamefont {J.}~\bibnamefont {Gong}},\ }\bibfield  {title}
  {\bibinfo {title} {Emergent {Fermi} surface in a many-body non-{Hermitian}
  fermionic chain},\ }\href {https://doi.org/10.1103/PhysRevB.102.081115}
  {\bibfield  {journal} {\bibinfo  {journal} {Phys. Rev. B}\ }\textbf {\bibinfo
  {volume} {102}},\ \bibinfo {pages} {081115} (\bibinfo {year}
  {2020})}\BibitemShut {NoStop}%
\bibitem [{\citenamefont {Kawabata}\ \emph {et~al.}(2022)\citenamefont
  {Kawabata}, \citenamefont {Shiozaki},\ and\ \citenamefont
  {Ryu}}]{kawabata2022many}%
  \BibitemOpen
  \bibfield  {author} {\bibinfo {author} {\bibfnamefont {K.}~\bibnamefont
  {Kawabata}}, \bibinfo {author} {\bibfnamefont {K.}~\bibnamefont {Shiozaki}},\
  and\ \bibinfo {author} {\bibfnamefont {S.}~\bibnamefont {Ryu}},\ }\bibfield
  {title} {\bibinfo {title} {Many-body topology of non-{Hermitian} systems},\
  }\href {https://doi.org/10.1103/PhysRevB.105.165137} {\bibfield  {journal}
  {\bibinfo  {journal} {Phys. Rev. B}\ }\textbf {\bibinfo {volume} {105}},\
  \bibinfo {pages} {165137} (\bibinfo {year} {2022})}\BibitemShut {NoStop}%
\bibitem [{\citenamefont {Alsallom}\ \emph {et~al.}(2022)\citenamefont
  {Alsallom}, \citenamefont {Herviou}, \citenamefont {Yazyev},\ and\
  \citenamefont {Brzezi\ifmmode~\acute{n}\else
  \'{n}\fi{}ska}}]{alsallom2022fate}%
  \BibitemOpen
  \bibfield  {author} {\bibinfo {author} {\bibfnamefont {F.}~\bibnamefont
  {Alsallom}}, \bibinfo {author} {\bibfnamefont {L.}~\bibnamefont {Herviou}},
  \bibinfo {author} {\bibfnamefont {O.~V.}\ \bibnamefont {Yazyev}},\ and\
  \bibinfo {author} {\bibfnamefont {M.}~\bibnamefont
  {Brzezi\ifmmode~\acute{n}\else \'{n}\fi{}ska}},\ }\bibfield  {title}
  {\bibinfo {title} {Fate of the non-{Hermitian} skin effect in many-body
  fermionic systems},\ }\href
  {https://doi.org/10.1103/PhysRevResearch.4.033122} {\bibfield  {journal}
  {\bibinfo  {journal} {Phys. Rev. Res.}\ }\textbf {\bibinfo {volume} {4}},\
  \bibinfo {pages} {033122} (\bibinfo {year} {2022})}\BibitemShut {NoStop}%
\bibitem [{\citenamefont {Shen}\ and\ \citenamefont {Lee}(2022)}]{shen2022non}%
  \BibitemOpen
  \bibfield  {author} {\bibinfo {author} {\bibfnamefont {R.}~\bibnamefont
  {Shen}}\ and\ \bibinfo {author} {\bibfnamefont {C.~H.}\ \bibnamefont {Lee}},\
  }\bibfield  {title} {\bibinfo {title} {Non-{Hermitian} skin clusters from
  strong interactions},\ }\href {https://doi.org/10.1038/s42005-022-01015-w}
  {\bibfield  {journal} {\bibinfo  {journal} {Communications Physics}\ }\textbf
  {\bibinfo {volume} {5}},\ \bibinfo {pages} {238} (\bibinfo {year}
  {2022})}\BibitemShut {NoStop}%
\bibitem [{\citenamefont {Tomasi}\ and\ \citenamefont
  {Khaymovich}(2023)}]{de2023stable}%
  \BibitemOpen
  \bibfield  {author} {\bibinfo {author} {\bibfnamefont {G.~D.}\ \bibnamefont
  {Tomasi}}\ and\ \bibinfo {author} {\bibfnamefont {I.~M.}\ \bibnamefont
  {Khaymovich}},\ }\bibfield  {title} {\bibinfo {title} {Stable many-body
  localization under random continuous measurements in the no-click limit},\
  }\href@noop {} {\bibfield  {journal} {\bibinfo  {journal} {arXiv e-prints}\ }
  (\bibinfo {year} {2023})},\ \Eprint {https://arxiv.org/abs/2311.00019}
  {arXiv:2311.00019 [cond-mat.dis-nn]} \BibitemShut {NoStop}%
\bibitem [{\citenamefont {Weinberg}\ and\ \citenamefont
  {Bukov}(2017)}]{weinberg2017quspin}%
  \BibitemOpen
  \bibfield  {author} {\bibinfo {author} {\bibfnamefont {P.}~\bibnamefont
  {Weinberg}}\ and\ \bibinfo {author} {\bibfnamefont {M.}~\bibnamefont
  {Bukov}},\ }\bibfield  {title} {\bibinfo {title} {{QuSpin: a Python Package
  for Dynamics and Exact Diagonalisation of Quantum Many Body Systems part I:
  spin chains}},\ }\href {https://doi.org/10.21468/SciPostPhys.2.1.003}
  {\bibfield  {journal} {\bibinfo  {journal} {SciPost Phys.}\ }\textbf
  {\bibinfo {volume} {2}},\ \bibinfo {pages} {003} (\bibinfo {year}
  {2017})}\BibitemShut {NoStop}%
\bibitem [{\citenamefont {Weinberg}\ and\ \citenamefont
  {Bukov}(2019)}]{weinberg2019quspin}%
  \BibitemOpen
  \bibfield  {author} {\bibinfo {author} {\bibfnamefont {P.}~\bibnamefont
  {Weinberg}}\ and\ \bibinfo {author} {\bibfnamefont {M.}~\bibnamefont
  {Bukov}},\ }\bibfield  {title} {\bibinfo {title} {{QuSpin: a Python Package
  for Dynamics and Exact Diagonalisation of Quantum Many Body Systems. Part II:
  bosons, fermions and higher spins}},\ }\href
  {https://doi.org/10.21468/SciPostPhys.7.2.020} {\bibfield  {journal}
  {\bibinfo  {journal} {SciPost Phys.}\ }\textbf {\bibinfo {volume} {7}},\
  \bibinfo {pages} {20} (\bibinfo {year} {2019})}\BibitemShut {NoStop}%
\bibitem [{\citenamefont {Zhou}\ \emph {et~al.}(2022)\citenamefont {Zhou},
  \citenamefont {Li}, \citenamefont {Yi},\ and\ \citenamefont
  {Cui}}]{zhou2021engineering}%
  \BibitemOpen
  \bibfield  {author} {\bibinfo {author} {\bibfnamefont {L.}~\bibnamefont
  {Zhou}}, \bibinfo {author} {\bibfnamefont {H.}~\bibnamefont {Li}}, \bibinfo
  {author} {\bibfnamefont {W.}~\bibnamefont {Yi}},\ and\ \bibinfo {author}
  {\bibfnamefont {X.}~\bibnamefont {Cui}},\ }\bibfield  {title} {\bibinfo
  {title} {Engineering non-{Hermitian} skin effect with band topology in
  ultracold gases},\ }\href {https://doi.org/10.1038/s42005-022-01021-y}
  {\bibfield  {journal} {\bibinfo  {journal} {Communications Physics}\ }\textbf
  {\bibinfo {volume} {5}},\ \bibinfo {pages} {252} (\bibinfo {year}
  {2022})}\BibitemShut {NoStop}%
\bibitem [{\citenamefont {Hu}\ \emph {et~al.}(2017)\citenamefont {Hu},
  \citenamefont {Urvoy}, \citenamefont {Vendeiro}, \citenamefont {Cr{\'e}pel},
  \citenamefont {Chen},\ and\ \citenamefont {Vuleti{\'c}}}]{hu2017creation}%
  \BibitemOpen
  \bibfield  {author} {\bibinfo {author} {\bibfnamefont {J.}~\bibnamefont
  {Hu}}, \bibinfo {author} {\bibfnamefont {A.}~\bibnamefont {Urvoy}}, \bibinfo
  {author} {\bibfnamefont {Z.}~\bibnamefont {Vendeiro}}, \bibinfo {author}
  {\bibfnamefont {V.}~\bibnamefont {Cr{\'e}pel}}, \bibinfo {author}
  {\bibfnamefont {W.}~\bibnamefont {Chen}},\ and\ \bibinfo {author}
  {\bibfnamefont {V.}~\bibnamefont {Vuleti{\'c}}},\ }\bibfield  {title}
  {\bibinfo {title} {Creation of a bose-condensed gas of 87rb by laser
  cooling},\ }\href@noop {} {\bibfield  {journal} {\bibinfo  {journal}
  {Science}\ }\textbf {\bibinfo {volume} {358}},\ \bibinfo {pages} {1078}
  (\bibinfo {year} {2017})}\BibitemShut {NoStop}%
\bibitem [{\citenamefont {Bravyi}\ \emph {et~al.}(2011)\citenamefont {Bravyi},
  \citenamefont {DiVincenzo},\ and\ \citenamefont
  {Loss}}]{bravyi2011schrieffer}%
  \BibitemOpen
  \bibfield  {author} {\bibinfo {author} {\bibfnamefont {S.}~\bibnamefont
  {Bravyi}}, \bibinfo {author} {\bibfnamefont {D.~P.}\ \bibnamefont
  {DiVincenzo}},\ and\ \bibinfo {author} {\bibfnamefont {D.}~\bibnamefont
  {Loss}},\ }\bibfield  {title} {\bibinfo {title} {Schrieffer--wolff
  transformation for quantum many-body systems},\ }\href@noop {} {\bibfield
  {journal} {\bibinfo  {journal} {Annals of physics}\ }\textbf {\bibinfo
  {volume} {326}},\ \bibinfo {pages} {2793} (\bibinfo {year}
  {2011})}\BibitemShut {NoStop}%
\bibitem [{\citenamefont {Heiss}(2012)}]{heiss2012physics}%
  \BibitemOpen
  \bibfield  {author} {\bibinfo {author} {\bibfnamefont {W.}~\bibnamefont
  {Heiss}},\ }\bibfield  {title} {\bibinfo {title} {The physics of exceptional
  points},\ }\href {https://doi.org/10.1088/1751-8113/45/44/444016} {\bibfield
  {journal} {\bibinfo  {journal} {Journal of Physics A: Mathematical and
  Theoretical}\ }\textbf {\bibinfo {volume} {45}},\ \bibinfo {pages} {444016}
  (\bibinfo {year} {2012})}\BibitemShut {NoStop}%
\bibitem [{\citenamefont {Klett}\ \emph {et~al.}(2017)\citenamefont {Klett},
  \citenamefont {Cartarius}, \citenamefont {Dast}, \citenamefont {Main},\ and\
  \citenamefont {Wunner}}]{klett2017relation}%
  \BibitemOpen
  \bibfield  {author} {\bibinfo {author} {\bibfnamefont {M.}~\bibnamefont
  {Klett}}, \bibinfo {author} {\bibfnamefont {H.}~\bibnamefont {Cartarius}},
  \bibinfo {author} {\bibfnamefont {D.}~\bibnamefont {Dast}}, \bibinfo {author}
  {\bibfnamefont {J.}~\bibnamefont {Main}},\ and\ \bibinfo {author}
  {\bibfnamefont {G.}~\bibnamefont {Wunner}},\ }\bibfield  {title} {\bibinfo
  {title} {Relation between {$\mathcal{PT}$}-symmetry breaking and
  topologically nontrivial phases in the {Su-Schrieffer-Heeger} and {Kitaev}
  models},\ }\href {https://doi.org/10.1103/PhysRevA.95.053626} {\bibfield
  {journal} {\bibinfo  {journal} {Phys. Rev. A}\ }\textbf {\bibinfo {volume}
  {95}},\ \bibinfo {pages} {053626} (\bibinfo {year} {2017})}\BibitemShut
  {NoStop}%
\bibitem [{\citenamefont {Longhi}(2019{\natexlab{b}})}]{longhi2019topological}%
  \BibitemOpen
  \bibfield  {author} {\bibinfo {author} {\bibfnamefont {S.}~\bibnamefont
  {Longhi}},\ }\bibfield  {title} {\bibinfo {title} {Topological phase
  transition in non-{Hermitian} quasicrystals},\ }\href
  {https://doi.org/10.1103/PhysRevLett.122.237601} {\bibfield  {journal}
  {\bibinfo  {journal} {Phys. Rev. Lett.}\ }\textbf {\bibinfo {volume} {122}},\
  \bibinfo {pages} {237601} (\bibinfo {year} {2019}{\natexlab{b}})}\BibitemShut
  {NoStop}%
\bibitem [{\citenamefont {Sakhdari}\ \emph {et~al.}(2019)\citenamefont
  {Sakhdari}, \citenamefont {Hajizadegan}, \citenamefont {Zhong}, \citenamefont
  {Christodoulides}, \citenamefont {El-Ganainy},\ and\ \citenamefont
  {Chen}}]{sakhdari2019experimental}%
  \BibitemOpen
  \bibfield  {author} {\bibinfo {author} {\bibfnamefont {M.}~\bibnamefont
  {Sakhdari}}, \bibinfo {author} {\bibfnamefont {M.}~\bibnamefont
  {Hajizadegan}}, \bibinfo {author} {\bibfnamefont {Q.}~\bibnamefont {Zhong}},
  \bibinfo {author} {\bibfnamefont {D.~N.}\ \bibnamefont {Christodoulides}},
  \bibinfo {author} {\bibfnamefont {R.}~\bibnamefont {El-Ganainy}},\ and\
  \bibinfo {author} {\bibfnamefont {P.-Y.}\ \bibnamefont {Chen}},\ }\bibfield
  {title} {\bibinfo {title} {Experimental observation of {$PT$} symmetry
  breaking near divergent exceptional points},\ }\href
  {https://doi.org/10.1103/PhysRevLett.123.193901} {\bibfield  {journal}
  {\bibinfo  {journal} {Phys. Rev. Lett.}\ }\textbf {\bibinfo {volume} {123}},\
  \bibinfo {pages} {193901} (\bibinfo {year} {2019})}\BibitemShut {NoStop}%
\bibitem [{\citenamefont {Miri}\ and\ \citenamefont
  {Alù}(2019)}]{miri2019exceptional}%
  \BibitemOpen
  \bibfield  {author} {\bibinfo {author} {\bibfnamefont {M.-A.}\ \bibnamefont
  {Miri}}\ and\ \bibinfo {author} {\bibfnamefont {A.}~\bibnamefont {Alù}},\
  }\bibfield  {title} {\bibinfo {title} {Exceptional points in optics and
  photonics},\ }\href {https://doi.org/10.1126/science.aar7709} {\bibfield
  {journal} {\bibinfo  {journal} {Science}\ }\textbf {\bibinfo {volume}
  {363}},\ \bibinfo {pages} {eaar7709} (\bibinfo {year} {2019})}\BibitemShut
  {NoStop}%
\bibitem [{\citenamefont {{\"O}zdemir}\ \emph {et~al.}(2019)\citenamefont
  {{\"O}zdemir}, \citenamefont {Rotter}, \citenamefont {Nori},\ and\
  \citenamefont {Yang}}]{ozdemir2019parity}%
  \BibitemOpen
  \bibfield  {author} {\bibinfo {author} {\bibfnamefont {{\c{S}}.~K.}\
  \bibnamefont {{\"O}zdemir}}, \bibinfo {author} {\bibfnamefont
  {S.}~\bibnamefont {Rotter}}, \bibinfo {author} {\bibfnamefont
  {F.}~\bibnamefont {Nori}},\ and\ \bibinfo {author} {\bibfnamefont
  {L.}~\bibnamefont {Yang}},\ }\bibfield  {title} {\bibinfo {title}
  {Parity--time symmetry and exceptional points in photonics},\ }\href
  {https://doi.org/10.1038/s41563-019-0304-9} {\bibfield  {journal} {\bibinfo
  {journal} {Nature Materials}\ }\textbf {\bibinfo {volume} {18}},\ \bibinfo
  {pages} {783} (\bibinfo {year} {2019})}\BibitemShut {NoStop}%
\bibitem [{\citenamefont {Lee}(2022)}]{lee2022exceptional}%
  \BibitemOpen
  \bibfield  {author} {\bibinfo {author} {\bibfnamefont {C.~H.}\ \bibnamefont
  {Lee}},\ }\bibfield  {title} {\bibinfo {title} {Exceptional bound states and
  negative entanglement entropy},\ }\href
  {https://doi.org/10.1103/PhysRevLett.128.010402} {\bibfield  {journal}
  {\bibinfo  {journal} {Phys. Rev. Lett.}\ }\textbf {\bibinfo {volume} {128}},\
  \bibinfo {pages} {010402} (\bibinfo {year} {2022})}\BibitemShut {NoStop}%
\bibitem [{\citenamefont {Ding}\ \emph {et~al.}(2022)\citenamefont {Ding},
  \citenamefont {Fang},\ and\ \citenamefont {Ma}}]{ding2022non}%
  \BibitemOpen
  \bibfield  {author} {\bibinfo {author} {\bibfnamefont {K.}~\bibnamefont
  {Ding}}, \bibinfo {author} {\bibfnamefont {C.}~\bibnamefont {Fang}},\ and\
  \bibinfo {author} {\bibfnamefont {G.}~\bibnamefont {Ma}},\ }\bibfield
  {title} {\bibinfo {title} {Non-hermitian topology and exceptional-point
  geometries},\ }\href {https://doi.org/10.1038/s42254-022-00516-5} {\bibfield
  {journal} {\bibinfo  {journal} {Nature Reviews Physics}\ }\textbf {\bibinfo
  {volume} {4}},\ \bibinfo {pages} {745} (\bibinfo {year} {2022})}\BibitemShut
  {NoStop}%
\bibitem [{\citenamefont {Yang}\ and\ \citenamefont
  {Lee}(2023)}]{yang2023percolation}%
  \BibitemOpen
  \bibfield  {author} {\bibinfo {author} {\bibfnamefont {M.}~\bibnamefont
  {Yang}}\ and\ \bibinfo {author} {\bibfnamefont {C.~H.}\ \bibnamefont {Lee}},\
  }\bibfield  {title} {\bibinfo {title} {Percolation-induced {PT} symmetry
  breaking},\ }\href@noop {} {\bibfield  {journal} {\bibinfo  {journal} {arXiv
  e-prints}\ } (\bibinfo {year} {2023})},\ \Eprint
  {https://arxiv.org/abs/2309.15008} {arXiv:2309.15008 [cond-mat.stat-mech]}
  \BibitemShut {NoStop}%
\bibitem [{\citenamefont {Meng}\ \emph {et~al.}(2024)\citenamefont {Meng},
  \citenamefont {Ang},\ and\ \citenamefont {Lee}}]{meng2024exceptional}%
  \BibitemOpen
  \bibfield  {author} {\bibinfo {author} {\bibfnamefont {H.}~\bibnamefont
  {Meng}}, \bibinfo {author} {\bibfnamefont {Y.~S.}\ \bibnamefont {Ang}},\ and\
  \bibinfo {author} {\bibfnamefont {C.~H.}\ \bibnamefont {Lee}},\ }\bibfield
  {title} {\bibinfo {title} {{Exceptional points in non-{Hermitian} systems:
  Applications and recent developments}},\ }\href
  {https://doi.org/10.1063/5.0183826} {\bibfield  {journal} {\bibinfo
  {journal} {Applied Physics Letters}\ }\textbf {\bibinfo {volume} {124}},\
  \bibinfo {pages} {060502} (\bibinfo {year} {2024})}\BibitemShut {NoStop}%
\bibitem [{\citenamefont {Orito}\ and\ \citenamefont
  {Imura}(2023)}]{orito2023entanglement}%
  \BibitemOpen
  \bibfield  {author} {\bibinfo {author} {\bibfnamefont {T.}~\bibnamefont
  {Orito}}\ and\ \bibinfo {author} {\bibfnamefont {K.-I.}\ \bibnamefont
  {Imura}},\ }\bibfield  {title} {\bibinfo {title} {Entanglement dynamics in
  the many-body {Hatano-Nelson} model},\ }\href
  {https://doi.org/10.1103/PhysRevB.108.214308} {\bibfield  {journal} {\bibinfo
   {journal} {Phys. Rev. B}\ }\textbf {\bibinfo {volume} {108}},\ \bibinfo
  {pages} {214308} (\bibinfo {year} {2023})}\BibitemShut {NoStop}%
\bibitem [{\citenamefont {Chang}\ \emph {et~al.}(2020)\citenamefont {Chang},
  \citenamefont {You}, \citenamefont {Wen},\ and\ \citenamefont
  {Ryu}}]{chang2020entanglement}%
  \BibitemOpen
  \bibfield  {author} {\bibinfo {author} {\bibfnamefont {P.-Y.}\ \bibnamefont
  {Chang}}, \bibinfo {author} {\bibfnamefont {J.-S.}\ \bibnamefont {You}},
  \bibinfo {author} {\bibfnamefont {X.}~\bibnamefont {Wen}},\ and\ \bibinfo
  {author} {\bibfnamefont {S.}~\bibnamefont {Ryu}},\ }\bibfield  {title}
  {\bibinfo {title} {Entanglement spectrum and entropy in topological
  non-{Hermitian} systems and nonunitary conformal field theory},\ }\href
  {https://doi.org/10.1103/PhysRevResearch.2.033069} {\bibfield  {journal}
  {\bibinfo  {journal} {Phys. Rev. Res.}\ }\textbf {\bibinfo {volume} {2}},\
  \bibinfo {pages} {033069} (\bibinfo {year} {2020})}\BibitemShut {NoStop}%
\bibitem [{\citenamefont {Hsieh}\ and\ \citenamefont
  {Chang}(2023)}]{hsieh2023relating}%
  \BibitemOpen
  \bibfield  {author} {\bibinfo {author} {\bibfnamefont {C.-T.}\ \bibnamefont
  {Hsieh}}\ and\ \bibinfo {author} {\bibfnamefont {P.-Y.}\ \bibnamefont
  {Chang}},\ }\bibfield  {title} {\bibinfo {title} {{Relating non-Hermitian and
  Hermitian quantum systems at criticality}},\ }\href
  {https://doi.org/10.21468/SciPostPhysCore.6.3.062} {\bibfield  {journal}
  {\bibinfo  {journal} {SciPost Phys. Core}\ }\textbf {\bibinfo {volume} {6}},\
  \bibinfo {pages} {062} (\bibinfo {year} {2023})}\BibitemShut {NoStop}%
\bibitem [{\citenamefont {Zou}\ \emph {et~al.}(2023)\citenamefont {Zou},
  \citenamefont {Chen}, \citenamefont {Meng}, \citenamefont {Ang},
  \citenamefont {Zhang},\ and\ \citenamefont {Lee}}]{zou2023experimental}%
  \BibitemOpen
  \bibfield  {author} {\bibinfo {author} {\bibfnamefont {D.}~\bibnamefont
  {Zou}}, \bibinfo {author} {\bibfnamefont {T.}~\bibnamefont {Chen}}, \bibinfo
  {author} {\bibfnamefont {H.}~\bibnamefont {Meng}}, \bibinfo {author}
  {\bibfnamefont {Y.~S.}\ \bibnamefont {Ang}}, \bibinfo {author} {\bibfnamefont
  {X.}~\bibnamefont {Zhang}},\ and\ \bibinfo {author} {\bibfnamefont {C.~H.}\
  \bibnamefont {Lee}},\ }\bibfield  {title} {\bibinfo {title} {Experimental
  observation of exceptional bound states in a classical circuit network},\
  }\href@noop {} {\bibfield  {journal} {\bibinfo  {journal} {arXiv e-prints}\ }
  (\bibinfo {year} {2023})},\ \Eprint {https://arxiv.org/abs/2308.01970}
  {arXiv:2308.01970 [quant-ph]} \BibitemShut {NoStop}%
\bibitem [{\citenamefont {Fossati}\ \emph {et~al.}(2023)\citenamefont
  {Fossati}, \citenamefont {Ares},\ and\ \citenamefont
  {Calabrese}}]{fossati2023symmetry}%
  \BibitemOpen
  \bibfield  {author} {\bibinfo {author} {\bibfnamefont {M.}~\bibnamefont
  {Fossati}}, \bibinfo {author} {\bibfnamefont {F.}~\bibnamefont {Ares}},\ and\
  \bibinfo {author} {\bibfnamefont {P.}~\bibnamefont {Calabrese}},\ }\bibfield
  {title} {\bibinfo {title} {Symmetry-resolved entanglement in critical
  non-{Hermitian} systems},\ }\href
  {https://doi.org/10.1103/PhysRevB.107.205153} {\bibfield  {journal} {\bibinfo
   {journal} {Phys. Rev. B}\ }\textbf {\bibinfo {volume} {107}},\ \bibinfo
  {pages} {205153} (\bibinfo {year} {2023})}\BibitemShut {NoStop}%
\bibitem [{\citenamefont {Rottoli}\ \emph {et~al.}(2024)\citenamefont
  {Rottoli}, \citenamefont {Fossati},\ and\ \citenamefont
  {Calabrese}}]{rottoli2024entanglement}%
  \BibitemOpen
  \bibfield  {author} {\bibinfo {author} {\bibfnamefont {F.}~\bibnamefont
  {Rottoli}}, \bibinfo {author} {\bibfnamefont {M.}~\bibnamefont {Fossati}},\
  and\ \bibinfo {author} {\bibfnamefont {P.}~\bibnamefont {Calabrese}},\
  }\bibfield  {title} {\bibinfo {title} {Entanglement {Hamiltonian} in the
  non-{Hermitian} {SSH} model},\ }\href@noop {} {\bibfield  {journal} {\bibinfo
   {journal} {arXiv e-prints}\ } (\bibinfo {year} {2024})},\ \Eprint
  {https://arxiv.org/abs/2402.04776} {arXiv:2402.04776 [quant-ph]} \BibitemShut
  {NoStop}%
\bibitem [{\citenamefont {Turner}\ \emph
  {et~al.}(2018{\natexlab{c}})\citenamefont {Turner}, \citenamefont
  {Michailidis}, \citenamefont {Abanin}, \citenamefont {Serbyn},\ and\
  \citenamefont {Papi{\'c}}}]{TurnerNature}%
  \BibitemOpen
  \bibfield  {author} {\bibinfo {author} {\bibfnamefont {C.~J.}\ \bibnamefont
  {Turner}}, \bibinfo {author} {\bibfnamefont {A.~A.}\ \bibnamefont
  {Michailidis}}, \bibinfo {author} {\bibfnamefont {D.~A.}\ \bibnamefont
  {Abanin}}, \bibinfo {author} {\bibfnamefont {M.}~\bibnamefont {Serbyn}},\
  and\ \bibinfo {author} {\bibfnamefont {Z.}~\bibnamefont {Papi{\'c}}},\
  }\bibfield  {title} {\bibinfo {title} {Weak ergodicity breaking from quantum
  many-body scars},\ }\href
  {https://doi.org/https://doi.org/10.1038/s41567-018-0137-5} {\bibfield
  {journal} {\bibinfo  {journal} {Nature Physics}\ }\textbf {\bibinfo {volume}
  {14}},\ \bibinfo {pages} {745} (\bibinfo {year}
  {2018}{\natexlab{c}})}\BibitemShut {NoStop}%
\end{thebibliography}%
\setcounter{equation}{0}
\setcounter{figure}{0}
\setcounter{table}{0}
\setcounter{page}{1}
\setcounter{section}{0}
\renewcommand{\theequation}{S\arabic{equation}}
\renewcommand{\thefigure}{S\arabic{figure}}
\renewcommand{\thesection}{S\arabic{section}}
\onecolumngrid
\flushbottom
\newpage
\appendix
\begin{center}
\textbf{\large Supplemental Online Material for ``Enhanced many-body scars from the non-Hermitian Fock space skin effect'' }
\end{center}

\begin{center}
 {\small Ruizhe Shen$^{1}$, Fang Qin$^{1}$, Jean-Yves Desaules$^{2,3}$, Zlatko Papi\'c$^{2}$, and Ching Hua Lee$^{1}$ }  
\end{center}

\begin{center}
{\sl \footnotesize

$^{1}$Department of Physics, National University of Singapore, Singapore 117542

$^{2}$School of Physics and Astronomy, University of Leeds, Leeds LS2 9JT, United Kingdom

$^{3}$Institute of Science and Technology Austria (ISTA), Am Campus 1, 3400 Klosterneuburg, Austria
}
\end{center}

\begin{quote}
	{\small
		This supplementary material contains:  (SI) a detailed description of the experimental setup for realizing an effective non-Hermitian PXP model using ultracold atoms in optical lattices; (SII) details of the quantum simulation of the non-Hermitian PXP model on IBM Q device;  (SIII) further characterization of the Fock skin accumulation in three variations of the model with different types of non-Hermiticity;  (SIV) analysis of the energy spectrum as a function of non-Hermiticity; 
		(SV) additional evidence for the enhancement of scarring by the non-Hermiticity in the fidelity dynamics and entanglement entropy of eigenstates;  SVI. statistic behaviors of the non-Hermitian PXP model. 
	}   
\end{quote}

\section{SI. Effective non-Hermitian PXP model from a non-Hermitian Bose-Hubbard model}

In the main text, we introduced a non-Hermitian generalization of the PXP model and showed that its non-Hermiticity enhances the quantum-many body scars (QMBSs). Here we show how that model can effectively emerge in a Bose-Hubbard quantum simulator in the presence of laser-induced loss. Below we provide details of our proposed experimental setup and measurement, building upon previous work in the Hermitian setting~\cite{su2022observation}. 

\begin{figure}
	\centering
	\includegraphics[width=0.8\linewidth]{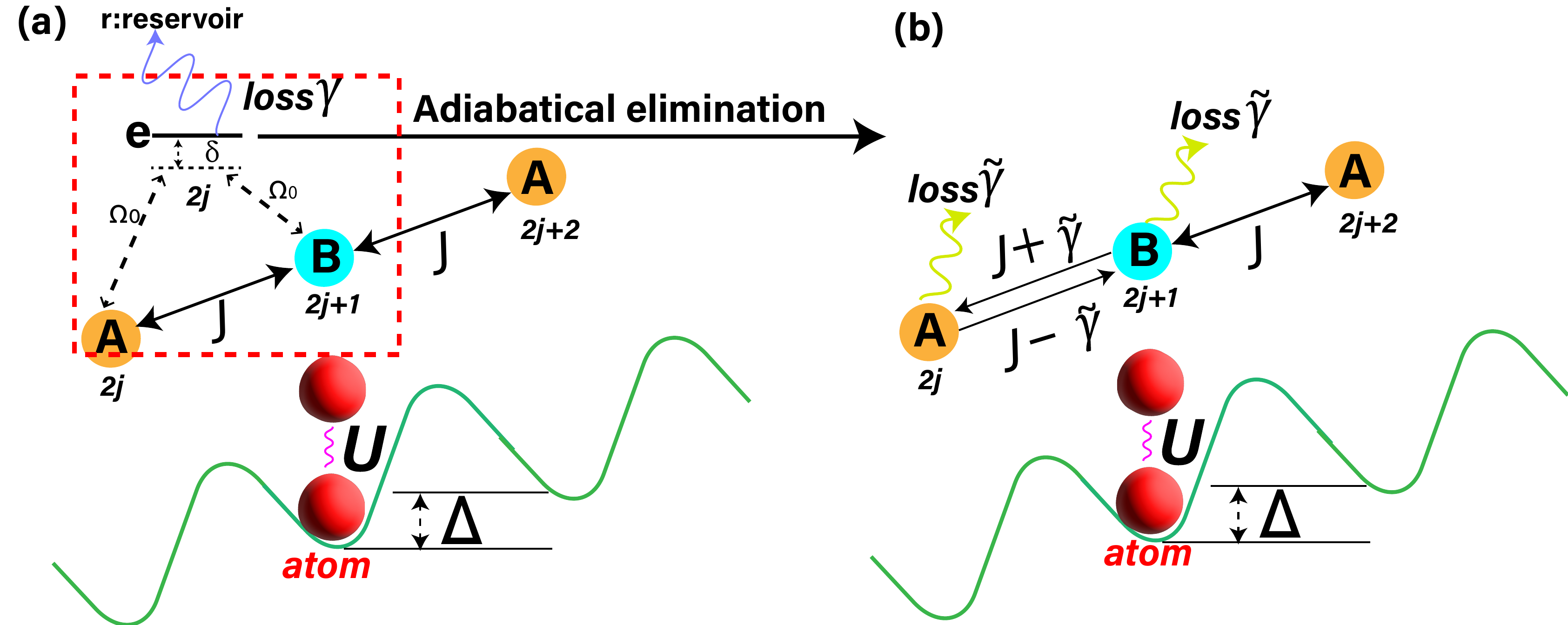}
	\caption{ (a) Schematic of the non-Hermitian Bose-Hubbard model in {\color{red} Eq.~\eqref{suphb}}. The atoms are trapped in a tilted optical lattice with an alternating chain of local ground states ``A'' and ``B'' (yellow and blue dots). $U$ is the strength of the on-site interaction, $\gamma$ is the laser-induced loss rate on the excited state ``e", and $\delta$ denotes the strength of detuning on the excited state~\cite{liang2022observation,qin2022non}. 
		(b) Adiabatic elimination of the excited state ``e" (red box in (a)) results in the model in {\color{red} Eq.~\eqref{supeff}~\cite{qin2022non}}. Effective non-reciprocal hoppings $J\pm \tilde \gamma$ are realized by coupling the two nearest ground states ``A'' and ``B'' via an excited state ``e''  with Rabi frequency $\Omega_{0}$ (black dashed arrows), with $\tilde\gamma = \Omega_0^2/(\gamma+i\delta)$.
	}
	\label{fig:ssh2}
\end{figure}

We consider $^{87}$Rb atoms in a one-dimensional (1D) optical lattice, FIG.~\ref{fig:ssh2}(a), described by a Bose-Hubbard type Hamiltonian defined in terms of standard boson annihilation and creation operators, $\hat{b}_j$ and $\hat{b}_j^{\dagger}$:
\begin{equation}\label{suphb}
	\hat{H}_{\rm BH}=\hat{H}_{\rm hopping}+\sum_{j=0}^{L-1} \Delta j \hat{n}_{j}+\frac{U}{2} \sum_{j=0}^{L-1} \hat{n_j}\left(\hat{n}_{j}-1\right),
\end{equation}
where $\hat{n}_{j}=\hat{b}^{\dagger}_{j}\hat{b}_{j}$ is the local boson number on site $j$. The three terms in the Hamiltonian stand for, respectively, the kinetic energy due to bosons hopping between lattice sites, the linear tilt potential of the optical lattice parametrized by $\Delta$, and the on-site interaction strength $U$ that penalizes the multiple occupancy of any site. 

The hopping term in the BH model in Eq.~\eqref{suphb} is of non-Hermitian form, reminsiscent of the Su-Schrieffer-Heeger model: 
\begin{equation}\label{suppssh}
	\begin{aligned}
		\hat{H}_{\mathrm{hopping}}&=-\sum_{j}\left\{\left(J+\tilde{\gamma}\right) \hat{b}_{2 j}^{\dagger} \hat{b}_{2 j+1} 
		+\left(J-\tilde{\gamma}\right) \hat{b}_{2 j+1}^{\dagger} \hat{b}_{2 j}+J \hat{b}_{2 j+2}^{\dagger} \hat{b}_{2 j+1}+J \hat{b}_{2 j+1}^{\dagger} \hat{b}_{2 j+2}\right\rbrace +\sum_{j}\tilde{\gamma}\hat{b}_{j}^{\dagger} \hat{b}_{j}, 
	\end{aligned}
\end{equation}
where $J$ is the hopping amplitude and the asymmetry characterized by $\bar{\gamma}$ is induced by atom loss, see FIG~\ref{fig:ssh2}.  The non-Hermitian asymmetric hoppings of strength $J\pm\tilde{\gamma}$ can be engineered as follows~\cite{liang2022observation,qin2022non}. For each unit cell shown in FIG.~\ref{fig:ssh2}(a), we couple two nearest ground-state levels (denoted by $A$ and $B$) via an excited state ($e$ state) with Rabi frequency $\Omega_{0}$ (dashed arrows).
The excitation towards state $e$ serves as atomic loss in the ground state manifold, which we quantify phenomenologically as $\gamma$. As sketched in FIG.~\ref{fig:ssh2}(b), this leads to effective non-reciprocal hoppings in Eq.\eqref{suppssh} between atoms on the nearest ground states, which can be rigorously shown by adiabatically eliminating the excited state (see details in Ref.~\cite{qin2022non}). The strength of $\tilde{\gamma}$ in {\color{red} Eq. \eqref{suppssh}} is given in terms of the Rabi frequency $\Omega_0$ by
\begin{equation}\label{suppga}
	\tilde{\gamma}=\frac{\Omega_{0}^{2}}{\gamma+i \delta}\approx\frac{\Omega_{0}^{2}}{\gamma},
\end{equation}
where $\delta$ is the detuning on the excited state and we assumed $\gamma\gg\delta$~\cite{qin2022non}. 

In experiments, dynamics is conveniently described by measuring the total occupation imbalance between odd and even sites:
\begin{equation}\label{imsupp}
	I(t)= \frac{\sum_{i} \left(n_{2i}(t)-n_{2i+1}(t)\right)}{\sum_{i} \left(n_{2i}(t)+n_{2i+1}(t)\right)},
\end{equation}
where $n_{j}= \bra{\psi(t)} \hat{b}_{j}^{\dagger} \hat{b}_{j} \ket{\psi(t)}/\bra{\psi(t)}\ket{\psi(t)}$ is the occupation density on site $j$. For this quantity, we can drop the global loss $\sum_{j}{\gamma}\hat{b}_{j}^{\dagger} \hat{b}_{j}$ in Eq.\eqref{suppssh}~\cite{zhou2021engineering}. This is justified as any loss that affects all sites similarly does not impact the imbalance.

\subsection{Effective non-Hermitian PXP model}

To induce kinetic constraints in the Bose-Hubbard model in Eq.~\eqref{suphb}, we tune the system to a resonance $U=\Delta \gg J, \tilde{\gamma}$ and assume the total number of bosons is equal to the number of lattice sites, corresponding to filling factor $\nu=N/L=1$. In the absence of atomic loss ($\tilde{\gamma}=0$), Ref.~\cite{su2022observation} observed that under this resonance condition, the two-body repulsion $U$ and linear potential $\Delta$ interplay in such a way that any triple occupancy on a single site is heavily suppressed, leaving only the transitions of the type``$11$''~$\leftrightarrow$~``$20$'', i.e., between two singly occupied adjacent sites and a doubly occupied site. This allows to map the dynamics onto a simpler PXP spin model. Below we will generalize this to the case where atomic loss is present. In terms of experimental implementation, the effective model derived below is expected to hold in the regime $U=\Delta =10J \approx 2.0$ kHz. The tilt potential $\Delta$ can be realized either with gravity or external magnetic fields; for instance, in the setup of Ref.~\cite{su2022observation} $\Delta_{g}=1632$ Hz from gravity (tilting degree $4^{\circ}$ and $767$ nm spacing for $^{87}$Rb atoms), and $\Delta_{B}=368$ Hz from external magnetic fields, resulting in the quoted value $\Delta = \Delta_g+\Delta_B \approx 2$ kHz.

\begin{figure}[h]
	\centering
	\includegraphics[width=0.9\linewidth]{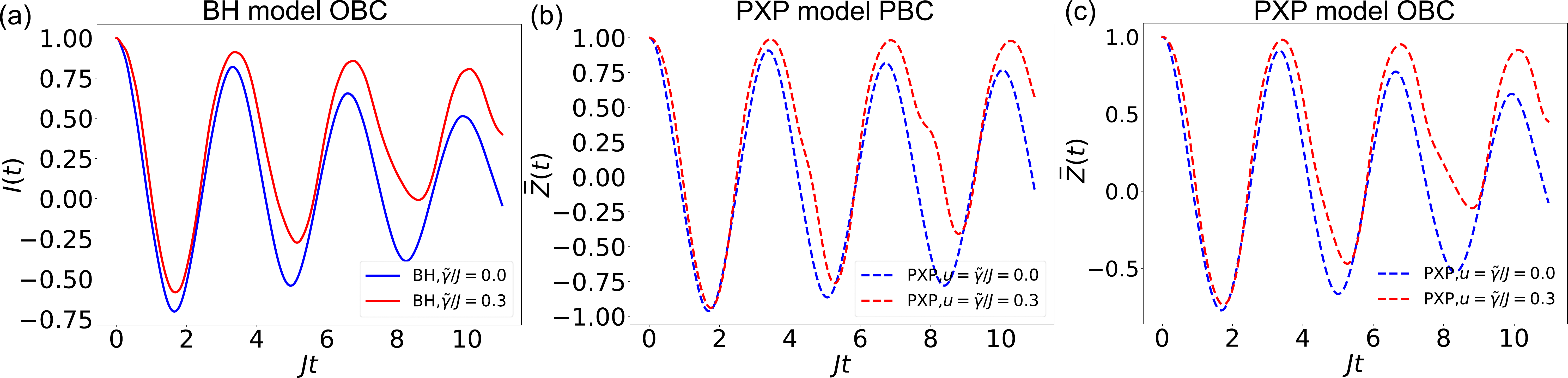}
	\caption{(a) The imbalance dynamics [Eq.~\eqref{imsupp}] for the  Bose-Hubbard model described by Eq.~\eqref{suphb}. (b) and (c) show the staggered magnetization dynamics [Eq.~\eqref{z}] for the effective PXP model described by Eq.~\eqref{suppxp}, with PBCs (b) and OBCs (c). The initial states are $\ket{20202020201}$ for (a) and $\ket{\uparrow\downarrow\uparrow\downarrow\uparrow\downarrow\uparrow\downarrow\uparrow\downarrow\uparrow\downarrow}$ for (b) and (c). For all results, we set $J$ ($\Omega=\sqrt{2}J$) as the unit for dynamics. For the two models, the enhanced revivals, induced by the Fock skin effect, emerge under the same condition of $\tilde{\gamma}/J=0.3$, and exhibit similar profiles.}
	\label{fig:suppfigure9}
\end{figure}

Working at the resonance $U=\Delta \gg J, \tilde{\gamma}$ and weak dissipation, the dynamics governed by the model in Eq.~\ref{suphb} can be described by the following effective Hamiltonian~\cite{su2022observation}

\begin{equation}\label{supeff}
		\begin{aligned}
			\hat{H}_{\mathrm{eff}}=&-\sum_{j\in \text{even} }(\left(J+\tilde{\gamma}\right)\underbrace{\hat{b}_{j}^{\dagger} \hat{b}_{j+1} \delta_{\hat{n}_{j}, 1} \delta_{\hat{n}_{j+1}, 1}}_{\sqrt{2} \hat{P}_{j-1} \hat{X}_{j}^{+} \hat{P}_{j+1}}+\left(J-\tilde{\gamma}\right)\underbrace{\hat{b}_{j+1}^{\dagger} \hat{b}_{j} \delta_{\hat{n}_{j}, 2} \delta_{\hat{n}_{j+1}, 0}}_{\sqrt{2} \hat{P}_{j-1} \hat{X}_{j}^{-} \hat{P}_{j+1}})\\
			&-\sum_{j\in \text{odd}}(J\underbrace{\hat{b}_{j}^{\dagger} \hat{b}_{j+1} \delta_{\hat{n}_{j}, 1} \delta_{\hat{n}_{j+1}, 1}}_{\sqrt{2} \hat{P}_{j-1} \hat{X}_{j}^{+} \hat{P}_{j+1}}+J\underbrace{\hat{b}_{j+1}^{\dagger} \hat{b}_{j} \delta_{\hat{n}_{j}, 2} \delta_{\hat{n}_{j+1}, 0}}_{\sqrt{2} \hat{P}_{j-1} \hat{X}_{j}^{-} \hat{P}_{j+1}}).
		\end{aligned}
\end{equation}
As indicated below each term in the effective Hamiltonian, different processes have a simple spin-1/2 representation with the Fock states mapped to spin product states as follows: 
\begin{equation}\label{suppmap}
	\begin{aligned}
		&``...\textcolor{red}{20}..."\rightarrow``...\downarrow\textcolor{red}{\uparrow}\downarrow...",\\
		&``...\textcolor{red}{11}..."\rightarrow``...\downarrow\textcolor{red}{\downarrow}\downarrow...".
	\end{aligned}
\end{equation}
Here, we map the occupation configuration ``$\textcolor{red}{20}$'' to the spin configuration ``$\downarrow\textcolor{red}{\uparrow}\downarrow$'' in the spin-1/2 basis, with the left- and right-most spins pointing down. In this case, the hopping from  ``$\textcolor{red}{20}$" to  ``$\textcolor{red}{11}$" in Eq.\eqref{supeff} is represented by the spin flip ``$\textcolor{red}{\uparrow}$" to  ``$\textcolor{red}{\downarrow}$". Then, as labeled in Eq.\eqref{supeff}, the effective hoppings under kinetic constraints can be expressed in terms of operators $\hat{X}^{\pm}$: 
\begin{equation}
	\hat{X}^{+}=\left[\begin{array}{ll}
		0 & 1 \\
		0 & 0
	\end{array}\right],	\;\;\; \hat{X}^{-}=\left[\begin{array}{ll}
		0 & 0 \\
		1 & 0
	\end{array}\right].
\end{equation}
Finally, the effective Hamiltonian in Eq.~\eqref{supeff} can be written as the following non-Hermitian PXP Hamiltonian
\begin{equation}\label{suppxp}
	\hat{H}_{\mathrm{NHPXP}}^{\prime}=\Omega\sum_{j\in {\rm even}}\hat{P}_{j-1} \hat{X}^{\prime}_{j} \hat{P}_{j+1}+\Omega\sum_{j\in {\rm odd}}\hat{P}_{j-1} \hat{X}_{j} \hat{P}_{j+1}
\end{equation}
where $\hat{P}=(1-\hat{Z})/2$, and we have introduced
\begin{equation}
	\hat{X}=\left[\begin{array}{ll}
		0 & 1 \\
		1 & 0
	\end{array}\right],
\end{equation}
\begin{equation}\label{X}
	\hat{X}^{\prime}=\left[\begin{array}{cc}
		0 & 1+u \\
		1-u & 0
	\end{array}\right]=\hat{X}+iu\hat{Y},
\end{equation}
\begin{equation}\label{ga}
	\Omega=\sqrt{2}J, \;\;\; u=\tilde{\gamma}/J.
\end{equation}
For the quench dynamics, we are primarily interested in the $\left|\mathbb{Z}_{2}\right\rangle=\left|\uparrow\downarrow\uparrow\downarrow\uparrow\downarrow\uparrow\downarrow\uparrow\downarrow......\right\rangle$ product state of spins, which corresponds to the product state of doublons, $\left|202020...... \right\rangle $, in the Bose-Hubbard model. The latter can be experimentally prepared using an optical superlattice~\cite{su2022observation}. 
Similarly, the occupation imbalance 
	in the initial $\left|202020...... \right\rangle $ state in the Bose-Hubbard model maps to the staggered magnetization in the effective PXP model:
\begin{equation}\label{z}
	\bar{Z}(t)=\sum_{i}(-1)^{i}\left\langle \hat{Z}_{i}(t)\right\rangle.
\end{equation}
FIG.\ref{fig:suppfigure9} shows a comparison of the imbalance dynamics in the Bose-Hubbard model (with OBCs) and its effective PXP counterpart (both PBCs and OBCs). Under the weak non-Hermitian condition $\tilde{\gamma}/J=0.3$, both models manifest pronounced revivals (red curves), a signature of enhanced scars. Importantly, these enhanced revivals occur at similar timescales in the natural unit $\hbar/J$.

\subsection{Coherence time}

In our experimental proposal, we set the coupling strength to $J=0.2$ kHz. This results in an observation duration of $t=10/t= 50$ ms for the results shown in  FIG.~\ref{fig:suppfigure9}. Additionally, the non-Hermitian term is given by $\frac{\Omega_{0}^{2}}{\gamma}\approx0.3 J= 60$ Hz for Eq.~\eqref{suppga}, with $\Omega_{0}=30$ Hz and atom loss $\gamma=15$ Hz.

The cold atom experiment in Ref.~\cite{liang2022observation} suggests the coherence time of $t^{\prime}=1$ ms under the atom loss of $\gamma^{\prime}= 1.3$ kHz, which is significantly stronger than our proposed value. Consequently, the condition $\gamma t < \gamma^{\prime} t^{\prime}$ in our setup implies that robust measurements and intrinsic physics of the scar enhancement can be observed in our framework. Generally, $^{87}$Rb atoms are known to have a long coherence time, exceeding  $100$~ms \cite{hu2017creation}. Therefore, our proposed time scale of $50$ ms is well within the reasonable range, ensuring that the enhanced scar we predict should be observable.

\section{SII. Digital quantum simulation of enhanced scars on the IBM Q processor}

Quantum computers can efficiently simulate the dynamics of many-body systems by representing quantum states using qubits~\cite{preskill2018quantum,preskill2023quantum}. By applying a series of quantum gates, one can then simulate the evolution of complex quantum states. We use the IBM Q platform \cite{santos2016ibm} to demonstrate enhanced scarring on a quantum device. We consider the following model
\begin{equation}\label{ibmqmodel}
	\hat{H}_{Q}=V\sum^{L-2}_{j=0}(\mathds{1}-\hat{P}_{j})(\mathds{1}-\hat{P}_{j+1}) + \sum_{j\in\mathrm{even}}\hat{X}^{\prime}_{j} + \sum_{j\in\mathrm{odd}} \hat{X}_{j},
\end{equation}
where we assume open boundary conditions and focus on the regime $V\gg 1$, which restricts the dynamics to the sector of the Hilbert space described by Eq.~\eqref{suppxp}. This can be shown formally by applying the Schrieffer-Wolff transformation~\cite{bravyi2011schrieffer} at the leading order in \jyd{$1/V$}, which results in the effective model $\mathcal{P}\left(\sum_{j\in\mathrm{even}}\hat{X}^{\prime}_{j} + \sum_{j\in\mathrm{odd}} \hat{X}_{j}\right)\mathcal{P}$, where we have defined the global projector $\mathcal{P}=\prod_j (\mathds{1}-\ket{\uparrow\uparrow}\bra{\uparrow\uparrow}_{j,j+1})$. The action of $\mathcal{P}$ can be equivalently expressed in terms of local projectors $\hat P_j$, resulting in the desired non-Hermitian PXP model from the main text. Thus, we expect enhanced scars to be present in the above model~\eqref{ibmqmodel}, which is easier to implement in terms of native gates on the IBM Q device.

In general, simulating quantum many-body models requires the design of a quantum circuit that represents the Hamiltonian. This setup involves choosing and arranging quantum gates which mimic coupling. However, most native gates on the IBM Q processor are unitary, so we can not directly decompose arbitrary non-unitary operations using these unitary gates. Therefore, to implement the dynamics of the model~\eqref{ibmqmodel}, we employ the ancilla-based method, demonstrated in Refs.~\cite{chen2022high,shen2023observation}.
The idea behind this method is to use additional ancilla qubits that will be measured at the end of the time evolution. While the evolution of the full system (physical and ancilla qubits) is unitary, the evolution of the physical subsystem does not have to be, since it is not a closed system but interacts with the ancilla qubits. The desired time evolution $U(t)=e^{-it\hat{H}_{Q}}$ is embedded in an extended unitary:
	\begin{equation}\label{unitary}
		U^{\prime}=\left[\begin{array}{cc}
			zU(t) & B \\                          
			C & D
		\end{array}\right] \quad \text{as} \quad U^{\prime}=\left[\begin{array}{cc}
			\text{physical}\to\text{physical} & \text{ancilla}\to\text{physical} \\                          
			\text{physical}\to \text{ancilla}  & \text{ancilla}\to\text{ancilla}
		\end{array}\right],
	\end{equation}
	where $z=1/\sqrt{\lambda_\mathrm{max}}$, with $\lambda_\mathrm{max}$ representing the maximum eigenvalue of $U^{\dagger}(t)U(t)$. This extended unitary $U^{\prime}$ represents the blue block shown in FIG.~4 (c) of the main text, which couples the physical and ancilla qubits. Here, the component $C$ is given by 
	$C=A\sqrt{I-z^{2}\Sigma^{2}}E$, where $A$, $\Sigma$ and $E$ are obtained through the singular value decomposition $U^{\dagger}U=A\Sigma E^{\dagger}$ ~\cite{lin2021real}. The components $B$ and $C$ can then be determined by numerically solving for the QR decomposition 
	\begin{equation}\label{QR}
		\left[\begin{array}{cc}
			zU & I \\
			C & I
		\end{array}\right]=U^{\prime}R,
	\end{equation}
	where $R$ is the upper triangular matrix of the QR decomposition. In this process, the block $zU$ and $C$ will not be modified, and the solved $B$ and $D$ can ensure the unitary nature of $U^{\prime}$

\begin{figure}
	\centering
	\includegraphics[width=0.9\linewidth]{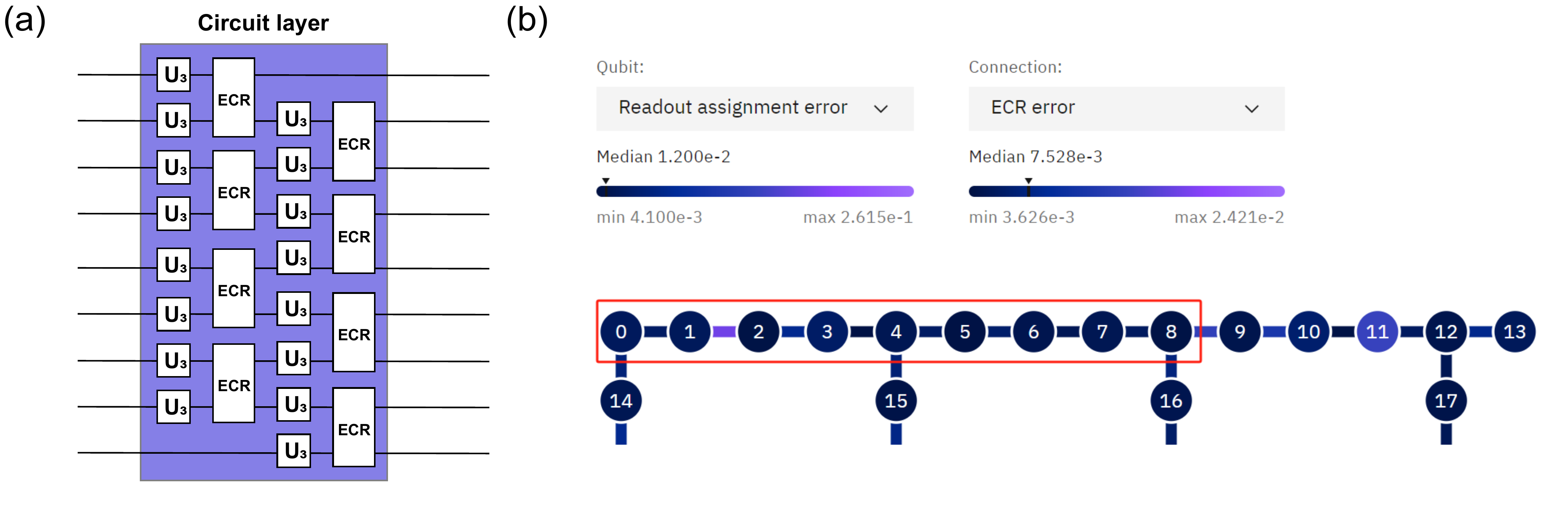}
	\caption{(a) The structure of the trainable circuit for Eq.~\eqref{cost}.  Each blue block represents a circuit layer. The trainable part is the $U_{3}$ gate described by Eq.~\eqref{u3}. The two-qubit gate is the Echoed Cross-Resonance (ECR) gate. (b) Noise conditions of the IBM Q Brisbane device on $2024/05/08$. We present the error rates of measurement and ECR gates. The simulation in this work was conducted on the chain of $0$-$8$ qubits.
	} 
	\label{fig:ibmq}
\end{figure}

We set the first site to be the ancilla qubit, which requires post-selection on $\ket{\uparrow}$, and the rest are for the physical system. In this subspace of the ancilla qubit, this initial state can be expressed as $(\ket{\psi_{\rm Physical}},0)^\mathsf{T}$, with the ancilla $\ket{\uparrow}=(1,0)^\mathsf{T}$. The extended unitary leads to the result: 
	\begin{equation}\label{nonunssh}
		U^{\prime}\ket{\psi_{\rm Physical}}\ket{\uparrow}=\left[\begin{array}{cc}
			zU & B \\
			C & D
		\end{array}\right]\left[\begin{array}{c}
			\ket{\psi_{\rm Physical}}\\
			0
		\end{array}\right]=\left[\begin{array}{c}
			zU\ket{\psi_{\rm Physical}}\\
			C\ket{\psi_{\rm Physical}}
		\end{array}\right].
	\end{equation}
	The post-selection of $\ket{\uparrow}$ on the ancilla qubit leads to the normalization as follows
	\begin{equation}
		(I\otimes\ket{\uparrow}\bra{\uparrow})(U^{\prime}\ket{\psi_{\rm Physical}}\ket{\uparrow})\rightarrow U(t)\ket{\psi_{\rm Physical}}/\left\|U(t)\ket{\psi_{\rm Physical}}\right\| .
	\end{equation}
	For measurements on quantum computers, we discard the string with the $\downarrow$ ancilla outcome to achieve the above   post-selection.
	In order to implement this unitary on quantum circuits, we prepare a trainable circuit $M$, as shown in FIG.~\ref{fig:ibmq} (a), and minimize the following cost function to precisely replicate the target dynamics
	\begin{equation}\label{cost}
		{\bf Cost}=1-|\bra{\psi_{0}}M^{\dagger}U_{\bf tar}\ket{\psi_{0}}|,
	\end{equation}
	where $\ket{\psi_{0}}$ is the initial state. Here, we use the initial state $\ket{\psi_{0}}=\ket{\psi_{\rm Physical}}\ket{\psi_{\rm Ancilla}}=\ket{\uparrow\downarrow\uparrow\downarrow\uparrow\downarrow\uparrow\downarrow}\ket{\uparrow}$ for a 9-qubit system. This state can be prepared through applying $X$ gates on  an initial state $\ket{\uparrow\uparrow\uparrow\uparrow\uparrow\uparrow\uparrow\uparrow}\ket{\uparrow}$.

The circuit $M$ is parameterized through $U_{3}$ gates 
	\begin{equation}\label{u3}
		U_{3}(\theta, \phi, \lambda)=\left[\begin{array}{cc}
			\cos \left(\frac{\theta}{2}\right) & -e^{i \lambda} \sin \left(\frac{\theta}{2}\right) \\
			e^{i \phi} \sin \left(\frac{\theta}{2}\right) & e^{i(\phi+\lambda)} \cos \left(\frac{\theta}{2}\right)
		\end{array}\right].
	\end{equation}
	The training process involves optimizing the angles $\theta$, $\phi$, and $\lambda$. The trainable circuits for simulations are optimized through the L-BFGS-B method, that can efficiently handle large-scale problems with bounds on the parameter range. The L-BFGS-B method iteratively adjusts the parameters to minimize the cost function towards zero, ensuring that the circuit can reach proper convergence.

In general, deeper trainable circuits can lead to faster convergence in optimization tasks, but they can also be more significantly impacted by noise. Therefore, in designing our trainable circuits, we aimed to reach a balance between achieving efficient convergence and maintaining a manageable circuit depth (i.e., the number of circuit layers).

Through empirical testing, we found that the optimal trainable circuits could be constructed within $8$ layers, shown in FIG.~\ref{fig:ibmq} (a). This ensures that the cost function can reach the convergence below $1\%$, and simulations are robust to noise. Each circuit layer is built by 16 $U_{3}$ gates and $8$ ECR gates. The ECR gate is defined by the matrix:
	\begin{equation}
		E C R=\frac{1}{\sqrt{2}}\left(\begin{array}{cccc}
			0 & 1 & 0 & i \\
			1 & 0 & -i & 0 \\
			0 & i & 0 & 1 \\
			-i & 0 & 1 & 0
		\end{array}\right)
	\end{equation}
	This specific gate is chosen to optimize the noise resilience of circuits. Our simulation results are shown in  FIG. 4 (d) (main text), conducted on the IBM Q Brisbane device.  For these simulations, we used the chain of $0$-$8$ qubits on this device. These results demonstrate that our approach successfully balances the circuit depth with the practical constraints imposed by noise and decoherence, ensuring robust performance on current quantum hardware.

\section{SIII. Fock skin accumulation and hopping inhomogeneity in the FSA chain}

As discussed in the main text, the origin of the enhanced scar in our model is the non-Hermitian skin accumulation occurring within the Fock space, which can be understood as the accumulation in the FSA chain. Contrary to the uniform non-reciprocity of the physical hoppings, the hopping asymmetry in the effective FSA chain is highly non-uniform, potentially even flipping at different parts of the chain. This is because the chain is along the Hamming distance $x$ from the $\ket{\mathbb{Z}_2}$ state, and each ``site'' actually represents an extensively large number of Fock basis states.

To better understand the nature of the Fock skin effect in our model, Eq.~\eqref{suppxp}, we compare three different variations of it, labeled by $a$, $b$, and $c$:
\begin{equation}\label{a}
	\hat{H}_{\mathrm{a}}=\sum_{j}\hat{P}_{j-1} \hat{X}^{\prime}_{j} \hat{P}_{j+1},
\end{equation}
\begin{equation}\label{b}
	{\bf Primary~model:} \hat{H}_{b}=\sum_{j\in {\rm even}}\hat{P}_{j-1} \hat{X}^{\prime}_{j} \hat{P}_{j+1}+\sum_{j\in {\rm odd}}\hat{P}_{j-1} \hat{X}_{j} \hat{P}_{j+1},
\end{equation}
\begin{equation}\label{c}
	\hat{H}_{c}=\sum_{j\in {\rm even}}\hat{P}_{j-1} \hat{X}^{\prime}_{j} \hat{P}_{j+1}+\sum_{j\in {\rm odd}}\hat{P}_{j-1} (\hat{X}^{\prime}_{j})^{\dagger} \hat{P}_{j+1}.
\end{equation}
Note that this model $b$ describes our experimental setup in Eq.~\ref{suppxp} with unit $\Omega=1$.

The Hamiltonian $\hat{H}_{a}$ describes the model considered in Ref.~\cite{chen2023weak}, which is invariant under translation. Model $b$ is the one we investigate in the main text, as $\hat{H}_{b}$ is equivalent to Eq.~\eqref{suppxp} with $\Omega=1$. Finally, we introduce the model $c$ as a more extreme version of the model $b$. Indeed, in the model $c$ the non-Hermitian elements are staggered to mimic the structure of the N\'eel state. This means that, in the graph representation of FIG.~1 in the main text, not only are all couplings asymmetric (unlike in our model $b$) but they \emph{all} point towards the N\'eel state (unlike in model $a$). This is expected to make the scarring more prominently linked to the Fock skin effect. However, the presence of $(\hat{X}^{\prime}_{j})^{\dagger}$ in $\hat{H}_c$ makes it much more complicated to implement in a Bose-Hubbard quantum simulator. For that reason, in the main text, we only consider model $b$. 

\begin{figure}[ht]
	\centering
	\includegraphics[width=0.9\linewidth]{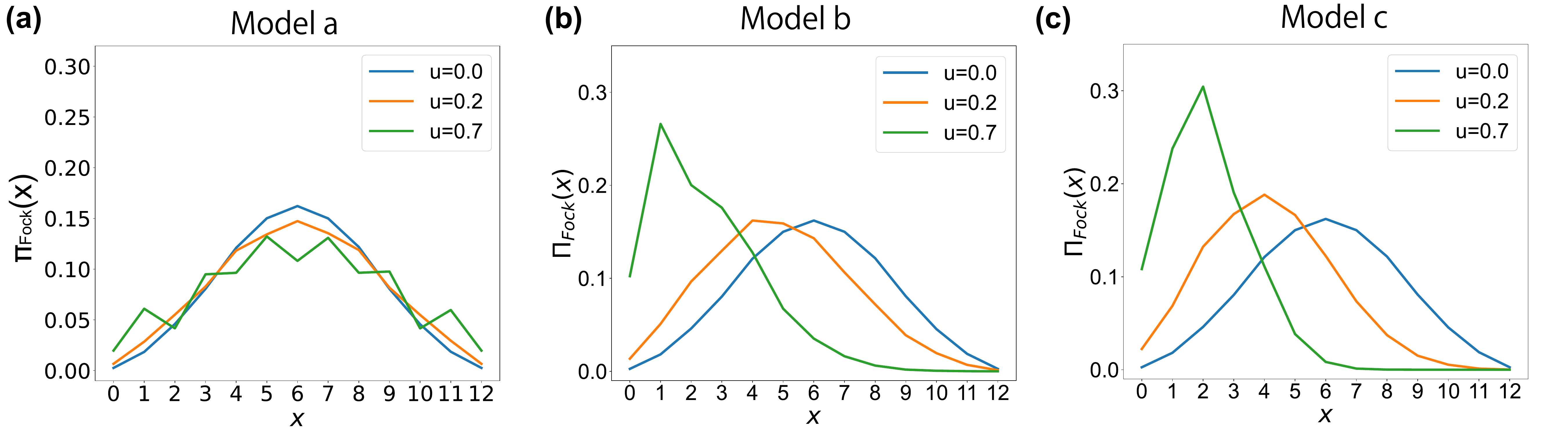}
	\caption{Fock skin accumulation in various models, Eqs.~(\ref{a})-(\ref{c}). All the panels show the Fock-space density, defined by Eq.~\eqref{suppipr}, for the system size $L=12$. 
		In the Hermitian case $u=0$, $\Pi_\mathrm{Fock}(x)$ is perfectly symmetric and there is no Fock skin accumulation. For the model $a$, there is still no accumulation even with non-zero $u$. On the other hand, as expected, models $b$ and $c$ show a strong bias towards $x=0$ when $u>0$.
	}
	\label{fig:suppfock}
\end{figure}
In the main text, we have introduced the density in Fock space
\begin{equation}\label{suppipr}
	\Pi_{\rm Fock}(x)=\frac{1}{\mathcal{D}}\sum_{i}\sum_{\ket{\varphi} \in \mathcal{L}(x)}|\bra{\varphi}\ket{\phi_{i}}|^{2},
\end{equation}
where $\mathcal{L}(x)$ denotes the set of states in the Hamming layer $x$~\cite{yao2023observation}, $\ket{\phi_{i}}$ is the $i$th eigenstate, and $\mathcal{D}$ is the total number of Fock states.  The Fock-space density for the above three models is shown in FIG.~\ref{fig:suppfock}. For our proposed improvement of QMBS through the Fock skin effect, a key criterion is that the non-Hermiticity must generate an overall asymmetric flow in the entire Fock space with respect to the scarred trajectory. However, here we find that the Fock skin modes only appear in models $b$ and $c$, as revealed by the asymmetric profiles in FIGs.~\ref{fig:suppfock}(b)-(c).

To understand why no Fock skin accumulation appears in the model $a$, we examine the symmetrized transition amplitudes between Hamming layers, as given by
\begin{equation}\label{fockh}
	T^{+}_{x}=\bra{\tilde x}H\ket{\tilde x+1} \quad \text{and} \quad T^{-}_{x}=\bra{\tilde x+1}H\ket{\tilde x},
\end{equation}
where $\ket{\tilde x}=\sum_{\ket{\varphi}\in \mathcal{L}(x)}\ket{\varphi}$ is a symmetric superposition of all the states in the Hamming layer $x$. It is important to note that $T^{\pm}_x$ differ from the $t^{\pm}_x$ FSA hoppings used in the main text. While $t^{\pm}_x$ are meant to capture the scarred dynamics, $T^\pm_x$ give a more general idea of a state flow between the Hamming layers.  Note that $T^\pm_x$ only meaningfully represent the individual physical Fock state transitions when $-1\leq u\leq 1$, where all non-zero matrix elements of the physical Hamiltonian are positive in the Fock basis, and hence there can be no cancellation between different terms.

\begin{figure}[ht]
	\centering
	\includegraphics[width=0.99\linewidth]{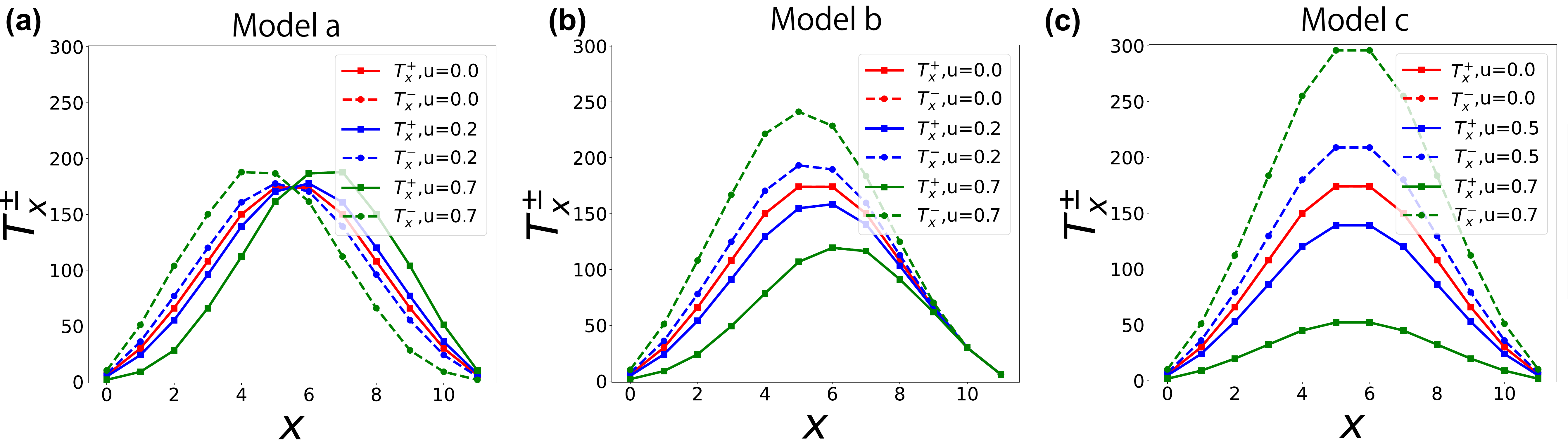}
	\caption{Symmetrized transition amplitudes $T^{\pm}_{x}$, Eq.~\eqref{fockh}, between the Hamming layers for the models in Eqs.~\eqref{a}-\eqref{c}.  The system size is $L=12$ in all the plots. Note that for the model $a$, $T^+_x$ is larger than $T^-_x$ on the right, and vice versa on the left, hence leading to no consistent Fock skin pumping.
	}
	\label{fig:suppfockhopping}
\end{figure}

To have the Fock skin effect for scar enhancement, instead of merely strong boundary localization, two conditions must be met: (i) consistent asymmetry of the transitions $T^{-}_{x}>T^{+}_{x}$, and  (ii) $T^{\pm}_{0}\rightarrow 0$ which prevents strong skin accumulation at the $x=0$ boundary. The latter condition is necessary to prevent outright exponential skin accumulation at $x=0$, which will largely prevent the dynamical state from exploring the other Hamming layers.  In light of these conditions, we first examine the hopping amplitudes of Fock-space Hamiltonians in FIG.~\ref{fig:suppfockhopping}. The model $a$ in Eq.~\eqref{a} fails to meet the condition  $T^{-}_{x}>T^{+}_{x}$, as it has opposite asymmetry in the left/right parts of the chain. This accounts for the previous lack of any Fock-space NHSE in FIG.~\ref{fig:suppfock}(a). Conversely, for models $b$ and $c$, the non-reciprocity, as demonstrated by blue and green curves in FIGs.~\ref{fig:suppfockhopping}(b)-(c), generates an overall non-Hermitian pumping in the Fock space. Indeed, the model $c$, with greater $T^\pm_x$ hopping asymmetry at some $x$, gives rise to a very slightly stronger Fock skin accumulation than model $b$. Importantly, the condition $T^{\pm}_{0}\rightarrow 0$ ensures that there is no robust boundary localization. 

\begin{figure}[htb]
	\centering
	\includegraphics[width=0.8\linewidth]{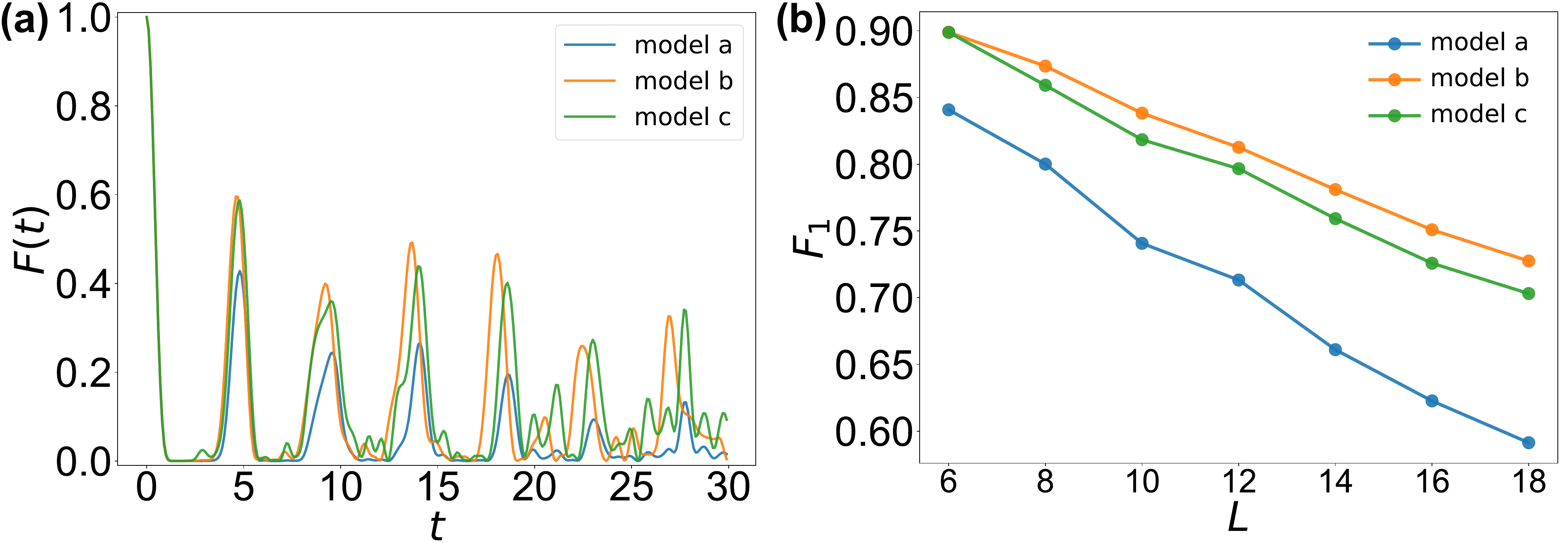}
	\caption{(a) Fidelity dynamics for the models in Eqs.~\eqref{a}-\eqref{c}, under the disorder of $W=1.0$ in Eq.~\eqref{suppdis}. We set $L=14$, $u=0.35$ and the $\ket{\mathbb{Z}_{2}}$ initial state. Due to the absence of the Fock skin effect, the QMBS revivals for the model $a$ (blue curve) rapidly collapse. (b) The scaling behavior of the first fidelity peak $F_{1}$ within $t\in[2.5,7.5]$, obtained by $200$ realizations. We set $W=0.8$ and $u=0.4$ for (b).}
	\label{fig:suppfigure8}
\end{figure}

As discussed in the main text, the resilience to perturbations stands out as a hallmark of the Fock skin effect. Thus, we examine the behavior of our models $a$, $b$ and $c$, subject to spatial disorder:
\begin{equation}\label{suppdis}
	\begin{aligned}
		\hat{H}_{\rm def}=\hat{H}+\sum_{j}w_{j}\hat{Z}_{j},
	\end{aligned}
\end{equation}
where $w_{j}\in [-W/2,W/2]$ is the strength of an on-site random potential~\cite{mondragon2021fate}. As shown in FIG.~\ref{fig:suppfigure8} (a), models $b$ and $c$ exhibit relatively robust QMBS revivals even at strong disorder. This is attributed to the Fock skin effect. In contrast, model $a$ shows a rapid decay, akin to the behavior observed in the unitary scenario shown in the main text. Moreover, the first fidelity peak $F_{1}$ across system sizes $L$, as depicted in FIG.~\ref{fig:suppfigure8} (b),  reveals that models b and c exhibit slower decay rates, compared to model a. Importantly, this shows the robustness of our enhanced scars, which are expected to be pronounced in physical experiments involving larger system sizes.

\begin{figure}[ht]
	\centering
	\includegraphics[width=0.6\linewidth]{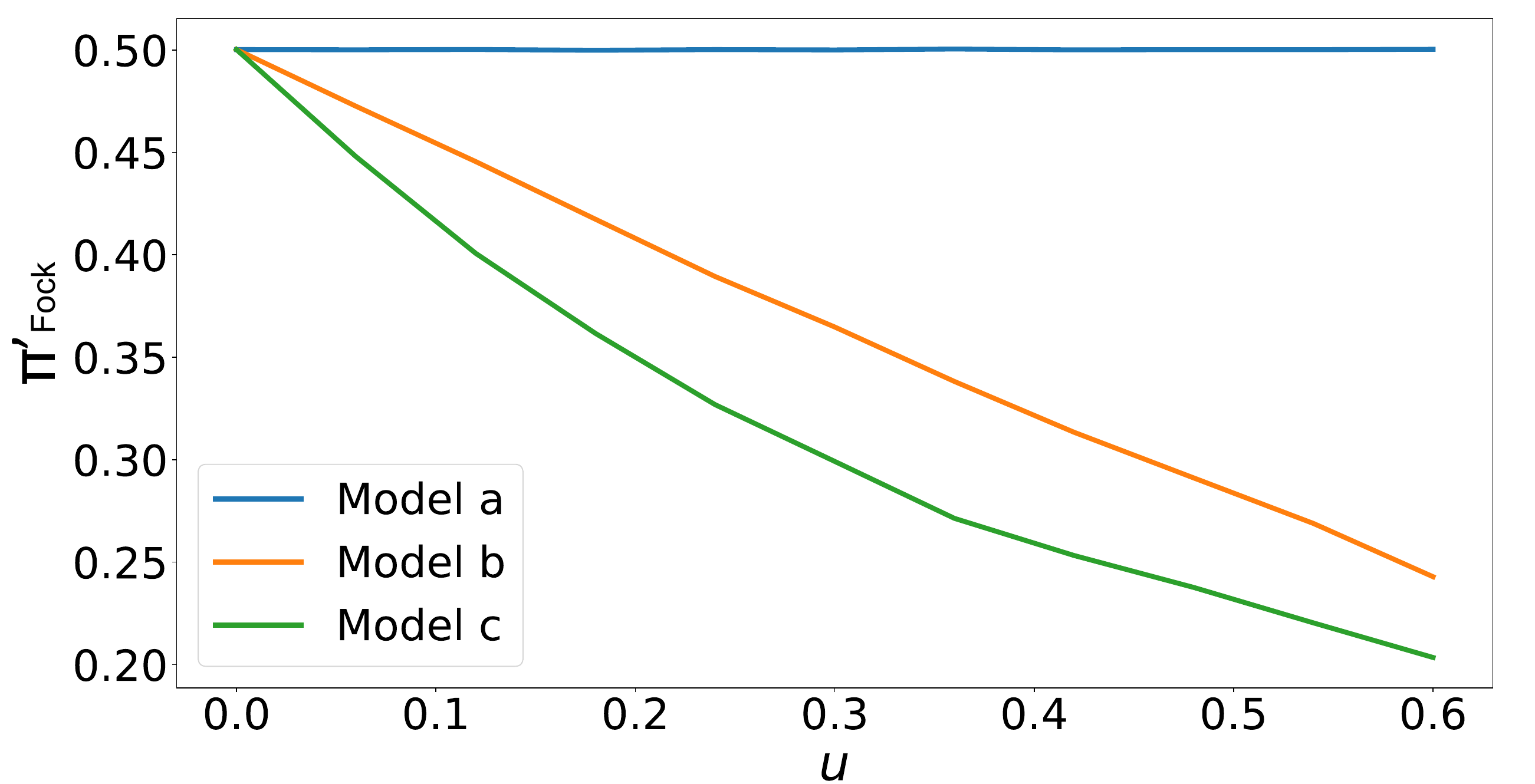}
	\caption{The center of Fock-space density, as defined in Eq.~\eqref{suppipr3} for $L=12$, compared across three models in Eqs.~\eqref{a}-\eqref{c}. Due to the absence of the Fock skin effect, the model $a$ does not exhibit a leftward shift. For models $b$  and $c$, the mean distance does not reach zero, indicating a lack of strong localization at the boundary of the Fock space. 
	} 
	
	\label{fig:supp-ipr}
\end{figure}

To quantify the extent of Fock skin accumulation, and also to clarify that there is no  boundary localization compared to the conventional NHSE, we define the mean position of the Fock-space density as follows
\begin{equation}\label{suppipr3}
	\begin{aligned}
		\Pi^{\prime}_{\rm Fock}=\frac{1}{L}\sum_{x}x\Pi_\mathrm{Fock}(x).
	\end{aligned}
\end{equation}
This quantity reveals which Hamming layers the state mostly resides in. As illustrated in FIG.~\ref{fig:suppfock}, a key factor linked to enhanced scar dynamics is the leftward shift of this density, due to the Fock skin effect. This leads to a shift of $\Pi^{\prime}$ towards $x=0$ for the models in Eq.~\eqref{b} and  Eq.~\eqref{c}, as shown in FIG.~\ref{fig:supp-ipr}.
However, importantly $\Pi^{\prime}_{\rm Fock}$ does not tend towards zero as $u$ increases to $0.6$. This is further supported by the effective hopping depicted in FIG.~\ref{fig:suppfockhopping}.

\section{SIV. Real-complex transition in the spectrum of the non-Hermitian PXP model}

Our PXP Hamiltonian can be expressed as follows
\begin{equation}\label{suppro}
	\hat{H}=\sum_{i=1}^{N} \mathcal{P} \hat{X}_{i} \mathcal{P},
\end{equation}
where the global projector is
$\mathcal{P}=\prod_{i}\left[1-\left(1+\hat{Z}_{i}\right)\left(1+\hat{Z}_{i+1}\right) / 4\right]$. Applying this global projector on the full Hilbert space, we can extract effective vectors, which lead to the following form
\begin{equation}
	H^{P}=\sum_{i=1}^{N}\hat{X}_{i},
\end{equation} 
Similarly, we can also express our non-Hermitian PXP in this sector
\begin{equation}\label{supppxpp}
	\hat{H}_{\mathrm{NHPXP}}^{P}=\sum_{j=2n} \hat{X}^{\prime}_{j}+\sum_{j=2n+1}\hat{X}_{j}.
\end{equation}

\begin{figure}[htb]
	\centering
	\includegraphics[width=0.75\linewidth]{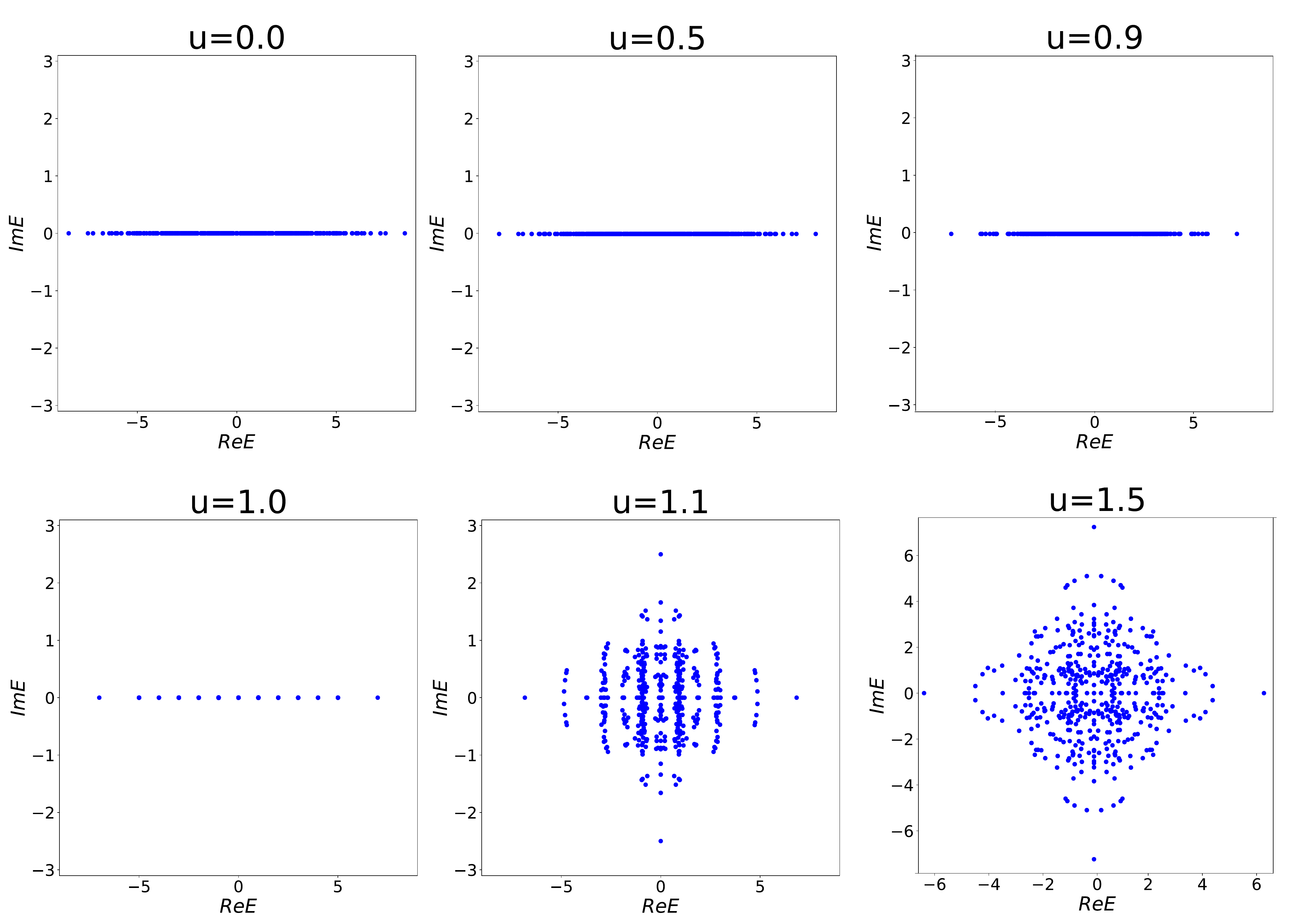}
	\caption{The real-complex transition at $u=1$ in our non-Hermitian PXP model in Eq.\eqref{b} under PBCs.}
	\label{fig:spectra}
\end{figure}

As such, the $\mathcal{P} \mathcal{T}$ symmetry breaking in our non-Hermitian setup appears at $u=1$ with the emergence of complex spectra, see FIG.~\ref{fig:spectra}~\cite{heiss2012physics,klett2017relation,longhi2019topological,sakhdari2019experimental,miri2019exceptional,ozdemir2019parity,lee2022exceptional,ding2022non,yang2023percolation,meng2024exceptional}. Given the following form of $\hat{X}^{\prime}$
\begin{equation}
	\hat{X}^{\prime}=\left[\begin{array}{cc}
		0 & 1+u \\
		1-u & 0
	\end{array}\right],
\end{equation}
its singularity appears at $u=1$, and complex eigenvalues emerge under $u>1$. 
In the main text, we thus only consider the case $u<1$, where the spectrum of our non-Hermitian PXP model is purely real.

\section{SV. Entanglement entropy}

In this section, we study the entanglement entropy of eigenstates of our non-Hermitian PXP model. As the focus of our work is state dynamics~\cite{kawabata2023entanglement,orito2023entanglement} (of the $\mathbb{Z}_2$ N\'eel state), we will consider the entanglement of the \emph{right} eigenvectors, rather than the biorthogonal entanglement entropy~\cite{chang2020entanglement,lee2022exceptional,hsieh2023relating,zou2023experimental,fossati2023symmetry,rottoli2024entanglement}.

\begin{figure}[htb]
	\centering
	\includegraphics[width=0.54\linewidth]{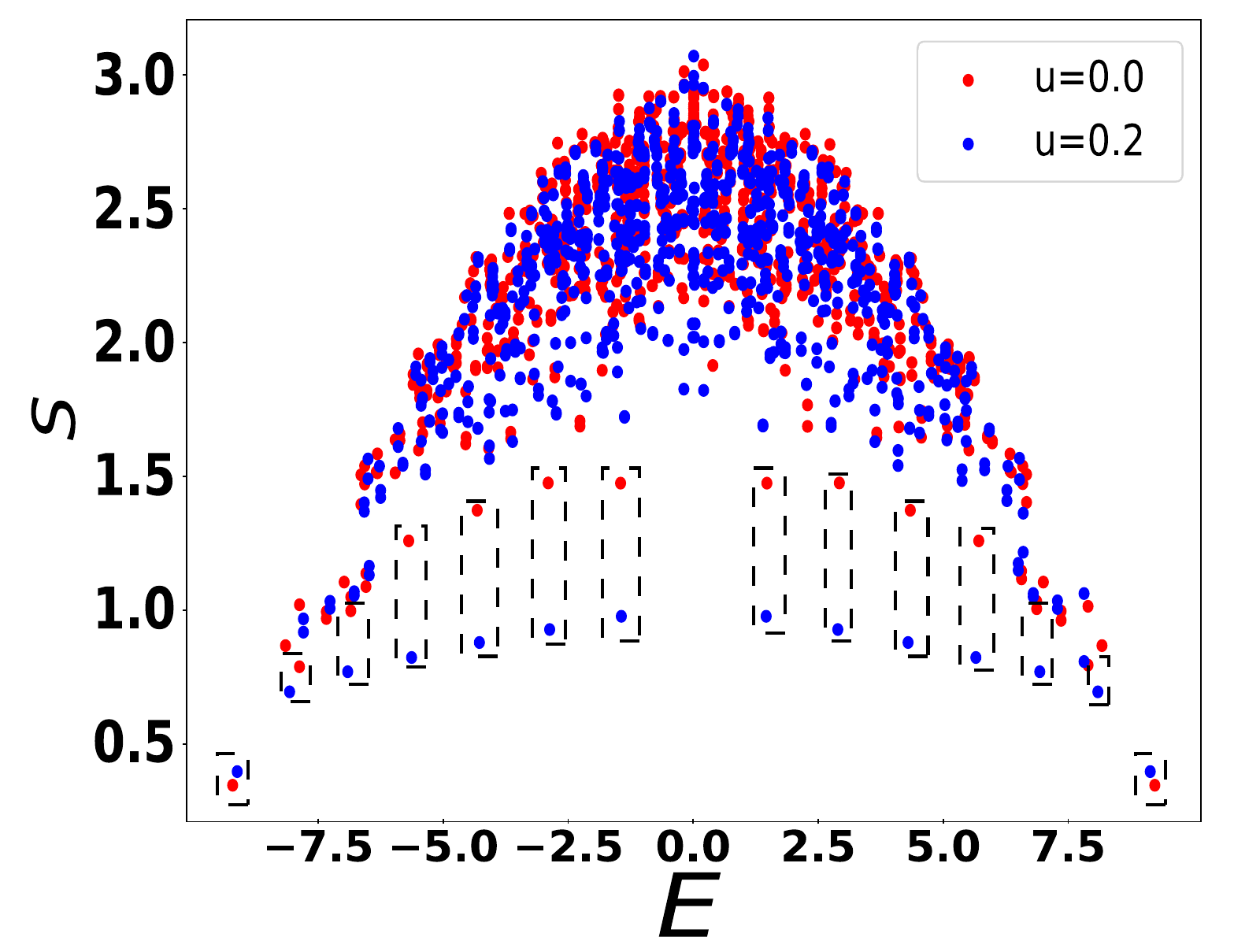}
	\caption{Half-chain entanglement entropy in Eq.~\eqref{b} . We set $u=0.0$ for red dots and $u=0.2$ for blue dots. Note the significantly separated branch of low $S$ for the non-Hermitian case $u=0.2$ (blue).
		The system size is $L=16$.}
	\label{fig:suppee}
\end{figure}

The evolution of a wavefunction can be expressed using only the right eigenstates $\left|R_{i}\right\rangle$ with eigenvalues $E_{i}$ as
\begin{equation}
	\ket{\psi(t)}=\sum_{i}C_{i}e^{-itE_{i}}\ket{R_{i}}.
\end{equation}
Moreover, there is no significant skin effect along physical spaces. Thus, to characterize such dynamics through entanglement entropy, here we express the density matrix of our non-Hermitan model in terms of right eigenstates. We consider the half-chain entanglement entropy given by
\begin{equation}\label{suppen}
	S\left(\rho^{RR}\right)=-\operatorname{Tr} [\rho^{RR} \log\rho^{RR}],
\end{equation}
where $\rho^{RR}$ is the reduced density matrix of the left-half chain.
The half-chain entanglement entropy $S$ is presented in FIG.~\ref{fig:suppee}. This allows us to identify and further characterize the scar eigenstates. Due to their large projection on a single product state $\ket{\mathbb{Z}_2}$, the entanglement entropy of scarred eigenstates is expected to be significantly lower compared to typical thermalizing eigenstates at the same energy density. This is confirmed in FIG.~\ref{fig:suppee} which shows the entanglement entropy of scarred eigenstates to be smaller and more separated from that of the thermal states for $u=0.2$ (blue) compared to $u=0$ (red), as highlighted by the dashed boxes. As these marked scar eigenstates are the main states underpinning the dynamics studied in the main text, the entanglement suppression can be related to the enhancement in the scarring dynamics. 

\begin{figure}
	\centering
	\includegraphics[width=0.7\linewidth]{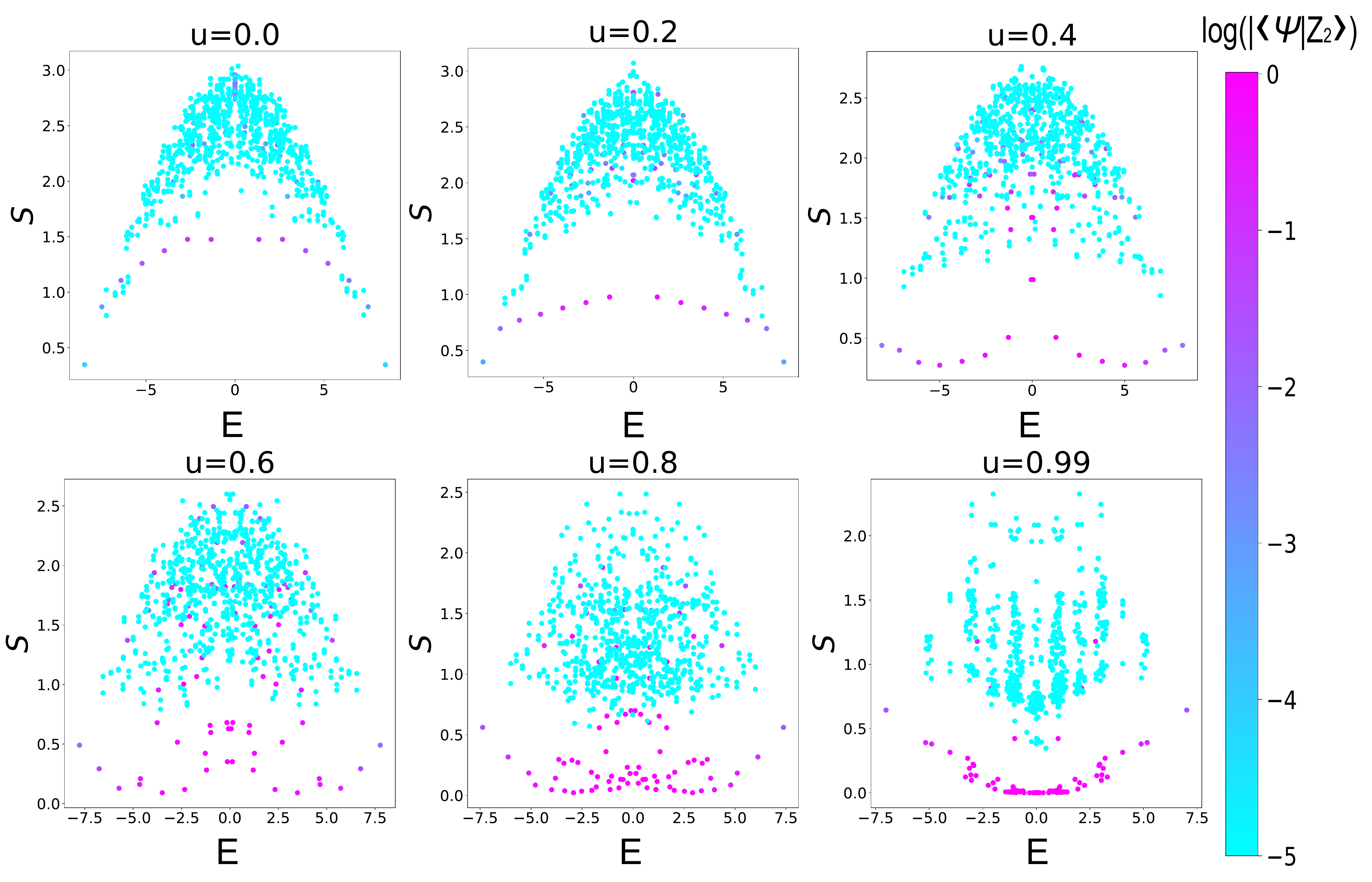}
	\caption{Half-chain entanglement entropy in Eq.~\eqref{b} under different values of $u$. The states with high $\mathbb{Z}_{2}$ overlap ($\log(|\left\langle\psi|\mathbb{Z}_{2} \right\rangle |)$) are colored in violet. Under the increasing  values of $u$, the intrinsic scarred states show lower entropy, and there emerge more low-entropy states. The system size is $L=14$.}
	\label{fig:suppen}
\end{figure}

Finally, we consider the entanglement entropy  under different values of $u$. As shown in FIG.~\ref{fig:suppen}, each eigenstate is identified through the overlap  $\log(|\left\langle\psi|\mathbb{Z}_{2} \right\rangle |)$.  It is evident that more low-entropy states emerge when $u>0.4$. When $u$ reaches $0.99$, these states accumulate near $S=0$. These results indicate a significant enhancement  under large $u$, which is further supported by the results in FIG.~\ref{fig:oscillation}.

\section{Structure of scarred revivals}
\begin{figure}[ht]
	\centering
	\includegraphics[width=1\linewidth]{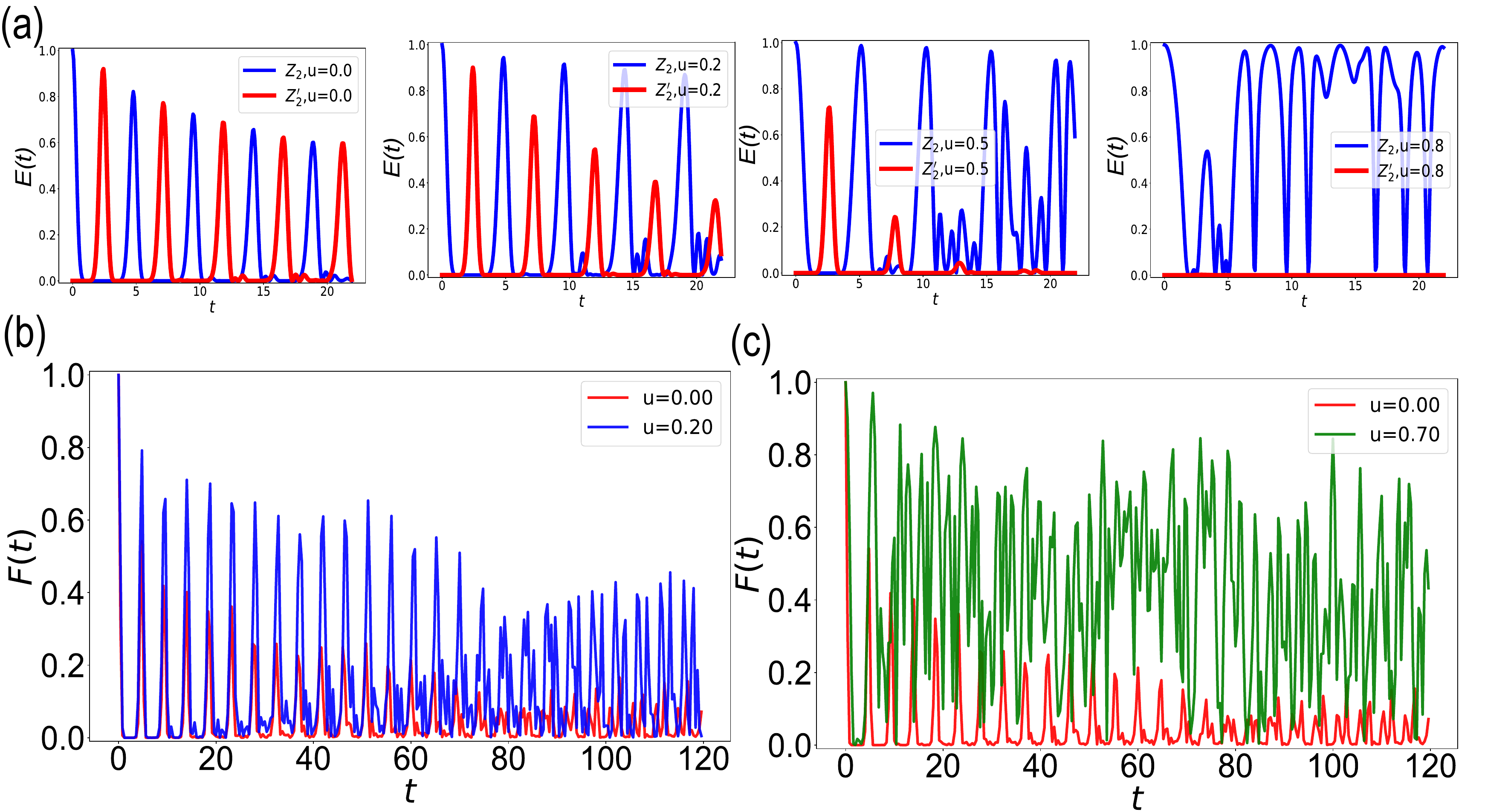}
	\caption{(a) Echoes in Eq.~\eqref{echo}, with $\ket{\phi}=\ket{\mathbb{Z}_{2}}$ for red curves and $\ket{\phi}=\ket{\mathbb{Z}^{\prime}_{2}}=\prod_{i} \hat{X}_{i}\ket{\mathbb{Z}_{2}}$ for blue curves. For the Hermitian case with $u=0.0$, the dynamics exhibits balanced oscillation between states $\ket{\mathbb{Z}_{2}}$ and $\ket{\mathbb{Z}^{\prime}_{2}}$. Under the weak non-Hermiticity with $u=0.2$, the echo from $\ket{\mathbb{Z}_{2}}$ state exhibits higher peaks, showing enhanced quantum scarring in our model. For the cases under strong non-Hermiticity under $u=0.5$ and $u=0.8$, more complicated oscillations emerge, and the echo of $\ket{\mathbb{Z}^{\prime}_{2}}$ vanishes. (b) Normalized fidelity $F(t)$ over long-time dynamics. Compared to the conventional scarring revivals (red curve), the non-Hermiticity of $u=0.2$ (dashed green curve) gives rise to robust enhanced revivals over an extended period of $t=120$. (c) Normalized fidelity $F(t)$ under the disorder $\sum_{j}w_{j}\hat{Z}_{j}$, with $w_{j}\in [-0.5,0.5]$. The green curve in (c) demonstrates the strong protection under disorders. System size is $L=14$ for all panels.
	}
	\label{fig:oscillation}
\end{figure}

To further characterize the QMBS revivals discussed in the main text, we plot the following echoes in FIG.~\ref{fig:oscillation} (a)
\begin{equation}\label{echo}
	E(t)=|\bra{\phi}\ket{\psi(t)}|^{2}/{\bra{\psi(t)}\ket{\psi(t)}},
\end{equation}
with $\bra{\psi(t)}=\exp(-i\hat H t)\ket{\psi(0)}$, states $\ket{\phi}=\ket{\mathbb{Z}_{2}}$ and $\ket{\phi}=\ket{\mathbb{Z}^{\prime}_{2}}=\prod_{i} \hat{X}_{i}\ket{\mathbb{Z}_{2}}$.
For $u=0.0$, we find that many-body revivals act as an oscillation between $\ket{\mathbb{Z}}$ and  $\ket{\mathbb{Z}^{\prime}_{2}}$, with the fidelity decaying with each state in-step. Meanwhile, for the non-Hermitian case $u=0.2$, these oscillations become unbalanced, with echoes to $\ket{\mathbb{Z}_{2}}$ being stronger than for the Hermitian $u=0.0$ case. This is further exemplified for $u=0.5$ and $u=0.8$, where the dynamics and revivals predominantly occur only for the $\ket{\mathbb{Z}_{2}}$ state. This leads to the more complex oscillations observed in our model. Nonetheless, we emphasize that the lifetime of the echoes to the $\ket{\mathbb{Z}_{2}}$ state are already greatly improved for $u=0.2$, where the dynamics still explore most of the Hilbert space. This can be extended to the long-time dynamics shown in FIG.~\ref{fig:oscillation} (b). These results indicate that the improvement is not merely due to the wave function localizing around the $\ket{\mathbb{Z}_{2}}$ state.

As previously demonstrated in the main text and FIG.~\ref{fig:oscillation}(a), there is a strong scar enhancement under large values of $u$, Building on this, we further investigated the dynamics in the presence of disorder $\sum_{j}w_{j}\hat{Z}_{j}$.  Compared to the dynamics under $u=0.2$ [blue curve in FIG.~\ref{fig:oscillation} (b)], the green curve for $u=0.7$ in FIG.~\ref{fig:oscillation}(c) shows robust revivals, while the red curve for $u=0.0$ decays rapidly. These results highlight the significant protection and stability of the enhanced scarred states.

\begin{figure}
	\centering
	\includegraphics[width=0.7\linewidth]{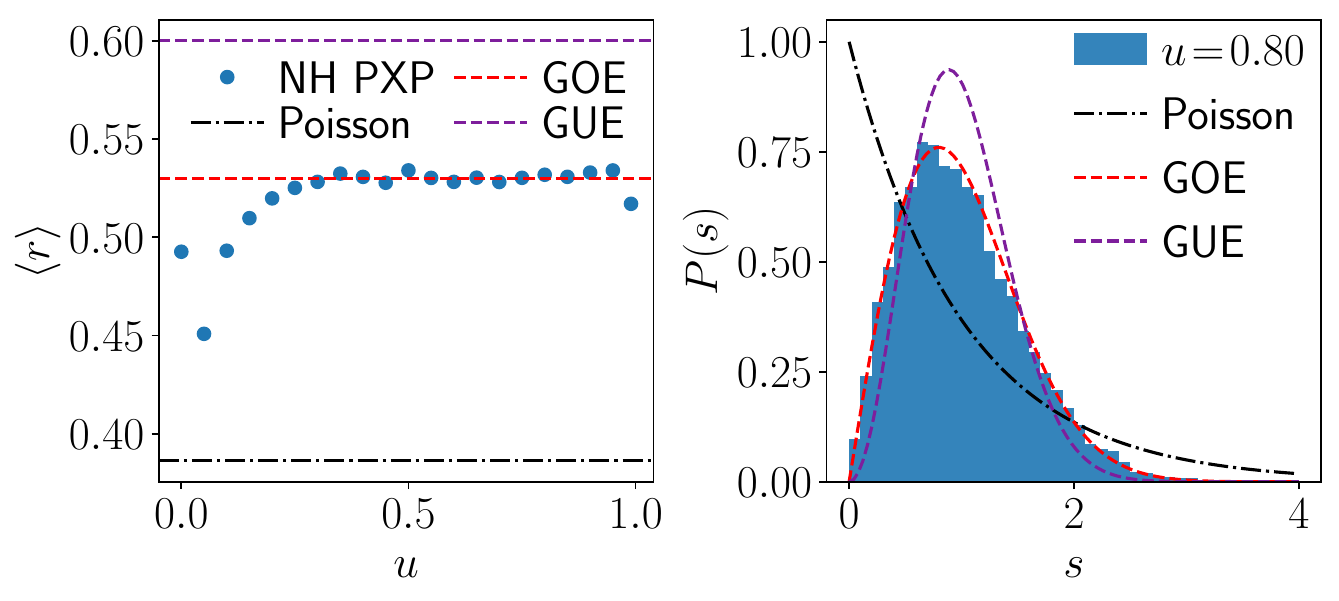}
	\caption{Left: The mean ratio of adjacent energy gaps $\left\langle r\right\rangle $. Right: The level spacing distribution. All data is for the fixed system size $L=28$. The red and violet dashed curves represent the the Gaussian orthogonal ensemble (GOE) and Gaussian unitary ensemble (GUE), respectively, while the black line is for the Poisson distribution of spacings, typical of integrable systems.}
	\label{fig:levelspacing}
\end{figure}

\section{SVI. Level spacing distribution}

As a probe of ergodicity, we compute the average ratio of consecutive level spacings 
	\begin{equation}\label{r}
		r_{n}={\rm Min}[\delta_{n},\delta_{n+1}]/{\rm Max}[\delta_{n},\delta_{n+1}],
	\end{equation}
	where $\delta_{n}=E_{n+1}-E_n$ is the energy difference between two adjacent energy levels in ascending order. We recall that for $|u|<1$ the eigenvalues of the Hamiltonian are all real, therefore they can be sorted without ambiguity. To avoid degeneracies (or near degeneracies) between different symmetry sectors, we always restrict our study to the maximally symmetric sector. For $u=0$, this is the sector symmetric under translation and spatial inversion. For $u>0$, it is instead the sector symmetric under translation by two sites (due to odd and even terms of the Hamiltonian being different) and under the combination of spatial inversion and translation by one site. As shown in Fig.~\ref{fig:levelspacing}, while for $u=0$ the average gap ratio  is not yet at the expected value for a chaotic system due to the slow convergence with system size of the PXP model~\cite{TurnerNature}, $\left\langle r\right\rangle $ quickly approaches $0.53$ upon increasing $u$. This matches the prediction for the Gaussian Orthogonal Ensemble (GOE). As expected, the gap ratio value then starts to dip again near the singular point $u=1$. We note that the dip of $\langle r \rangle$  shortly after $u=0$ is due to the translation symmetry being only weakly broken and is expected to disappear in the thermodynamic limit. 

In addition to the level spacing ratio, we also show the full distribution of level spacings after unfolding for $u=0.8$ in Fig.~\ref{fig:levelspacing}, which again aligns well with the GOE expectation. In both panels of Fig.~\ref{fig:levelspacing}, we also show the prediction for an integrable system (Poisson) and for the Gaussian unitary ensemble (GUE), neither aligning with our data. Overall, the data is strongly suggestive that our model is robustly chaotic in the non-Hermitian regime as long as $u$ is not close to 1. The match with the GOE prediction also indicates that the Hermitian matrix with the same spectrum as our non-Hermitian one is invariant under time-reversal.

\end{document}